\documentclass[10pt,journal]{IEEEtran}
\usepackage{graphicx}
\usepackage{subfigure}
\usepackage{amsmath}
\usepackage{mathrsfs}
\usepackage{amssymb}
\usepackage{bm}
\usepackage{color}
\usepackage{enumerate}
\usepackage{algorithm}
\usepackage{algorithmic}
\DeclareMathOperator*{\argmin}{arg\,min}

\newtheorem{theorem}{Theorem}

\ifCLASSOPTIONcompsoc
\else
\fi

\ifCLASSINFOpdf
\else
\fi

\begin{document}
%
\title{SLRMA: Sparse Low-Rank Matrix Approximation for Data Compression}

\author{Junhui Hou,~\IEEEmembership{Student Member,~IEEE,}
        Lap-Pui Chau,~\IEEEmembership{Senior Member,~IEEE,}
        Nadia Magnenat-Thalmann, \\
        and Ying He,~\IEEEmembership{Member,~IEEE}
\thanks{This research, which is carried out at BeingThere Centre, collaboration among IMI of Nanyang Technological University (NTU) Singapore, ETH Zürich, and UNC Chapel Hill, is supported by the Singapore National Research Foundation (NRF) under its International Research Centre @ Singapore Funding Initiative and administered by the Interactive Digital Media Programme Office (IDMPO).
Ying He is partially supported by MOE2013-T2-2-011 and MOE RG23/15.}
\thanks{J. Hou and L.-P. Chau are with the School of Electrical and Electronics Engineering,
Nanyang Technological University, Singapore, 639798. Email:
\{houj0001$|$elpchau\}@ntu.edu.sg}
\thanks{N. Magnenat-Thalmann is with the Institute
for Media Innovation, Nanyang Technological University, Singapore,
639798. Email: nadiathalmann@ntu.edu.sg}
\thanks{Y. He is with the
School of Computer Engineering, Nanyang Technological University,
Singapore, 639798. Email:yhe@ntu.edu.sg}}

%
%

\markboth{}%
{\MakeLowercase{\textit{et al.}}: }

\IEEEtitleabstractindextext{%
\begin{abstract}
  Low-rank matrix approximation (LRMA) is a powerful technique for signal processing and pattern analysis.
  However, its potential for data compression has not yet been fully investigated in the literature.
  In this paper, we propose \emph{sparse} low-rank matrix approximation (SLRMA), an effective computational tool for data compression.
  SLRMA extends the conventional LRMA by exploring both the intra- and inter-coherence of data samples \emph{simultaneously}.
  With the aid of prescribed orthogonal transforms (e.g., discrete cosine/wavelet transform and graph transform),
  SLRMA decomposes a matrix into a product of two smaller matrices,
  where one matrix is made of extremely sparse and orthogonal column vectors, and the other consists of the transform coefficients.
  Technically, we formulate SLRMA as a constrained optimization problem, i.e., minimizing the approximation error in the least-squares sense regularized by $\ell_0$-norm and orthogonality,
  and solve it using the inexact augmented Lagrangian multiplier method.
  Through extensive tests on real-world data, such as 2D image sets and 3D dynamic meshes,
  we observe that
  (\lowercase\expandafter{\romannumeral1}) SLRMA empirically converges well;
  (\lowercase\expandafter{\romannumeral2}) SLRMA can produce approximation error comparable to LRMA but in a much sparse form;
  (\lowercase\expandafter{\romannumeral3}) SLRMA-based compression schemes significantly outperform the state-of-the-art in terms of rate-distortion performance.
\end{abstract}

}

\maketitle \IEEEdisplaynontitleabstractindextext
\IEEEpeerreviewmaketitle

\section{Introduction}
\begin{figure}
\centering
 \includegraphics[width=3.2in]{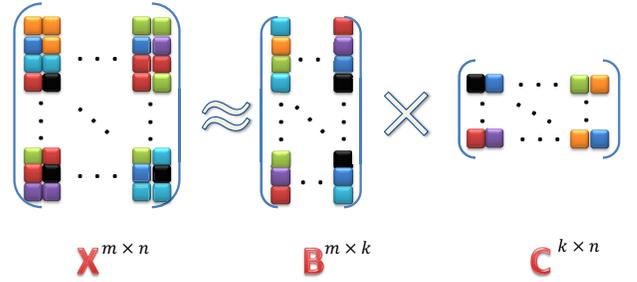}
  \caption{Illustration of low-rank matrix approximation.
  $k\ll\min(m,n)$ and
  $\mathbf{B}^\textsf{T}\mathbf{B}=\mathbf{I}_k$; Each column of $\mathbf{X}$
  corresponds to one data sample.}
  \label{fig:LRMA}
\end{figure}

  \IEEEPARstart{G}{iven} a matrix $\mathbf{X}\in \mathbb{R}^{m\times n}$ corresponding to $n$ data samples in $\mathbb{R}^m$,
  low-rank matrix approximation (LRMA) (a.k.a. principal component analysis (PCA) and subspace factorization)
  seeks a matrix $\widehat{\mathbf{X}}\in \mathbb{R}^{m\times n}$ of rank $k\ll\min(m,n)$ that best approximates $\mathbf{X}$ in the least-squares sense \cite{halko2011finding}.
  Alternatively, the rank constraint can be implicitly expressed in a factored form,
  i.e., $\mathbf{X}\approx \widehat{\mathbf{X}}= \mathbf{B}\mathbf{C}$ where $\mathbf{B}\in \mathbb{R}^{m\times k}$ and $\mathbf{C}\in \mathbb{R}^{k\times n}$ (see Figure \ref{fig:LRMA}).
  This decomposition is not unique, since $(\mathbf{BA})(\mathbf{A}^{-1}\mathbf{C})=\mathbf{BC}$ holds for any invertible matrix $\mathbf{A}\in \mathbb{R}^{k\times k}$.
  To reduce the solution space, one often requires the decomposed matrix $\mathbf{B}$ to be column-orthogonal.

  Since LRMA is able to reveal the inherent structure of the input data, it has been applied to a wide spectrum of engineering applications,
  including data compression \cite{gu20122dsvd},
  \cite{Hou2015compressing}, background subtraction \cite{candes2011robust}, \cite{wen2014joint}, classification and clustering \cite{liu2013robust}, \cite{zhang2013learning},
  image/video restoration and denoising \cite{ji2011robust}, \cite{wang2013robust}, image alignment and interpolation \cite{peng2012rasl}, \cite{cao2015image},  structure from motion \cite{dai2013projective}, \cite{dai2014simple}, \cite{meng2013cyclic}, etc.
  We refer the readers to the survey paper \cite{zhou2014low} for more details.

\begin{figure*}
\centering \makebox[1.2in]{\footnotesize
1$^{st}$}\makebox[1.2in]{\footnotesize
2$^{nd}$}\makebox[1.2in]{\footnotesize
3$^{rd}$}\makebox[1.2in]{\footnotesize
4$^{th}$}\makebox[1.2in]{\footnotesize 5$^{th}$} \subfigure[Column
vectors of $\mathbf{B}$ in LRMA]{
\includegraphics[width=1.2in]{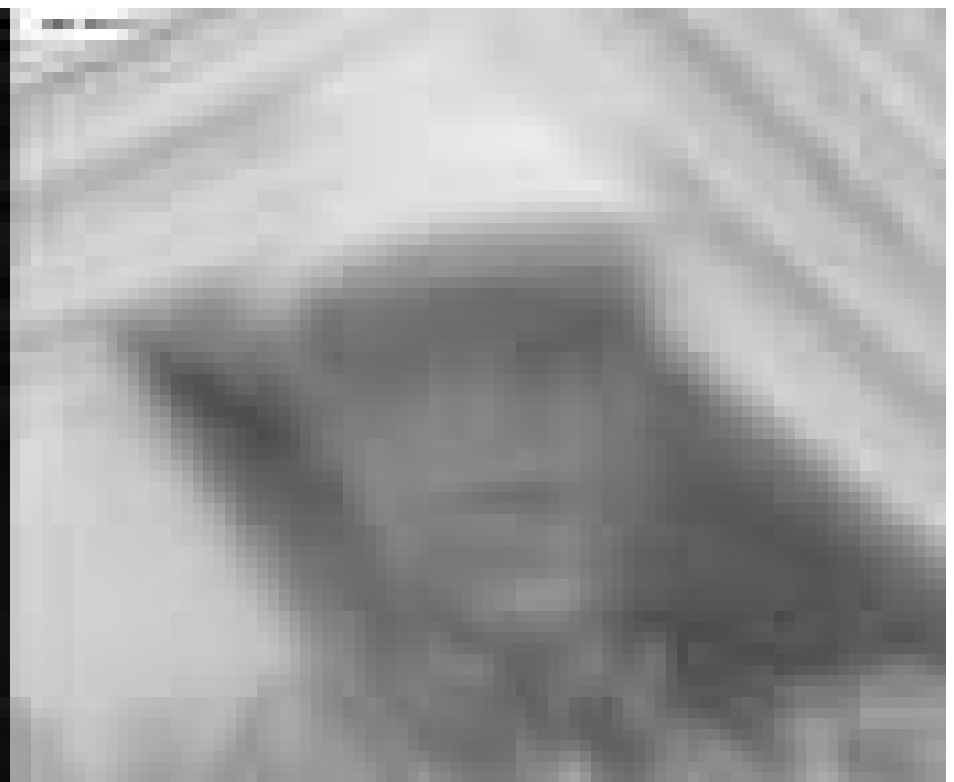}
\includegraphics[width=1.2in]{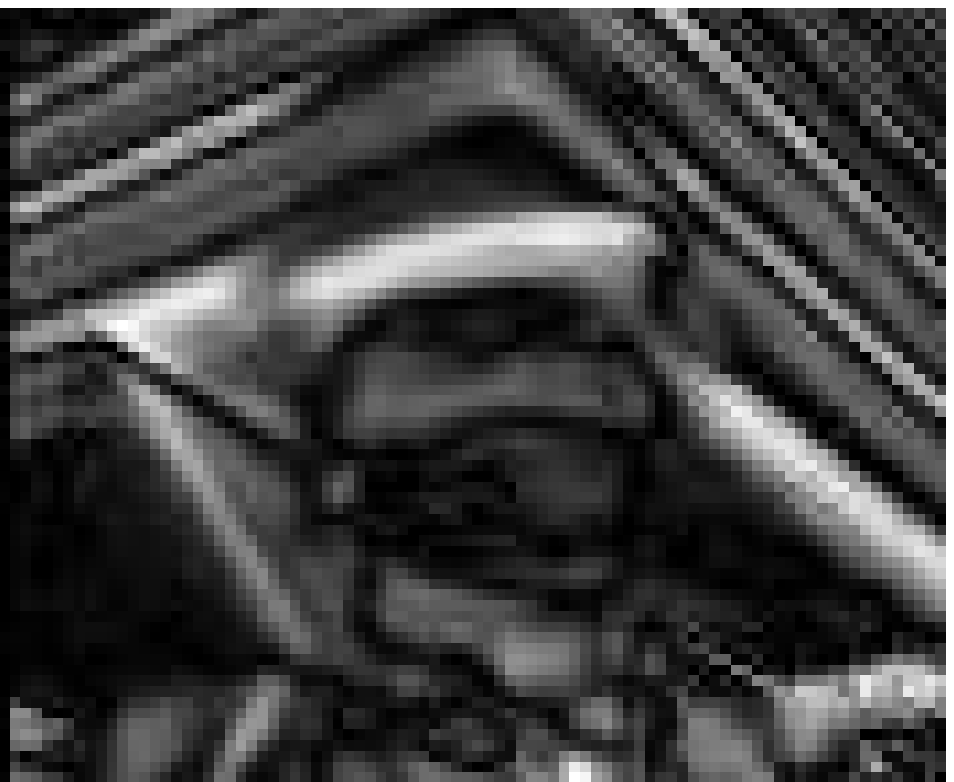}
\includegraphics[width=1.2in]{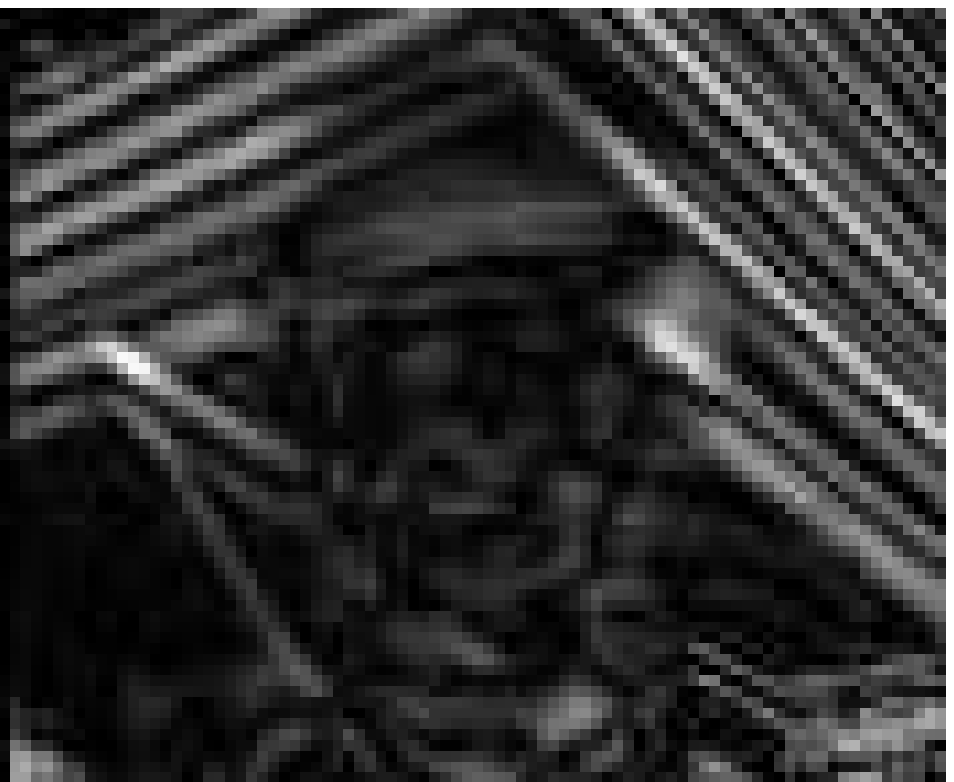}
\includegraphics[width=1.2in]{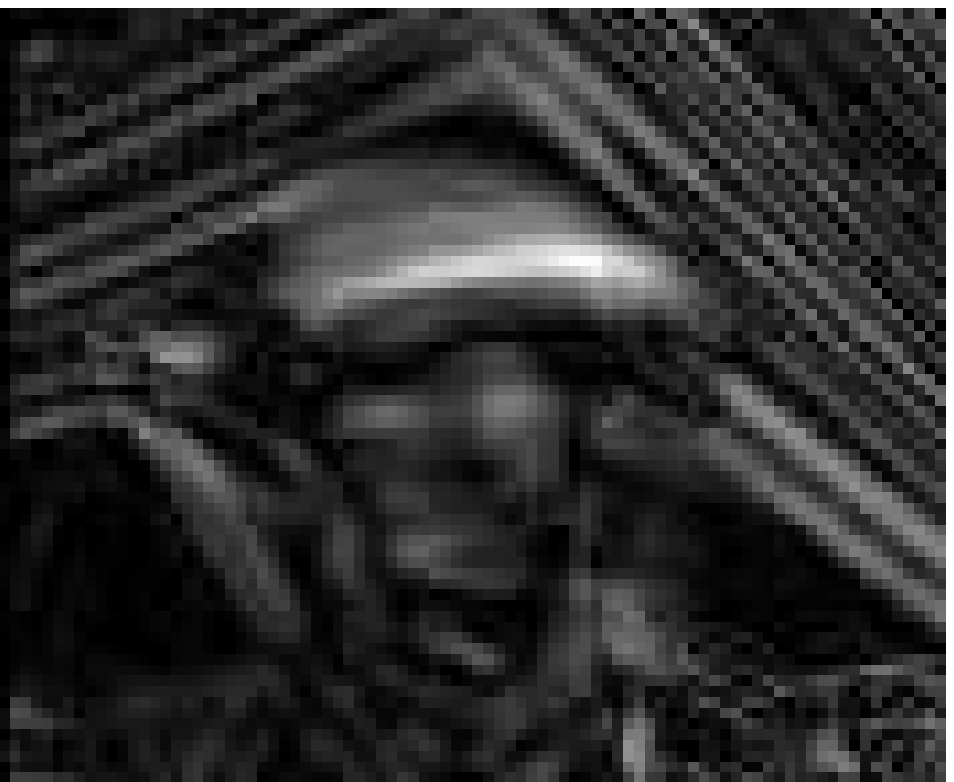}
\includegraphics[width=1.2in]{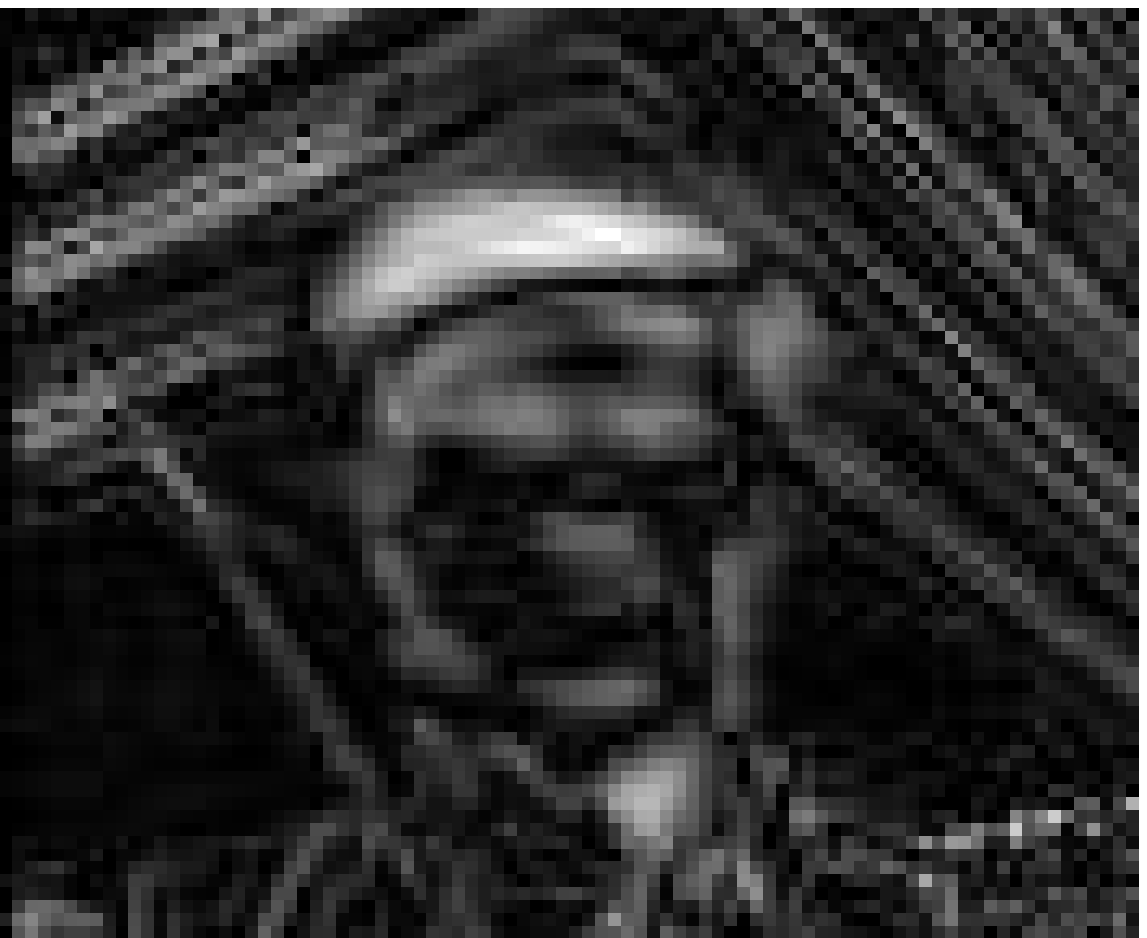}}
\makebox[1.2in]{\footnotesize 1$^{st}$}\makebox[1.2in]{\footnotesize
2$^{nd}$}\makebox[1.2in]{\footnotesize
3$^{rd}$}\makebox[1.2in]{\footnotesize
4$^{th}$}\makebox[1.2in]{\footnotesize 5$^{th}$} \subfigure[Column
vectors of $\mathbf{B}$ in SLRMA]{
\includegraphics[width=1.2in]{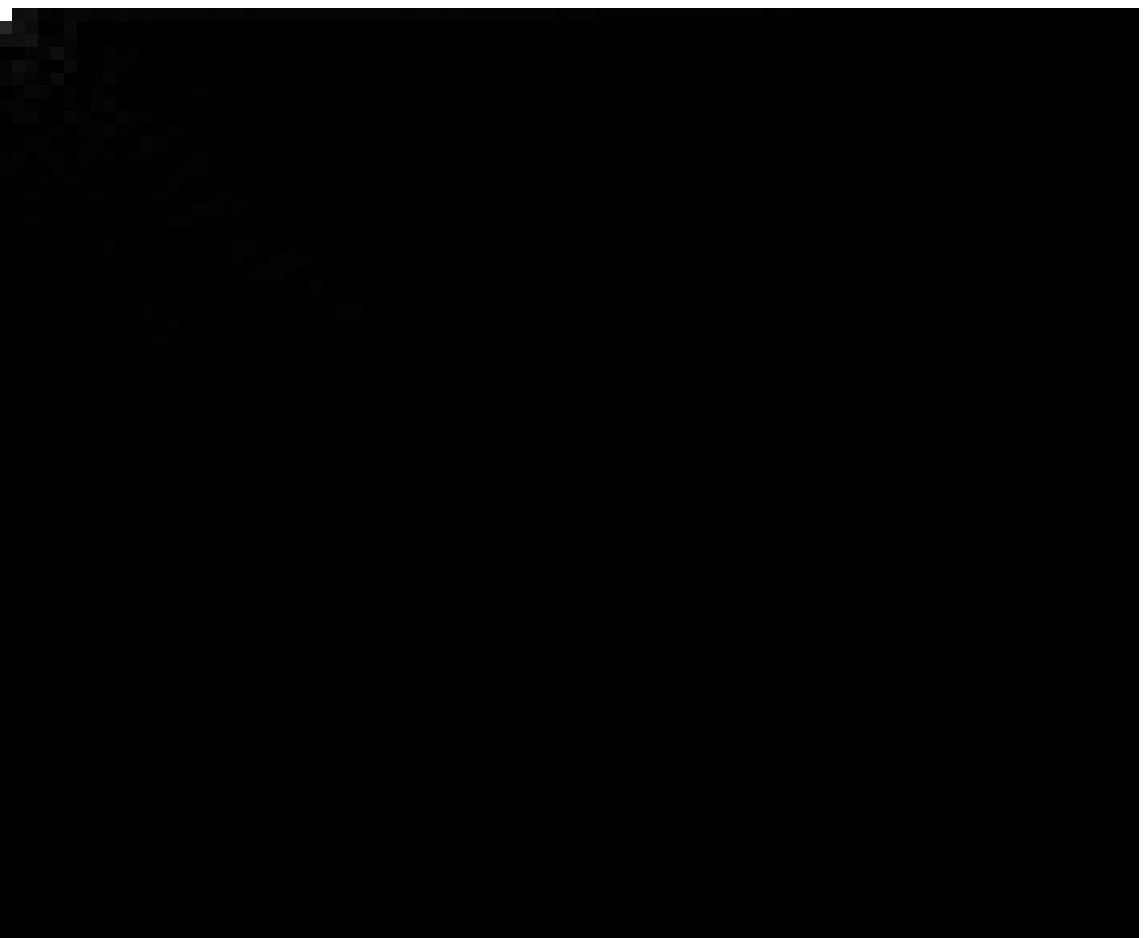}
\includegraphics[width=1.2in]{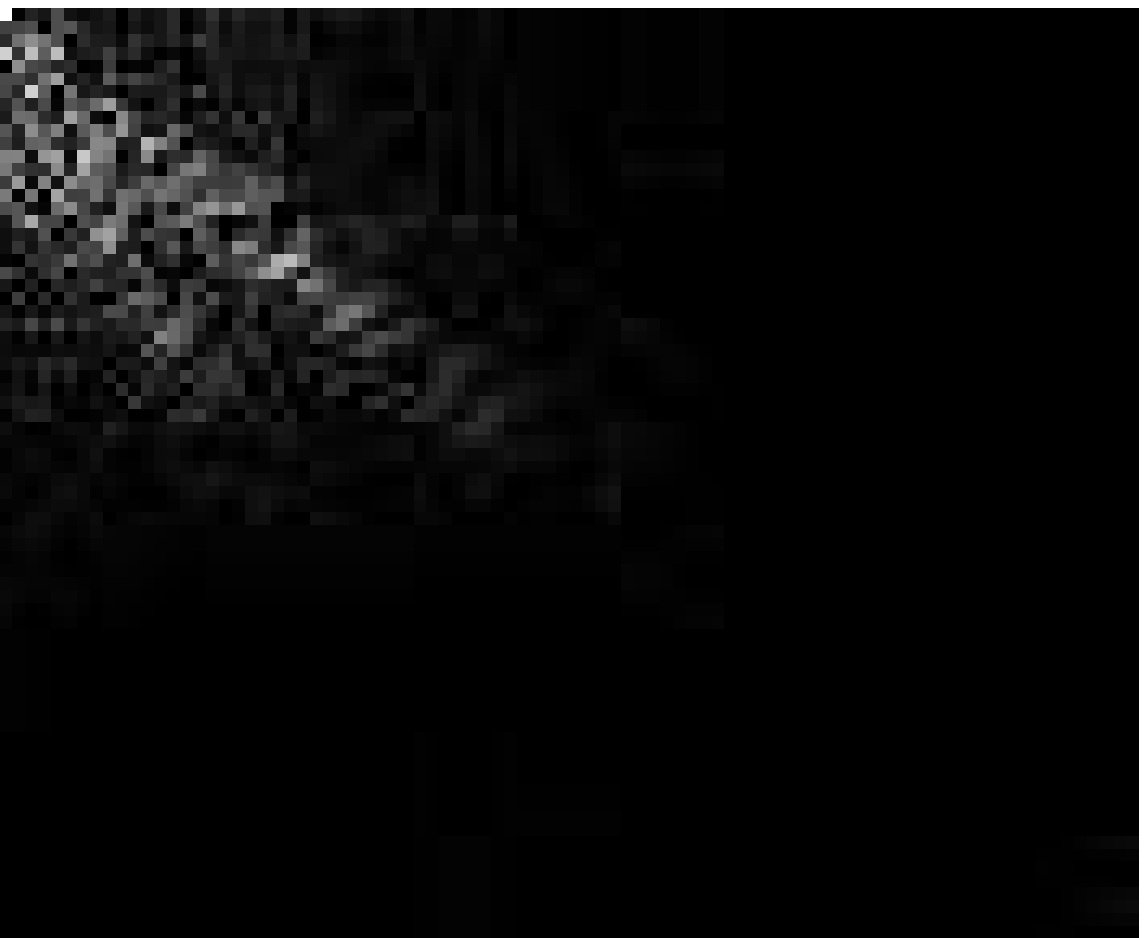}
\includegraphics[width=1.2in]{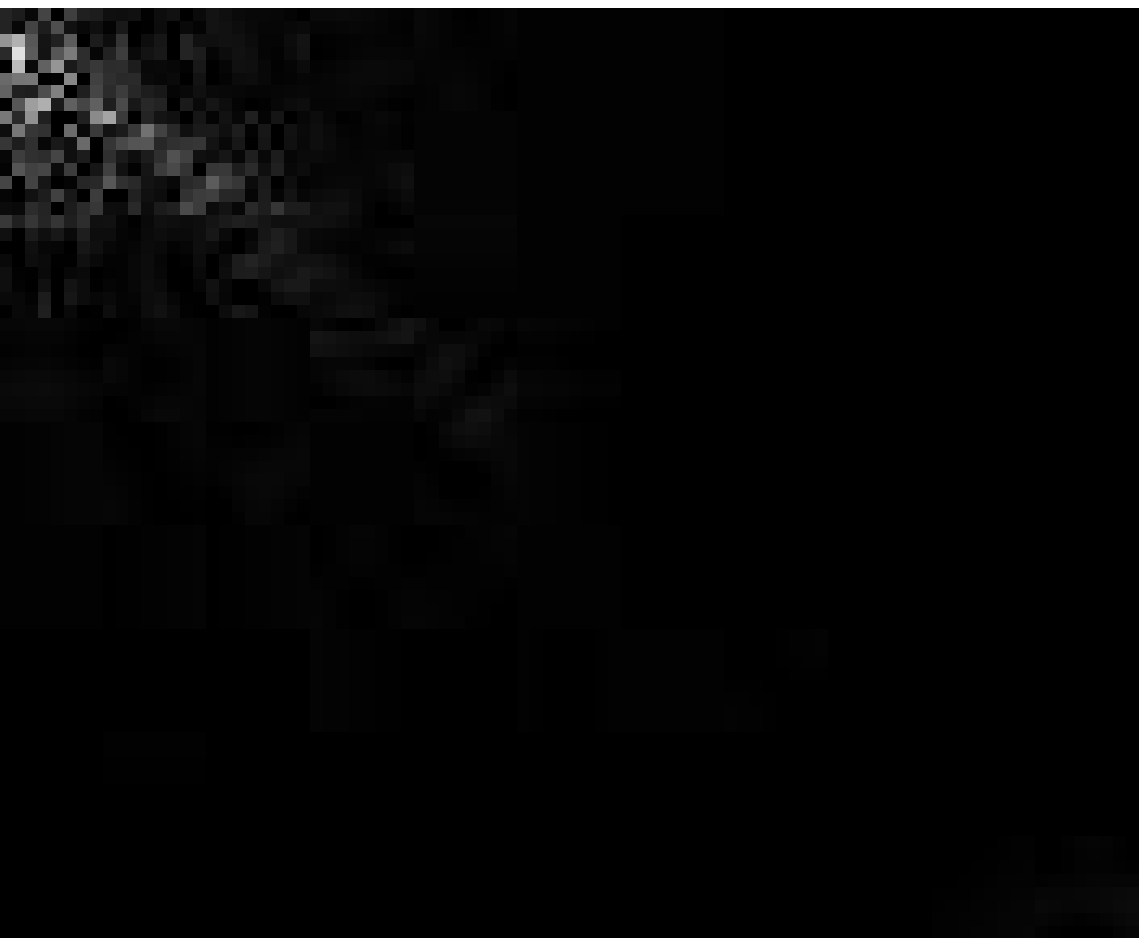}
\includegraphics[width=1.2in]{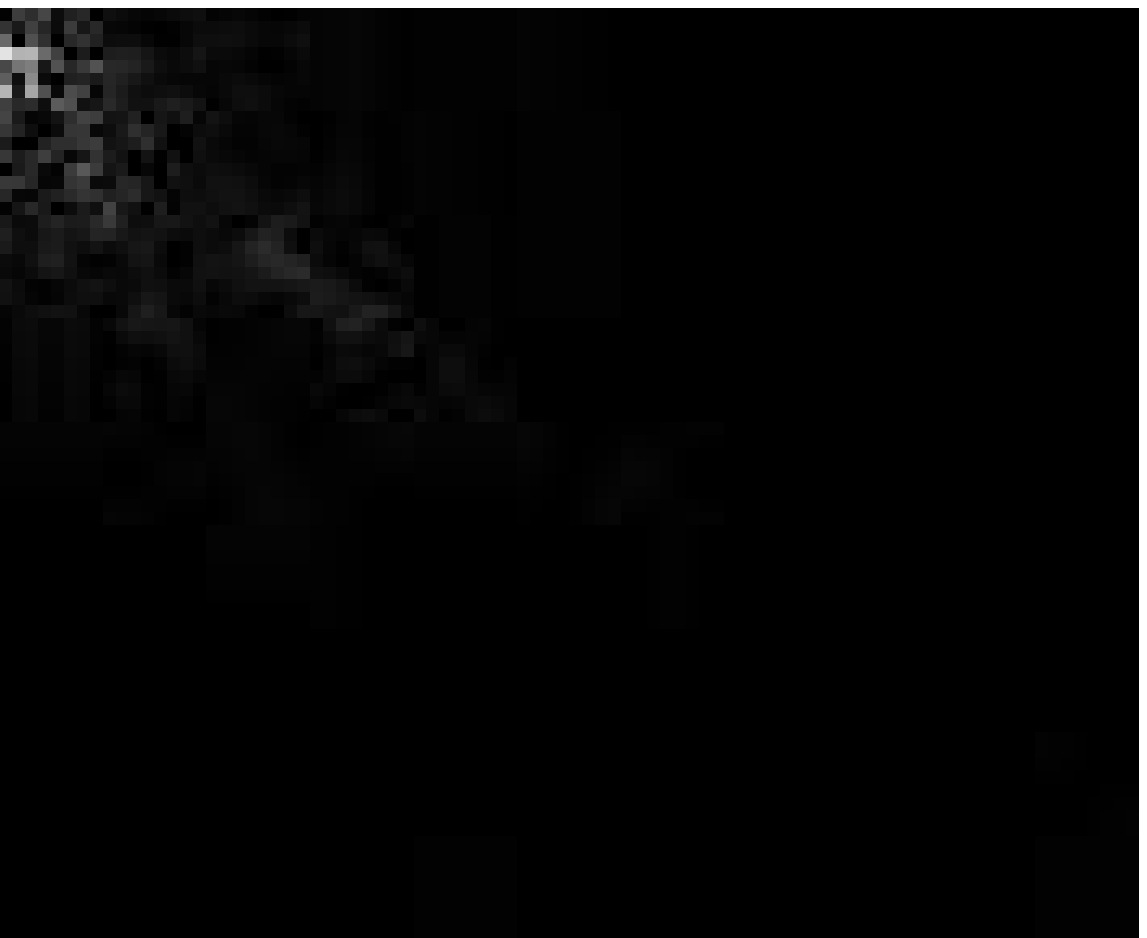}
\includegraphics[width=1.2in]{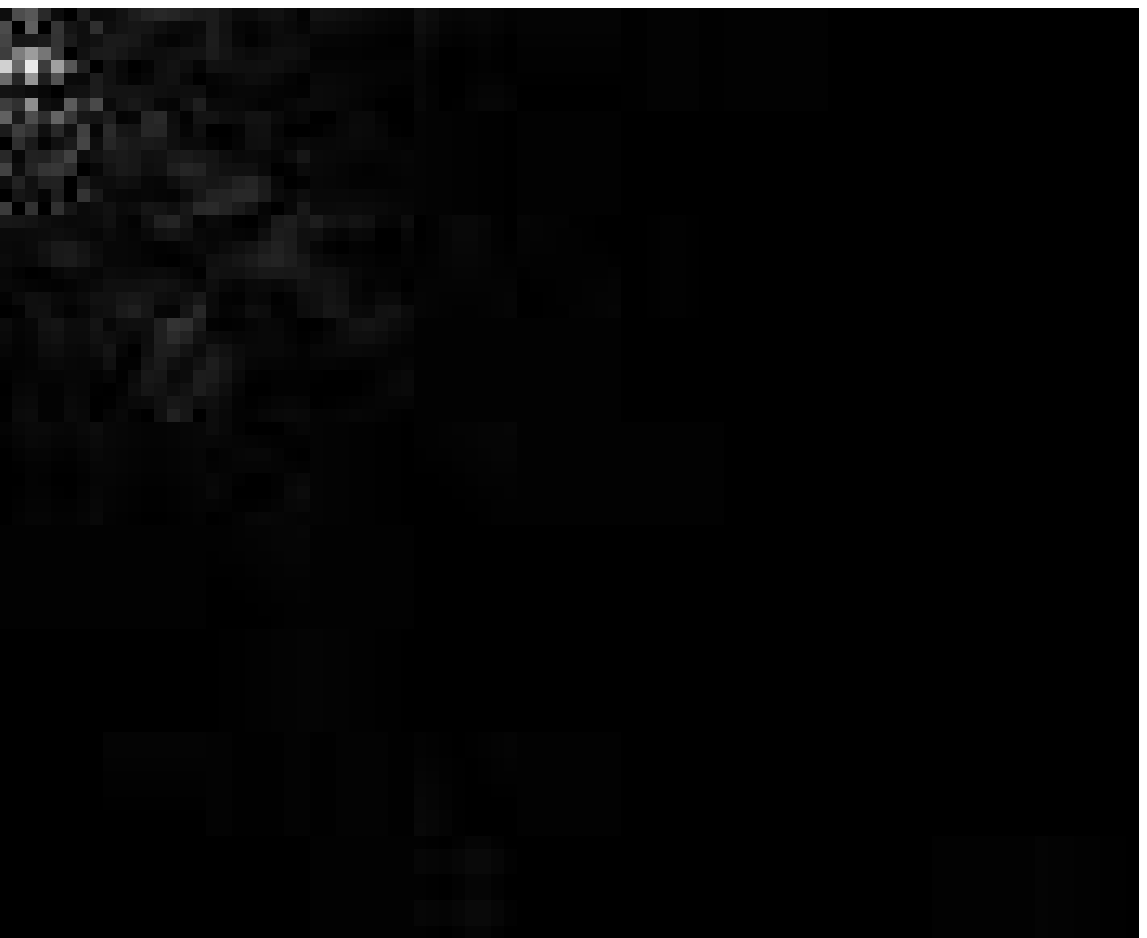}}
\makebox[1.2in]{\footnotesize 1$^{st}$}\makebox[1.2in]{\footnotesize
2$^{nd}$}\makebox[1.2in]{\footnotesize
3$^{rd}$}\makebox[1.2in]{\footnotesize
4$^{th}$}\makebox[1.2in]{\footnotesize 5$^{th}$} \subfigure[Column
vectors of $\breve{\mathbf{B}}$ in Equation (\ref{equ:original 1D CLRMA}).]{
\includegraphics[width=1.2in]{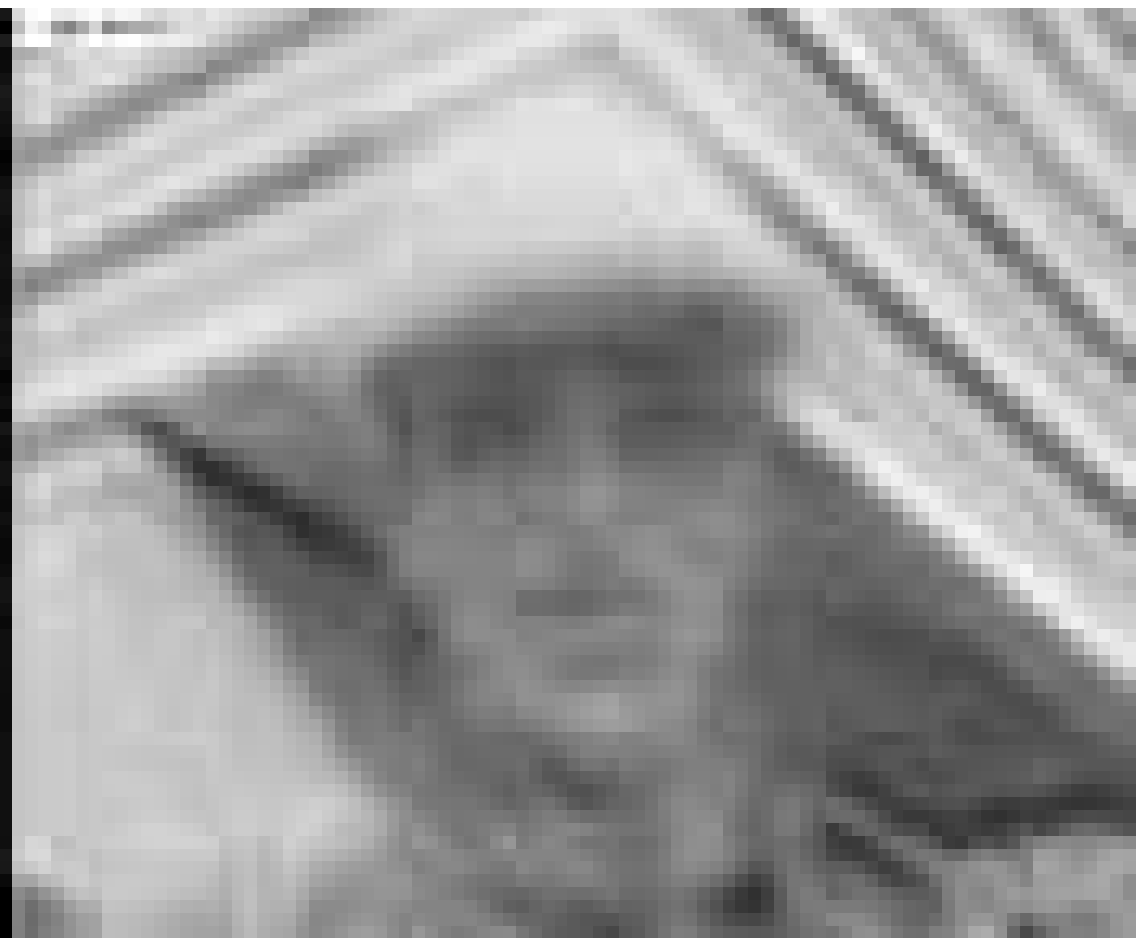}
\includegraphics[width=1.2in]{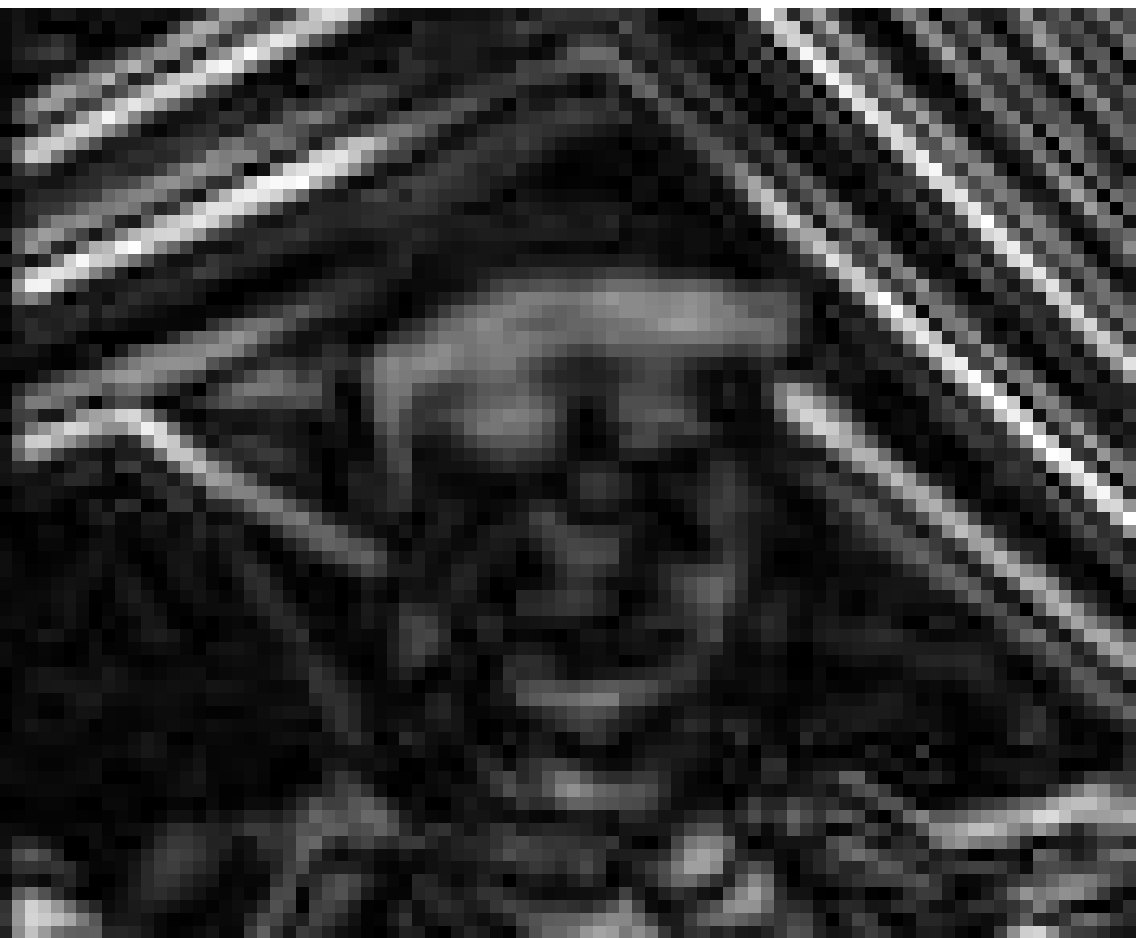}
\includegraphics[width=1.2in]{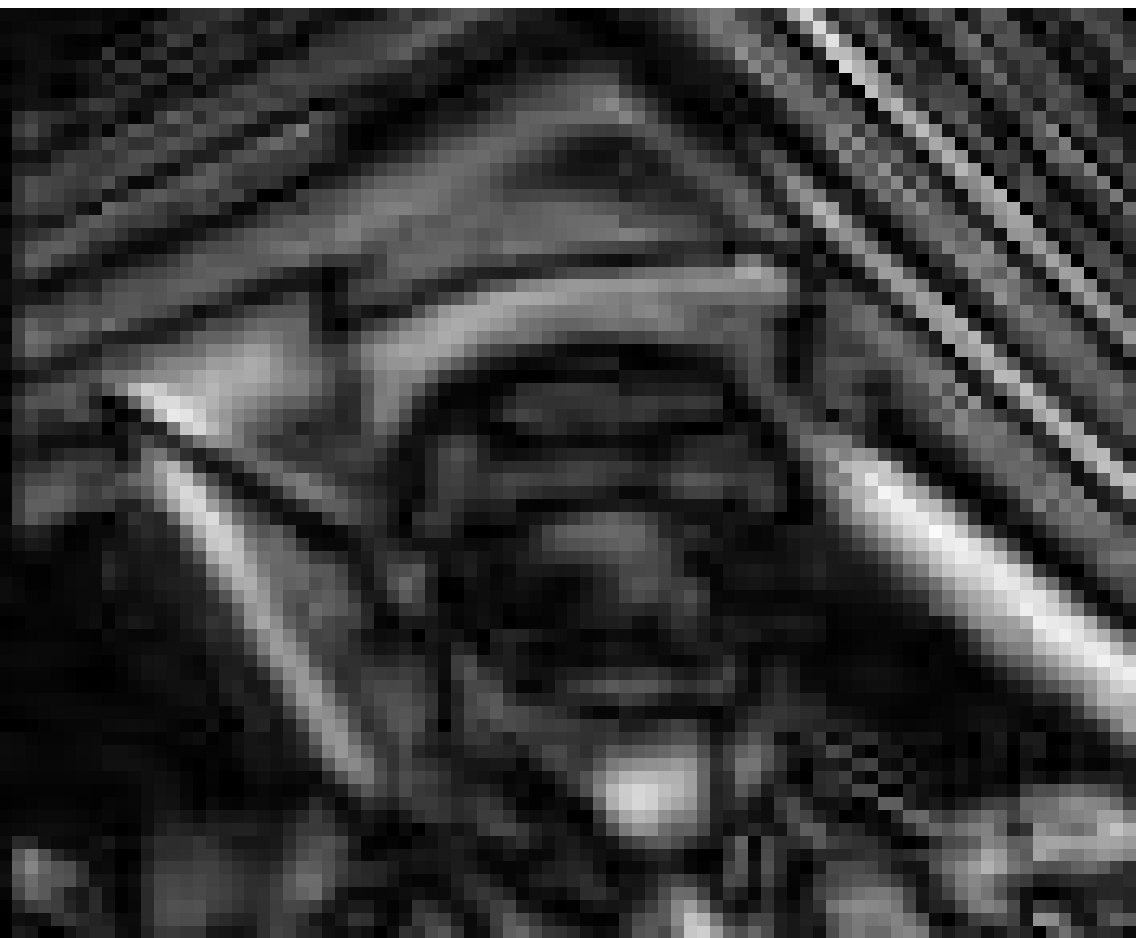}
\includegraphics[width=1.2in]{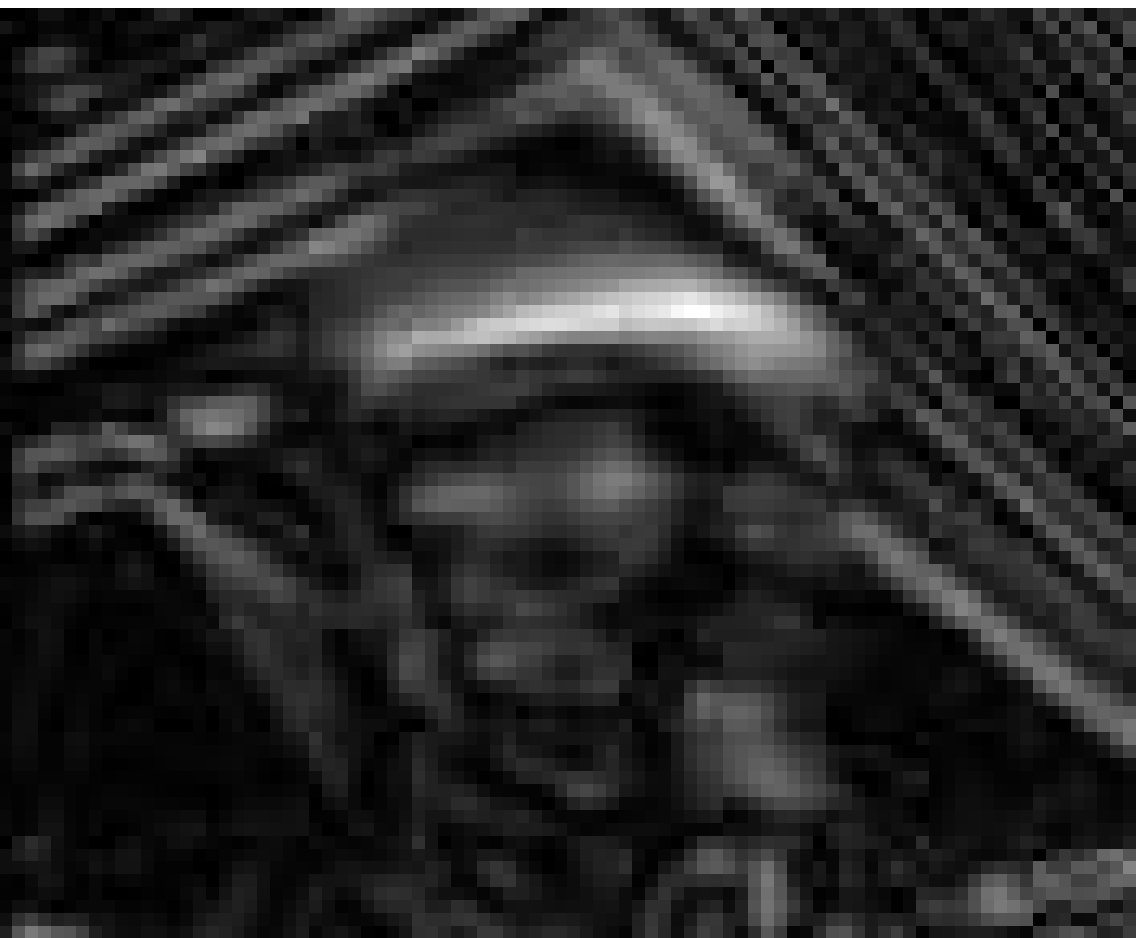}
\includegraphics[width=1.2in]{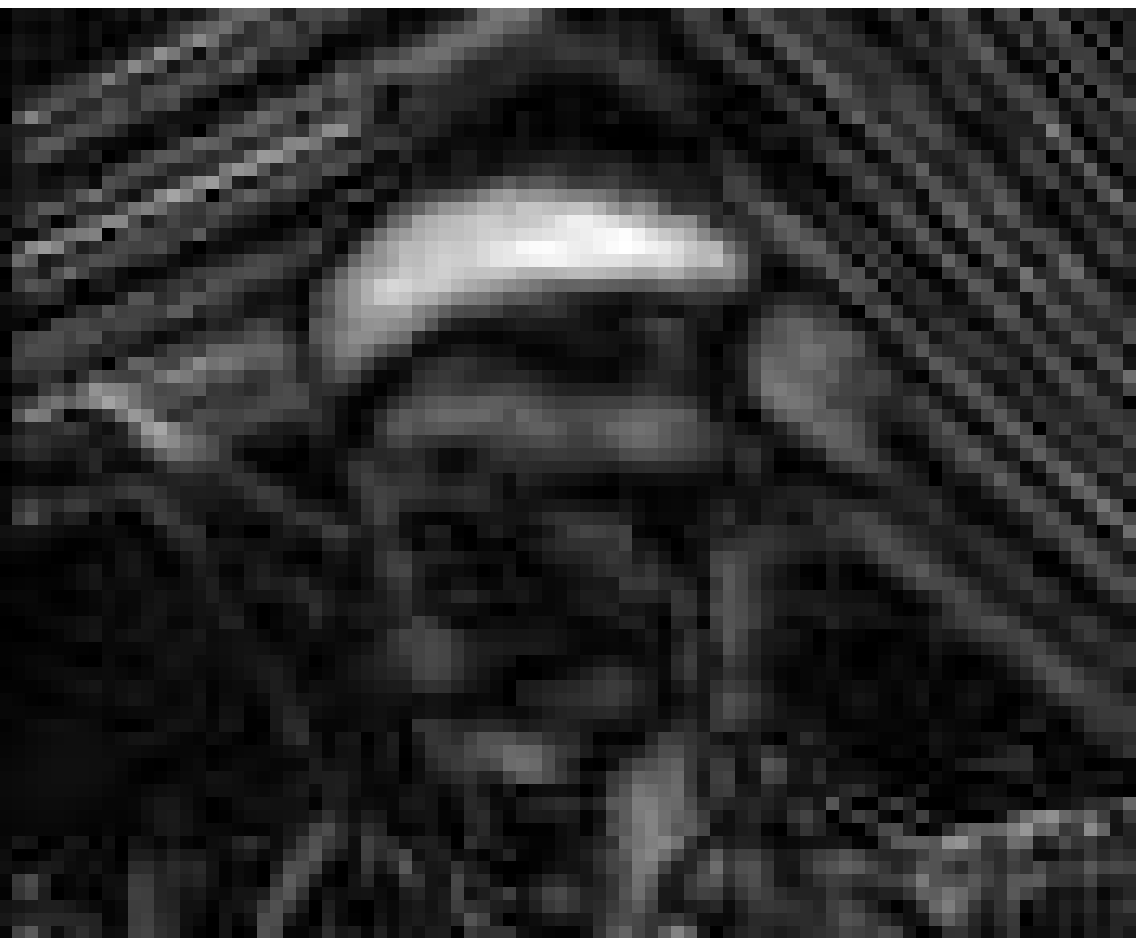}}
  \caption{Comparison of LRMA and SLRMA in terms of intra-correlation in $\bf B$.
  The test image set contains 150 frames of the ``foreman" video of resolution $88\times 72$,
  representing in a matrix $\mathbf{X}\in\mathbb{R}^{6336\times
  150}$. Some samples are shown in Figure \ref{fig:datasamples}(a).
  We visualize the first five column vectors of $\mathbf{B}$ for LRMA and SLRMA in (a) and (b).
  Both LRMA and SLRMA approximate matrix $\mathbf X$ with the same approximation error.
  In SLRMA, we set $\bf\Phi$ the 2D DCT matrix.
  } \label{fig:problem visual}
\end{figure*}

  Among the above-mentioned applications, we are particularly interested in data compression.
  At the first glance, applying LRMA to data compression seems to be straightforward,
  since one only needs to store $k(m+n)$ elements, with small approximation error introduced in LRMA.
  Such an idea has been used extensively to compress various types of data,
  e.g., images/videos~\cite{gu20122dsvd,yang1995combined,ochoa2003hybrid,du2007hyperspectral,chen2015incremental,HoufacialGV,Hou2015compressing},
  3D motion data~\cite{alexa2000representing,karni2004compression,vavsa2010geometry,vavsa2014dynamiccompressing,Hou2014tvcg},
  traffic data \cite{li2007flow,asif2013data,Asif2014near}.
  However, data samples usually exhibit both \emph{intra}-coherence (i.e., coherence within each data sample) and \emph{inter}-coherence
  (i.e., coherence among different data samples).
  LRMA can exploit the inter-coherence well, i.e., using $\mathbf{B}$ with much smaller orthogonal columns to represent $\mathbf{X}$,
  but it fails to address the intra-coherence in the columns of $\bf B$,
  and hereby compromises the compression performance~\cite{alexa2000representing}, \cite{karni2004compression}, \cite{Hou2014tvcg}, \cite{li2007flow}, \cite{asif2013data}, \cite{Asif2014near}.
  Figure \ref{fig:problem visual}(a) shows the problem using a 2D image set.
  One can clearly see that the columns of $\mathbf{B}$ produced by LRMA are locally smooth (the bright areas), indicating the strong coherence.

\begin{figure*}[t]
\centering
\includegraphics[width=4.5in]{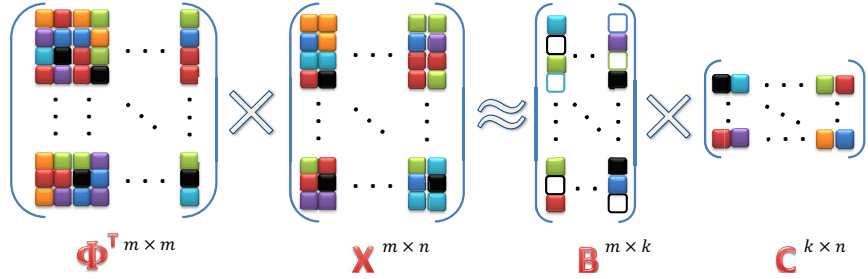}
  \caption{Illustration of sparse LRMA.
  ${\bf\Phi}$ is a prescribed orthogonal transform, such as DCT, DWT and GT, where each column corresponds to a frequency.
  The column vectors in matrix $\mathbf{B}$ are sparse and orthogonal, i.e., $\mathbf{B}^\textsf{T}\mathbf{B}=\mathbf{I}_k$.
  The empty boxes in $\mathbf{B}$ are zeros.}
\label{fig:CLRMA}
\end{figure*}

  In this paper, we propose sparse low-rank matrix approximation (SLRMA) for data compression.
  In contrast to the existing methods, which usually explore intra- and inter-coherence separately (see Section \ref{sec:related}),
  SLRMA is able to explore both the intra- and inter-coherence of data samples simultaneously from the perspective of optimization and transform.
  As Figure \ref{fig:CLRMA} shows, SLRMA multiplies a prescribed
  orthogonal matrix $\mathbf{\Phi}\in \mathbb{R}^{m\times m}$ (such as
  discrete consine/wavelet transform (DCT/DWT) and graph transform) to the input
  matrix $\bf X$ and then factors $\mathbf{\Phi}^\textsf{T}\mathbf{X}$
  into a product of the extremely sparse and column-orthogonal matrix
  $\mathbf{B}$ and the coefficient matrix $\mathbf{C}$.
  We formulate SLRMA as a constrained optimization problem, i.e., minimizing the approximation error in least-squares sense under the $\ell_0$-norm and orthogonality constraints,
  and solve it using the inexact augmented Lagrangian multiplier method.

  Through extensive tests on real-world data, such as 2D image sets and 3D dynamic meshes,
  we observe that
  (\lowercase\expandafter{\romannumeral1}) SLRMA empirically converges well;
  (\lowercase\expandafter{\romannumeral2}) SLRMA can produce comparable approximation error as LRMA but in a much sparser form;
  (\lowercase\expandafter{\romannumeral3}) SLRMA-based compression schemes outperform the state-of-the-art scheme to a large
  extent in terms of rate-distortion performance.
Figure~\ref{fig:problem visual}(b) visualizes the column vectors of
$\mathbf{B}$ of the proposed SLRMA which produces the same
approximation error as that of LRMA in Figure~\ref{fig:problem
visual}(a), where one can clearly see that column vectors of SLRMA
do not exhibit such intra-coherence.

  The rest of this paper is organized as follows:
  Section~\ref{sec:related} reviews the related work on LRMA-based data compression.
  Section~\ref{sec:preliminaries} briefly introduces LRMA and graph transform.
  Sections \ref{sec:CLRMA} presents the sparse LRMA algorithm followed by experimental results in Section \ref{sec:verification}.
  Finally, Section \ref{sec:con} concludes this paper and points out several promising future directions.

  \textbf{Notations.} Throughout this paper, scalars are denoted by italic lower-case letters, vectors by bold lower-case letters and matrices by upper-case ones, respectively.
  For instance, we consider a matrix $\mathbf{A}\in\mathbb{R}^{m\times n}$. The $i$-th row and $j$-th column of $\bf A$ are represented by $\mathbf{a}^i\in \mathbb{R}^{1\times n}$ and $\mathbf{a}_j\in \mathbb{R}^{m\times 1}$, respectively.
  Let $a_{ij}$ be the $(i,j)$-th entry of $\mathbf{A}$ with its absolute being $|a_{ij}|$. $\mathbf{A}^\textsf{T}$ and $\mathbf{A}^\dag$ are the transpose and pseudoinverse of $\mathbf{A}$.
  We denote by $\|\mathbf{A}\|_F=\sqrt{\sum_{i=1}^m\sum_{j=1}^na_{ij}^2}$ and $\|\mathbf{A}\|_{\infty}=\max_{i,j}{|a_{ij}|}$ the Frobenious norm and the $\ell_{\infty}$ norm of $\mathbf{A}$,
  respectively.
  The $\ell_0$-norm $\|\mathbf{A}\|_0$ counts the number of nonzero entries in $\mathbf{A}$.
  Let ${\rm Tr}(\mathbf{A})=\sum_{i}a_{ii}$ be the trace of $\bf A$.
  $\mathbf{I}_k$ is the identity matrix of size $k\times k$.

\section{Related Work}
\label{sec:related}
  In this section, we briefly review the LRMA-based methods for data compression.
  Yang and Lu \cite{yang1995combined} represented the basis vectors using vector quantization for image coding.
  Similar techniques were also adopted in video compression \cite{gu20122dsvd}, \cite{ochoa2003hybrid}.
  Note that training coodbooks is computationally expensive due to the large amount of training data.
  Moreover, the trained codebooks may be biased, leading to large approximation error to the data if they are fundamentally different than the training data.
  Du \emph{et al}. \cite{du2007hyperspectral} applied LRMA to hyperspectral images to remove the spectral coherence, then adopted JP2K \cite{skodras2001jpeg},
  the standard image encoder, to further compress the basis vectors.
  Hou \emph{et al}. \cite{Hou2015compressing} extracted a few keyframes from geometry videos and compressed them using a standard video encoder H.264/AVC
\cite{H264overview}, and the whole geometry video was reconstructed by linearly interpolating the decompressed keyframes.
  Chen \emph{et al}. \cite{chen2015incremental} and Hou \emph{et al}. \cite{HoufacialGV} also presented robust LRMA \cite{candes2011robust} and video encoders combined compression
 frameworks for surveillance videos and geometry videos, respectively.

  3D triangle meshes are a simple and flexible tool to represent complex 3D objects for graphics applications ranging from modeling, animation to rendering.
  A mesh consists of structural and geometric data.
  Dynamic meshes are the sequence of triangle meshes with the same connectivity, representing a deformable model with time-varying geometry.
  Alex and M\"{u}ller \cite{alexa2000representing} adopted LRMA to compress 3D dynamic meshes,
  representing each frame as the linear combination of few basis vectors.
  Karni and Gostamn \cite{karni2004compression} improved this scheme by applying second-order linear predictive coding to coefficients to further remove the temporal coherence.
  Heu \emph{et al}. \cite{heu2009snr} proposed an adaptive bit plane coder to encode the basis vectors to achieve progressive transmission.
  However, these methods cannot eliminate the spatial (or intra-) coherence.
  Unlike \cite{alexa2000representing,karni2004compression,heu2009snr}, \emph{trajectory-LRMA} was employed in \cite{vavsa2010geometry,vavsa2014dynamiccompressing,sattler2005simple,libor2007coddyac,vavsa2009cobra},
  which represents trajectories of vertices as a linear combination of few basis vectors.
  Sattler \emph{et al}. \cite{sattler2005simple} proposed clustered LRMA to compress 3D dynamic meshes,
  in which clustering of trajectories and trajectory-LRMA were performed simultaneously to explore the spatial and temporal correlation.
  V\'{a}s\v{a} \emph{et al}. ~\cite{vavsa2009cobra} studied the trajectory-LRMA-based compression for 3D dynamic meshes systematically.
  For example, to remove the coherence among basis vectors, they proposed predictive coding to encode them.
  They also introduced three types of predictions to exploit the coherence among coefficients,
  namely parallelogram-based prediction \cite{libor2007coddyac}, least squares prediction, and radial basis function-based prediction \cite{vavsa2010geometry}.
  Recently, they used the discrete geometric Laplacian of a computed average surface to encode the coefficients \cite{vavsa2014dynamiccompressing},
  and achieved the state-of-the-art rate-distortion performance. But
  this method requires the models of sequences to be manifolds due to the process of computing an average surface,
  limiting its range of applications.

\section{Preliminaries}
\label{sec:preliminaries}

\subsection{Low-Rank Matrix Approximation}
\label{subsec:LRMA}
  Given $\mathbf{X}\in \mathbb{R}^{m\times n}$, the LRMA problem can be mathematically formulated as
  \begin{align}
  &\min_{\mathbf{B}\in \mathbb{R}^{m\times k}\atop \mathbf{C}\in \mathbb{R}^{k\times n}}
  \|\mathbf{X}-\mathbf{B}\mathbf{C}\|_F^2\nonumber \\
  &\mathrm{subject}~\mathrm{to}~~\mathbf{B}^\textsf{T} \mathbf{B}=\mathbf{I}_k. \label{equ:LRA2}
  \end{align}
  Setting the derivative of the objective function of $\bf C$ to zero, we obtain $\mathbf{C}=\mathbf{B}^\textsf{T} \mathbf{X}$.
  Since $\|\mathbf{A}\|_F^2={\rm Tr}(\mathbf{A}\mathbf{A}^\textsf{T})$ and ${\rm Tr}(\mathbf{A})={\rm Tr}(\mathbf{A}^\textsf{T})$, we can expand the objective function as
  \begin{align}
  &\|\mathbf{X}-\mathbf{B}\mathbf{C}\|_F^2\nonumber \\
  &={\rm Tr}\left\{(\mathbf{X}-\mathbf{B}\mathbf{C})(\mathbf{X}-\mathbf{B}\mathbf{C})^\textsf{T}\right\}\nonumber\\
  &={\rm Tr}\left(\mathbf{X}\mathbf{X}^\textsf{T}\right)-2{\rm Tr}\left(\mathbf{B}\mathbf{C}\mathbf{X}^\textsf{T}\right)+{\rm Tr}\left(\mathbf{B}\mathbf{C}\mathbf{C}^\textsf{T}\mathbf{B}^\textsf{T}\right).
  \label{equ:extension of LRMA}
  \end{align}
  Substituting $\mathbf{C}=\mathbf{B}^\textsf{T} \mathbf{X}$ into (\ref{equ:extension of LRMA}) and dropping the constant term,
  we obtain the equivalent problem of (\ref{equ:LRA2}), i.e.,
  \begin{align}
  &\max_{\mathbf{B}} {\rm
  Tr}\left(\mathbf{B}^\textsf{T}\mathbf{X}\mathbf{X}^\textsf{T}\mathbf{B}\right)\nonumber  \\
  &\mathrm{subject}~\mathrm{to}
  ~~\mathbf{B}^\textsf{T} \mathbf{B}=\mathbf{I}_k.
  \label{equ:equivalent LRMA}
  \end{align}
  It is well known that the problem in (\ref{equ:equivalent LRMA}) has an optimal solution
  \cite{kokiopoulou2011trace},
   which consists of $k$ eigenvectors of $\mathbf{X}\mathbf{X}^\textsf{T}$,
  corresponding to the $k$ largest eigenvalues. We refer the readers to ~\cite{markovsky2012low} for more technical details about LRMA.
 \subsection{Graph Transform}
  \label{subsec:GT}
  Let $\mathbf{s}\in \mathbb{R}^{n\times 1}$ be a signal defined on an undirected, connected and unweighted graph with $n$ vertices denoted by
  $\mathcal{G}=(\mathcal {V},\mathcal {E})$,
  where $\mathcal{V}$ and $\mathcal{E}$ are the sets of vertices and edges, respectively.
  The graph's adjacency matrix $\mathbf{E}\in \mathbb{R}^{n\times n}$ is given by
  \begin{equation}
   e_{ij}=\left\{
  \begin{array}{cc}
  1 & {\rm if}~  (i, j)\in\mathcal{E} \\
  0 & {\rm otherwise},
  \end{array} \right.
  \end{equation}
  and its degree matrix $\mathbf{F}\in \mathbb{R}^{n\times n}$, a diagonal
  matrix, is defined as
  \begin{equation}
  f_{ij}=\left\{
  \begin{array}{cc}
  \sum_{l=1}^n e_{il} & {\rm if}~ i=j\\
  0 & {\rm otherwise.}
  \end{array} \right.
  \end{equation}
  Then the graph Laplacian matrix $\mathbf{L}\in \mathbb{R}^{n\times n}$ is computed as
  \begin{equation}
  \centering  \mathbf{L}=\mathbf{F}- \mathbf{E}.
  \end{equation}

Since $\mathbf{L}$ is a real symmetric matrix, it has a set of real
and non-negative eigenvalues denoted by
$\{\lambda_i\}_{i=1,2,\cdots,n}$ associated with a complete set of
orthogonal eigenvectors denoted by $\{\mathbf{u}_i\in
\mathbb{R}^{n\times 1}\}_{i=1,2,\cdots,n}$, i.e.,
\begin{equation}
\mathbf{L}=\mathbf{U}_{gt}{\bf\Lambda} \mathbf{U}_{gt}^\textsf{T},
\end{equation}
where
$\mathbf{U}_{gt}=[\mathbf{u}_1~\mathbf{u}_2~\cdots~\mathbf{u}_n]\in
\mathbb{R}^{n\times n}$ and ${\bf\Lambda}={\rm
diag}(\lambda_1,\cdots, \lambda_n)\in \mathbb{R}^{n\times n}$.

  Similar to the Fourier bases, which are the eigenvectors of the one-dimensional Laplace
operator, the eigenvectors of $\mathbf{L}$ also possess harmonic
behavior~\cite{zhang2010spectral}.
  Following~\cite{zhang2013analyzing}, we call the eigenvector matrix based transformation~\emph{graph transform} (GT).
  Thus, one can decorrelate the signal as follows:
\begin{equation}
\centering \mathbf{r}=\mathbf{U}_{gt}^\textsf{T}\mathbf{s},
\end{equation}
where $\mathbf{r}\in \mathbb{R}^{n\times 1}$ consists of sparse or
approximately sparse transform coefficients.

\section{Sparse Low-Rank Matrix Approximation}
\label{sec:CLRMA}

 \subsection{Problem Statement}
  Recall that LRMA can effectively exploit the
  \textit{inter}-coherence of data samples,
  but it fails to exploit their \textit{intra}-coherence.
  Thus, the intra-coherence is delivered into the columns of
  $\mathbf{B}$. The issue to be addressed is how to effectively
  exploit such intra-coherence.

  It is well-known that some prescribed orthogonal transforms, e.g, DCT, DWT, and GT denoted as ${\bf \Phi}\in
   \mathbb{R}^{m\times m}$ (${\bf\Phi}^\textsf{T}{\bf\Phi}={\bf\Phi}{\bf\Phi}^\textsf{T}=\mathbf{I}_m$),
  can decorrelate signals \cite{zhang2010spectral}, \cite{wang2012introduction}, i.e., producing approximately sparse transform coefficients.
  In order to explore the intra-coherence inheriting from the data samples, one may consider  applying $\bf\Phi$ to columns of $\mathbf{B}$ following
  LRMA, which is similar to techniques in \cite{du2007hyperspectral}
  and \cite{Hou2015compressing},
  but such stepwise manner induces large reconstruction error (See results in Section
  \ref{sec:verification}).
  The reasons are: (\lowercase\expandafter{\romannumeral1})
  LRMA suppresses the spatial characteristics of data,
  so they cannot explore the
  intra-coherence very well in a direct way; (\lowercase\expandafter{\romannumeral2}) columns of $\mathbf{B}$ are separately decorrelated,
  which cannot preserve the relationship among them (e.g.,
  orthogonality).
  Therefore, we propose SLRMA to explore the intra- and inter- coherence
  simultaneously and integrally, and it is cast as the following optimization
  problem:
\begin{align}
&\min_{\breve{\mathbf{B}}, \mathbf{C}}
\left\|\mathbf{X}-\breve{\mathbf{B}}\mathbf{C}\right\|_F^2\nonumber \\
&{\rm subject~to~~}\breve{\mathbf{B}}^\textsf{T}
\breve{\mathbf{B}}=\mathbf{I}_k\nonumber \\
&~~~~~\sum_{i=1}^k
\left\|{\bf\Phi}^\textsf{T}\breve{\mathbf{b}}_i\right\|_0=\left\|{\bf\Phi}^\textsf{T}\breve{\mathbf{B}}\right\|_0\leq
s. \label{equ:original 1D CLRMA}
\end{align}
  Note that we constrain $\breve{\mathbf{B}}$ to be sparse with respect to $\bf\Phi$, making the derived $\breve{\mathbf{B}}$ via optimization adapted to $\bf\Phi$.
  The difference between $\breve{\mathbf{B}}$ and $\mathbf{B}$ by LRMA can be observed by comparing Figure \ref{fig:problem visual}(a) and Figure \ref{fig:problem visual}(c).
  Furthermore, taking $\|{\bf\Phi}^\textsf{T}\mathbf{A}\|_F^2=\|\mathbf{A}\|_F^2$ and with a slight abuse of notation (i.e, replacing ${\bf\Phi}^\textsf{T}\breve{\mathbf{B}}$ by $\mathbf{B}$),
  we can rewrite the problem in (\ref{equ:original 1D CLRMA}) as
  \begin{align}
  &\min_{\mathbf{B}, \mathbf{C}}~ \|{\bf\Phi}^\textsf{T}\mathbf{X}-\mathbf{B}\mathbf{C}\|_F^2 \nonumber\\
  &{\rm subject~to}~~\mathbf{B}^\textsf{T} \mathbf{B}=\mathbf{I}_k
  ~~{\rm and}~~\|\mathbf{B}\|_0\leq s.  \label{equ:CLRMA}
   \end{align}

 In all, the proposed SLRMA can be interpreted as follows: the
 transform coefficients of data samples are low-rank approximated
 under the sparsity constraint, in which $\bf\Phi$ and the sparsity constraint can explore the intra-coherence integrally,
 and the low-rank representation can explore the inter-coherence. These two types of decorrelation are simultaneously
 performed through optimization.
 Regarding $\bf\Phi$, it can be set according to data
 samples. For example, we can adopt DCT or DWT as the transform matrix for natural images, and the GT matrix for data defined on general graphs.

\subsection{Numerical Solver}

Similar to the process of (\ref{equ:LRA2}) to (\ref{equ:equivalent
LRMA}), the problem in (\ref{equ:CLRMA}) can be simplified as
\begin{align}
&\min_{\mathbf{B}} -\|\mathbf{Z}^\textsf{T}\mathbf{B}\|_F^2
\nonumber\\
&{\rm subject~to}~~\mathbf{B}^\textsf{T} \mathbf{B}=\mathbf{I}_k~~
{\rm and}~~ \|\mathbf{B}\|_0\leq s, \label{equ:simplified SLRMA}
\end{align}
where $\mathbf{Z}={\bf\Phi}^\textsf{T}\mathbf{X}$. Instead of solving (\ref{equ:simplified SLRMA}) directly, we solve its Lagrangian version:
\begin{align}
&\min_{\mathbf{B}}~ -\|\mathbf{Z}^\textsf{T}\mathbf{B}\|_F^2+\gamma\|\mathbf{B}\|_0\nonumber \\
&{\rm subject~to}~~\mathbf{B}^\textsf{T} \mathbf{B}=\mathbf{I}_k,
\label{equ:CLRMA2}
\end{align}
  where the regularization parameter $\gamma$ controls the sparsity of $\mathbf{B}$:
  the larger the value of $\gamma$ is, the sparser the matrix $\mathbf{B}$ is.

  We adopt the inexact augmented Lagrangian multiplier method (IALM) \cite{lin2010augmented} to solve (\ref{equ:CLRMA2}).
  Introducing two auxiliary matrices $\mathbf{P}\in \mathbb{R}^{m\times k}$ and $\mathbf{Q}\in \mathbb{R}^{m\times k}$,
  we rewrite (\ref{equ:CLRMA2}) equivalently  as
  \begin{align}
&\min_{\mathbf{B},\mathbf{P},\mathbf{Q}}~ -\|\mathbf{Z}^\textsf{T}\mathbf{B}\|_F^2+\gamma\|\mathbf{P}\|_0\nonumber \\
& {\rm subject~to}~~\mathbf{P}=\mathbf{B},~~\mathbf{Q}=\mathbf{B},
~~\mathbf{Q}^\textsf{T} \mathbf{Q}=\mathbf{I}_k.
 \label{equ:CLRMA ADMM}
\end{align}
  The augmented Lagrangian form of (\ref{equ:CLRMA ADMM}) is given
  by
\begin{align}
&\argmin_{\mathbf{B},
\mathbf{P},\atop\mathbf{Q}^\textsf{T}\mathbf{Q}=\mathbf{I}_k}-\|\mathbf{Z}^\textsf{T}\mathbf{B}\|_F^2+\gamma\|\mathbf{P}\|_0+{\rm
Tr}\left(\mathbf{Y}_P^\textsf{T}(\mathbf{B}-\mathbf{P})\right)\nonumber
\\ &~+{\rm
Tr}\left(\mathbf{Y}_Q^\textsf{T}(\mathbf{B}-\mathbf{Q})\right)+\frac{\rho}{2}\left(\|\mathbf{B}-\mathbf{P}\|_F^2+\|\mathbf{B}-\mathbf{Q}\|_F^2\right),
\label{equ:ALF of CLRMA}
\end{align}
  where $\mathbf{Y}_p$ and $\mathbf{Y}_q\in \mathbb{R}^{m\times k}$ are the Lagrange multipliers,
  and the regularization parameter $\rho$ is a positive scalar. 

  With initialized $\bf P$, $\bf Q$, $\mathbf{Y}_P$, and $\mathbf{Y}_Q$,  the IALM solves the optimization problem (\ref{equ:ALF of CLRMA}) in an iterative manner
  (see Algorithm~\ref{Alg:1D CLRMA} for the pseudocode).
  Each iteration alternatively solves the following four subproblems:

\emph{1) The $\mathbf{B}$-Subproblem:} The $\bf B$-subproblem is
with a quadratic form:
\begin{align}
\min_{\mathbf{B}}~& -\|\mathbf{Z}^\textsf{T}\mathbf{B}\|_F^2
+\frac{\rho}{2}\Big(\|\mathbf{B}-\mathbf{P}+\mathbf{Y}_P/\rho\|_F^2\nonumber\\
&+\|\mathbf{B}-\mathbf{Q}+\mathbf{Y}_Q/\rho\|_F^2\Big). \label{equ:
B}
\end{align}
   Equation (\ref{equ: B}) reaches the minimal when the first-order derivative to $\bf B$ vanishes:
\begin{equation}
\centering \mathbf{B}=\left(2\rho
\mathbf{I}_m-2\mathbf{Z}\mathbf{Z}^\textsf{T}\right)^{-1}\Big(\rho(\mathbf{P}+\mathbf{Q})-\mathbf{Y}_P-\mathbf{Y}_Q\Big).
\label{equ:optB 1DCLRMA}
\end{equation}

 \emph{2) The $\mathbf{P}$-Subproblem:} The $\mathbf{P}$-subproblem is written as
\begin{equation}
\centering \min_{\mathbf{P}}
\gamma\|\mathbf{P}\|_0+\frac{\rho}{2}\|\mathbf{P}-(\mathbf{B}+\mathbf{Y}_P/\rho)\|_F^2.
\label{equ:P}
\end{equation}
  Let $\widetilde{\mathbf{B}}=\mathbf{B}+\mathbf{Y}_P/\rho$, and Equation (\ref{equ:P}) can be rewritten in element-wise as
\begin{equation}
 \centering \min_{\{p_{ij}\}}\sum_{i,j}\gamma \mathbf{1}_{(p_{ij}\neq 0)}+\frac{\rho}{2}\left(p_{ij}-\widetilde{b}_{ij}\right)^2, \label{equ:P elementwise}
\end{equation}
where $\mathbf{1}_{(p_{ij}\neq 0)}$ is a indicator function, i.e.,
$\mathbf{1}_{(p_{ij}\neq 0)}=1$, if $p_{ij}\neq 0$ and 0 otherwise.
Furthermore, the objective function (\ref{equ:P elementwise}) is minimized if each
independent univariate in the summarization is minimized, i.e.,
\begin{equation}
 \centering  \min_{p_{ij}} \gamma \mathbf{1}_{(p_{ij}\neq 0)}+\frac{\rho}{2}\left(p_{ij}-\widetilde{b}_{ij}\right)^2.
\end{equation}
 It can be easily checked that $p_{ij}$ has an unique solution given by the hard thresholding
 operator, i.e.,
  \begin{equation}
   p_{ij}=\left\{
  \begin{array}{cc}
  \widetilde{b}_{ij} & {\rm if}~  |\widetilde{b}_{ij}|> \sqrt{2\gamma/\rho}\\
  0 & {\rm otherwise}.
  \end{array} \right.
  \label{equ:optP 1DCLRMA}
  \end{equation}

\emph{3) The $\mathbf{Q}$-Subproblem:} It is expressed as
\begin{equation}
\centering \min_{\mathbf{Q}^\textsf{T}\mathbf{Q}=\mathbf{I}_k}
\frac{\rho}{2}\|\mathbf{Q}-(\mathbf{B}+\mathbf{Y}_Q/\rho)\|_F^2.
\label{equ: Q}
\end{equation}
  According to Theorem \ref{the2}, the problem (\ref{equ: Q}) has a closed-form
  solution:
\begin{equation}
\centering
\mathbf{Q}=\widehat{\mathbf{B}}\mathbf{V}\mathbf{D}^{-1/2}\mathbf{V}^\textsf{T},
\label{equ:optQ 1DCLRMA}
\end{equation}
  where $\mathbf{\widehat{B}}=\mathbf{B}+\mathbf{Y}_Q/\rho$, and the orthogonal matrix $\mathbf{V}\in \mathbb{R}^{k\times k}$ and the diagonal matrix $\mathbf{D}\in \mathbb{R}^{k\times k}$
  satisfy the eigendecomposition of $\widehat{\mathbf{B}}^\textsf{T}\widehat{\mathbf{B}}$,
  i.e., $\widehat{\mathbf{B}}^\textsf{T}\widehat{\mathbf{B}}=\mathbf{V}\mathbf{D}\mathbf{V}^\textsf{T}$.

\begin{theorem}[\cite{lai2014splitting}]\label{the2}
  \emph{Given $\mathbf{X}\in \mathbb{R}^{m\times n}$ and ${\rm rank}(\mathbf{X})=n$, the constrained quadratic problem:}
  \begin{equation}
  \centering
  \mathbf{X}^*=\argmin_{\mathbf{X}\in \mathbb{R}^{m\times n}} \frac{1}{2}\|\mathbf{X}-\mathbf{A}\|_F^2,~~{\rm
  subject~to}~\mathbf{X}^\textsf{T}\mathbf{X}=\mathbf{I}_n \nonumber
  \end{equation}
  \emph{has the closed-form solution,
  i.e.,$\mathbf{X}^*=\mathbf{A}\widetilde{\mathbf{V}}\widetilde{\mathbf{D}}^{-1/2}\widetilde{\mathbf{V}}^\textsf{T}$,
  where $\widetilde{\mathbf{V}}\in \mathbb{R}^{n\times n}$ is an orthogonal matrix and $\widetilde{\mathbf{D}}\in \mathbb{R}^{n\times n}$ is a diagonal matrix satisfying the eigendecomposition of
  $\mathbf{A}^\textsf{T}\mathbf{A}$.} 
\end{theorem}

\emph{5) Updating $\mathbf{Y}_P$, $\mathbf{Y}_Q$ and $\rho$:}
Finally, we update the Largrange multipliers $\mathbf{Y}_P$ and
$\mathbf{Y}_Q$, and the parameter $\rho$ as
\begin{align}
&\mathbf{Y}_P^{iter+1}=\mathbf{Y}_P^{iter}+\rho(\mathbf{B}-\mathbf{P}), \label{equ:Yp} \\
&\mathbf{Y}_Q^{iter+1}=\mathbf{Y}_Q^{iter}+\rho(\mathbf{B}-\mathbf{Q}), \label{equ:Yq} \\
&\rho^{iter+1}=\min(\rho^{iter}\alpha, \rho_{max}), \label{equ:rho}
\end{align}
where the parameter $\alpha>1$ improves the convergence rate and \textit{iter} is the iteration index.

  Lin \emph{et al}. \cite{lin2010augmented} proved that IALM guarantees convergence to an optimal solution on a convex optimization problem~\cite{candes2011robust}.
  However, the objective function in (\ref{equ:CLRMA2}) and the orthogonality constraint are both non-convex.
  To the best of our knowledge, there is no theoretical evidence on the global convergence of IALM on a non-convex problem.
  Fortunately, thanks to the closed-form solution for each subproblem,
  we observe that SLRMA via IALM empirically converges well (i.e., the objective function is reduced to a stable value after a few iterations) and produces satisfactory performance
  on real-world datasets (see Section \ref{sec:verification}).

\begin{algorithm}
\caption{Computing SLRMA via IALM}
 \textbf{Input}:  $\mathbf{X}$, ${\bf\Phi}$, $\gamma$, $k$, $\rho$, $\alpha$,
 $\rho_{max}$\\
 \textbf{Output}: $\mathbf{B}$, $\mathbf{C}$
\begin{algorithmic}[1]
 \STATE initialize $\mathbf{P}$ and
 $\mathbf{Q}$ using the leftmost $k$ columns of $\mathbf{I}_m$
 \STATE  $\mathbf{Y}_P=\mathbf{Y}_Q=\mathbf{0}$
 \STATE $\mathbf{Z}={\bf\Phi}^\textsf{T} \mathbf{X}$
   \WHILE{not convergence}
   \STATE update $\mathbf{B}$ using (\ref{equ:optB 1DCLRMA})
   \STATE update $\mathbf{P}$ using (\ref{equ:optP 1DCLRMA})
   \STATE update $\mathbf{Q}$ using (\ref{equ:optQ 1DCLRMA})
   \STATE update $\mathbf{Y}_P$, $\mathbf{Y}_Q$ and $\rho$ using
   (\ref{equ:Yp})-(\ref{equ:rho}), respectively
   \STATE check the convergence conditions \\ $\|\mathbf{B}-\mathbf{P}\|_{\infty}<10^{-6}$ and $\|\mathbf{B}-\mathbf{Q}\|_{\infty}<10^{-6}$
 \ENDWHILE
 \STATE update $\mathbf{C}$ using $\mathbf{B}^\textsf{T}\mathbf{Z}$
  \end{algorithmic}
\label{Alg:1D CLRMA}
\end{algorithm}

\section{SLRMA-based Data Compression}
\label{sec:verification}

In this section, we develop and evaluate the SLRMA-based compression
schemes for 3D dynamic meshes and 2D image sets consisting of facial
images or frames of video with slow motions and nearly stationary
background. These types of data exhibit intra- (or spatial)
coherence. Besides, the inter-coherence among them makes the
constructed matrices possess approximate low-rank characteristics.
Thus, it is suitable to compress them using LRMA- and SLRMA-based
methods.

\begin{figure}
\centering
\subfigure[]{\includegraphics[width=3.5in]{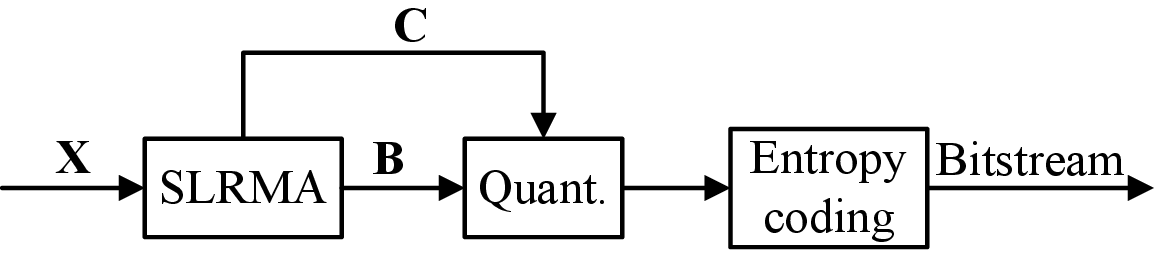}}
\subfigure[]{\includegraphics[width=3.5in]{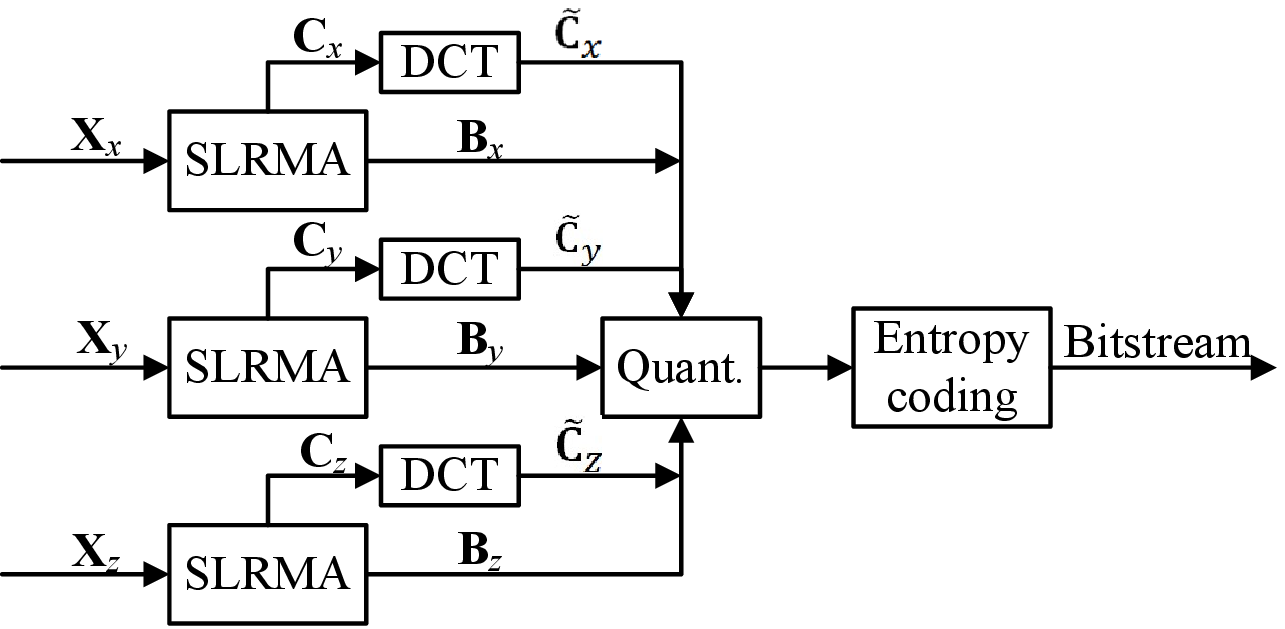}}
\caption{The flowcharts of the proposed SLRMA-based compression
schemes for image sets (a) and 3D dynamic meshes (b).}
\label{fig:flowchart}
\end{figure}

\begin{figure}[t]
\centering \subfigure[Image set used in Figure \ref{fig:problem
visual}]{
\includegraphics[width=3.1in]{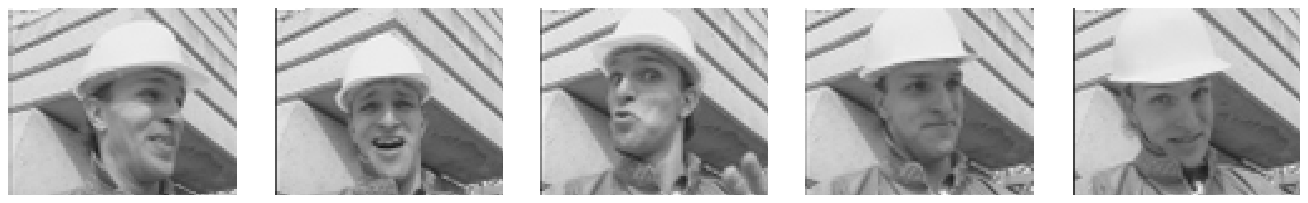}}
\subfigure[Image set \uppercase\expandafter{\romannumeral1}]{
\includegraphics[width=3.1in]{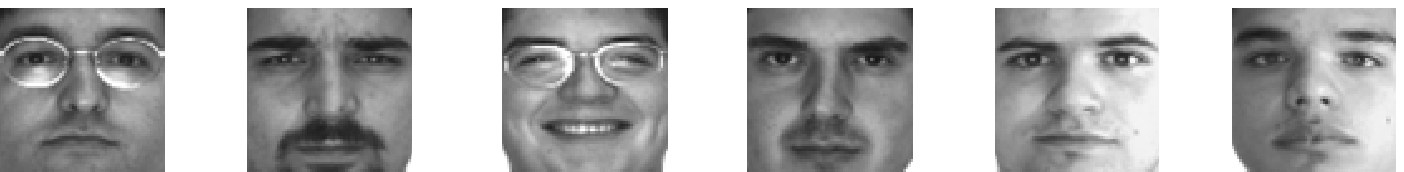}}
\subfigure[Image set \uppercase\expandafter{\romannumeral2}]{
\includegraphics[width=3.1in]{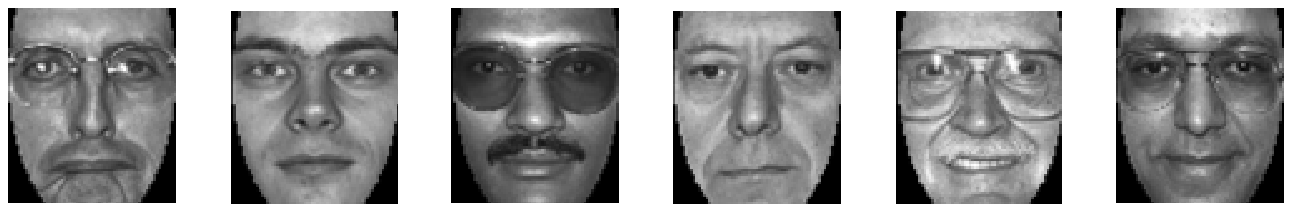}}
\subfigure[Image set \uppercase\expandafter{\romannumeral3}]{
\includegraphics[width=3.1in]{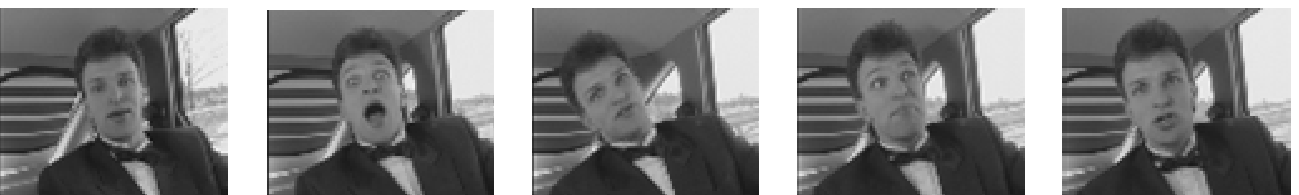}}
\subfigure[Image set \uppercase\expandafter{\romannumeral4}]{
\includegraphics[width=3.1in]{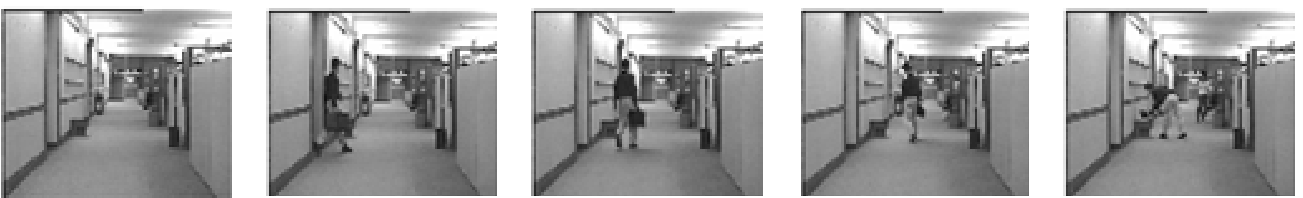}}
\subfigure[3D dynamic meshes]{
\includegraphics[width=3.1in]{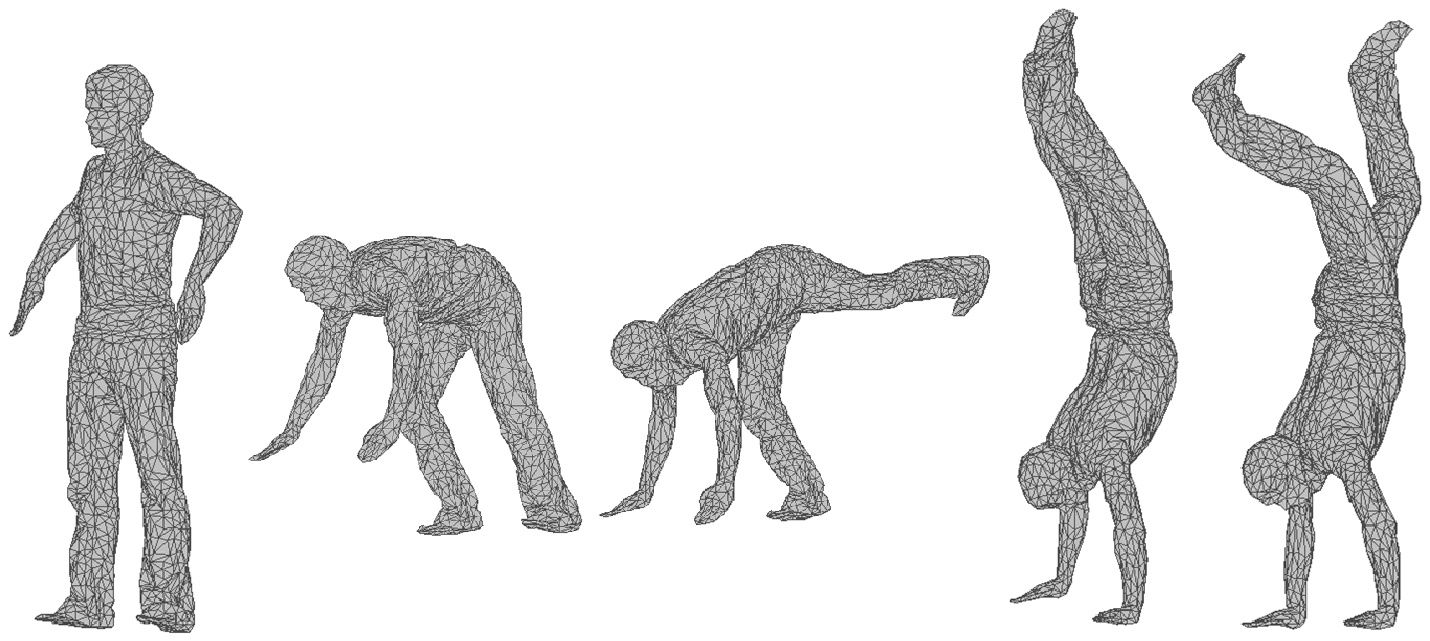}}
\caption{Examples of test datasets.} \label{fig:datasamples}
\end{figure}

\subsection{Image Sets}
\label{subsec:evaluation image and video}

Given a grayscale image set with $n$ images denoted by
$\{\mathbf{M}_i\in \mathbb{R}^{w\times h}\}_{i=1, 2, \cdots, n}$
where $w$ and $h$ are the resolution of the images, respectively, we
reshape each image as a column vector and stack them into a matrix
denoted by $\mathbf{X}\in \mathbb{R}^{wh\times n}$. As shown in
Figure \ref{fig:flowchart}(a), our SLRMA-based compression scheme is
very simple. It decomposes $\mathbf{X}$ into an extremely sparse
matrix $\mathbf{B}$ and a coefficients matrix $\mathbf{C}$. They are
then uniformly quantized before the nonzero elements, as well as
their locations, are entropy-coded into a bitstream using arithmetic
coding.

We tested four grayscale image sets: Image set
\uppercase\expandafter{\romannumeral1}: 150 facial images
(resolution: $64\times 64$) extracted from the AR dataset
\cite{martinez1998ar}; Image set
\uppercase\expandafter{\romannumeral2}: 150 facial images
(resolution: $65\times 75$) extracted from the Fa of FERET dataset
\cite{phillips2000feret}; Image set
\uppercase\expandafter{\romannumeral3}: 150 images (resolution:
$88\times 72$) extracted from the ``carphone" video
sequence\footnote{http://trace.eas.asu.edu/yuv/}; Image set
\uppercase\expandafter{\romannumeral4}: 150 images (resolution:
$88\times 72$) extracted from the ``hall" video sequence. Some data
samples are respectively shown in Figures
\ref{fig:datasamples}(b)-(e). The values of $\rho$, $\alpha$, and
$\rho_{max}$ are set to $10^{-4}$, 1.05, and $10^{10}$,
respectively. Note that these parameters are insensitive to the size
of image sets. The percentage of zero elements of $\mathbf{B}$ is
denoted by $p_B$, and various $p_B$ is obtained by adjusting
$\gamma$.

\begin{figure*}
\centering
\includegraphics[width=1.7in]{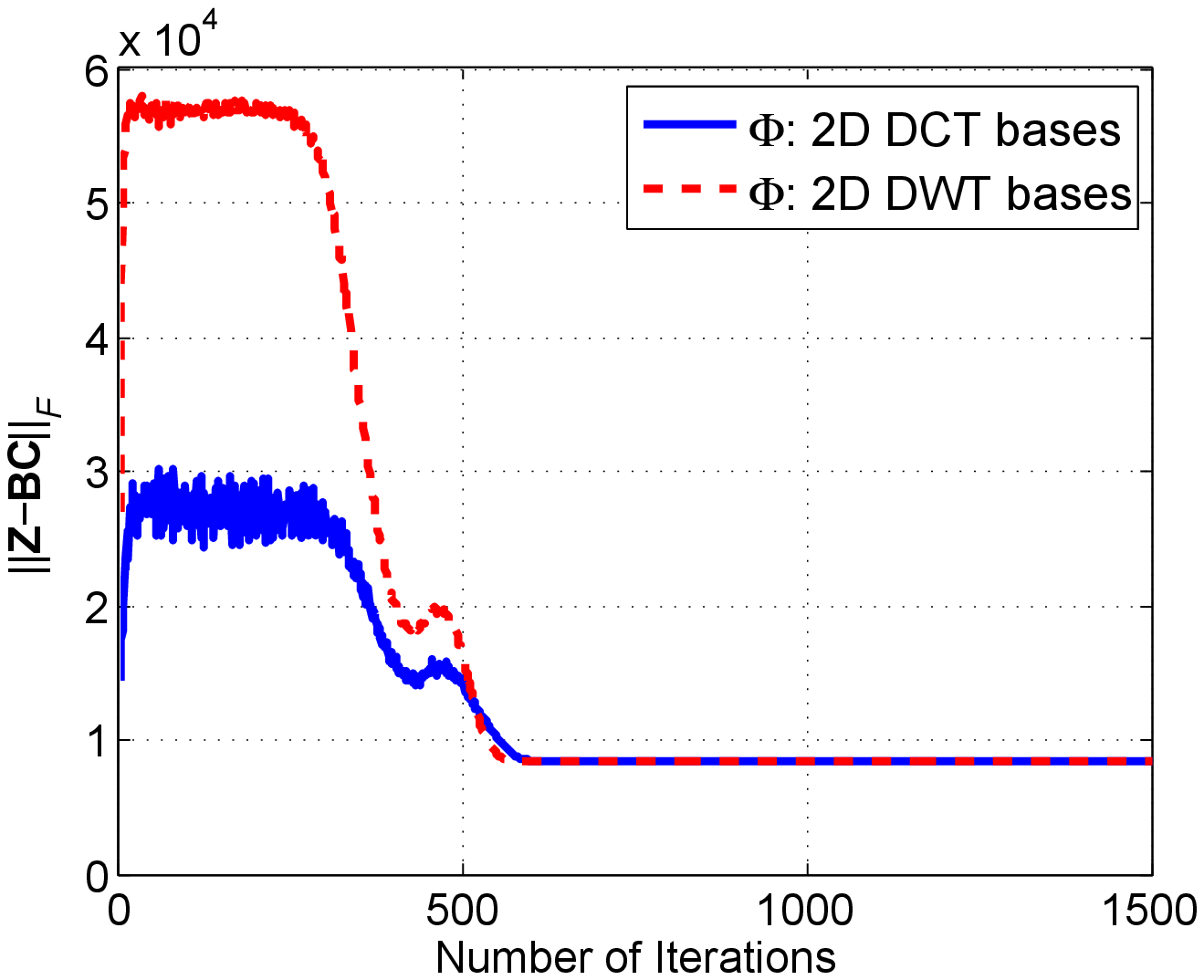}
\includegraphics[width=1.7in]{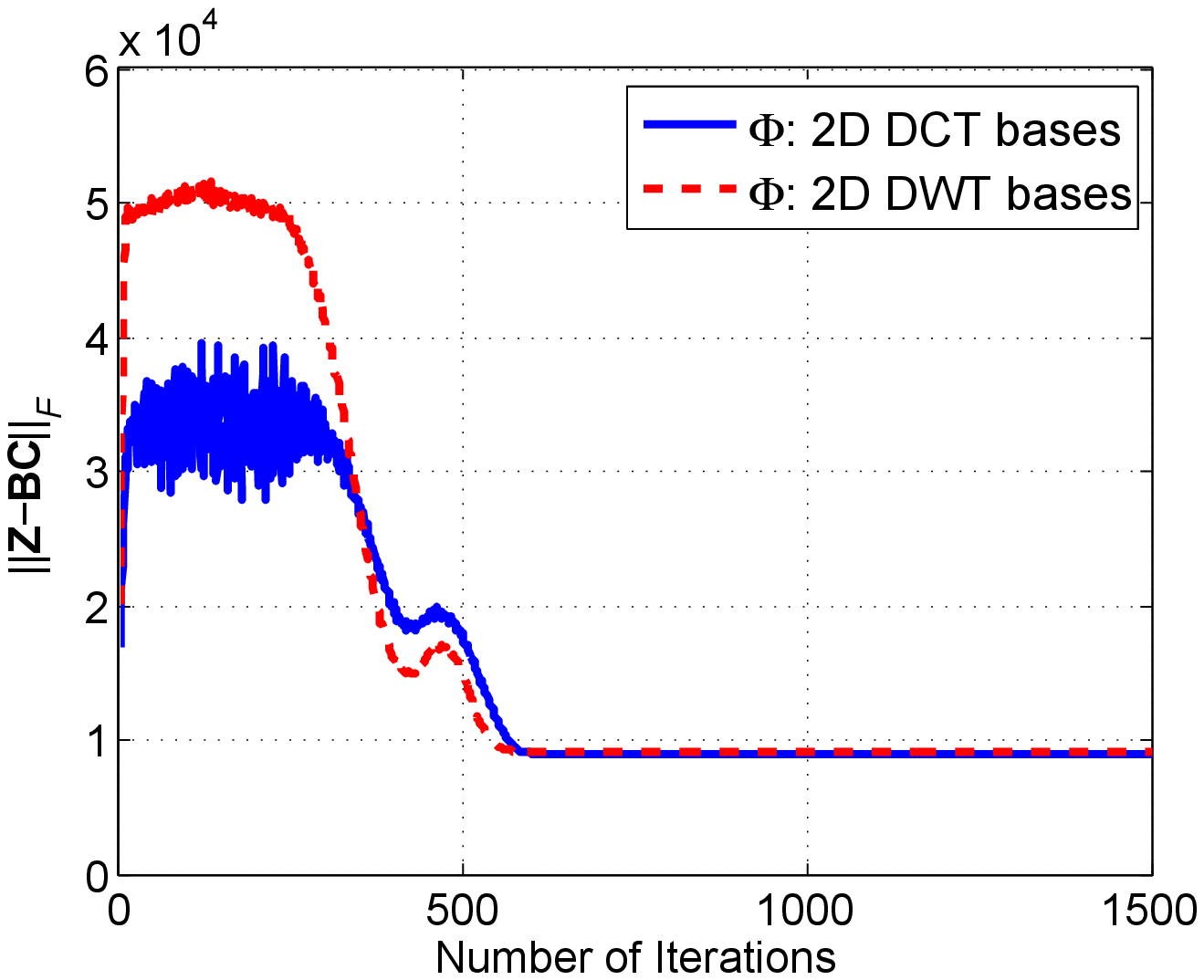}
\includegraphics[width=1.7in]{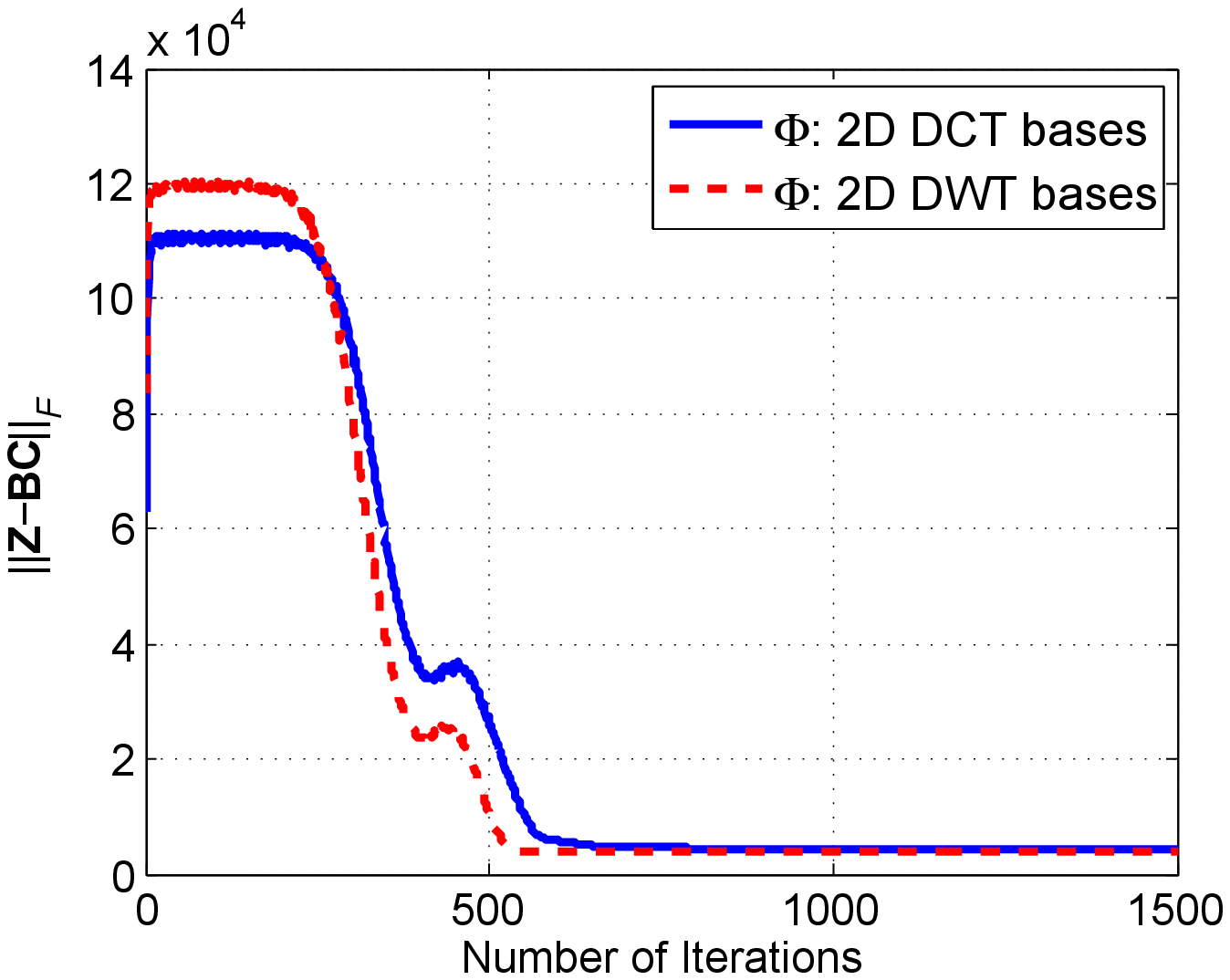}
\includegraphics[width=1.7in]{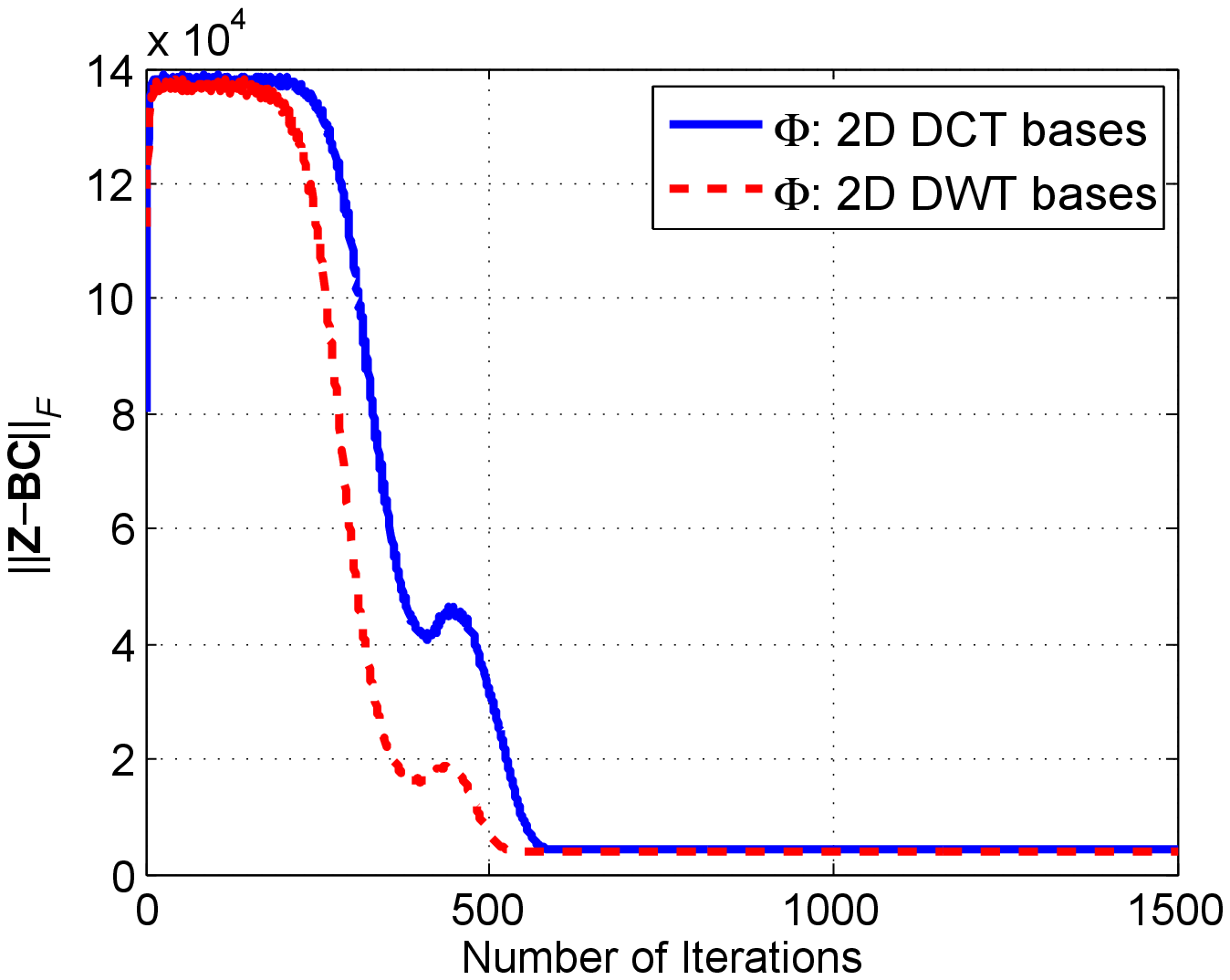}\\
\includegraphics[width=1.7in]{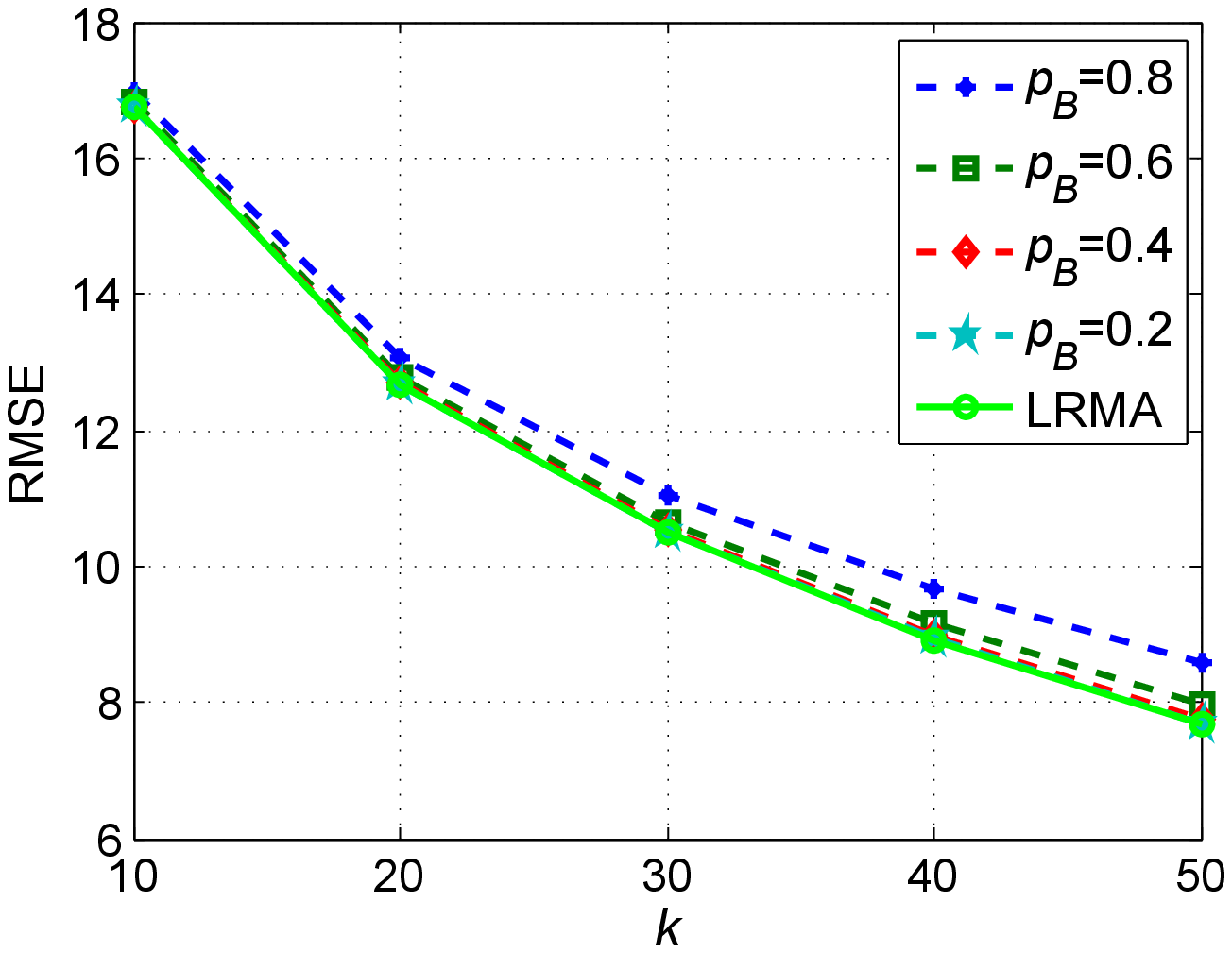}
\includegraphics[width=1.7in]{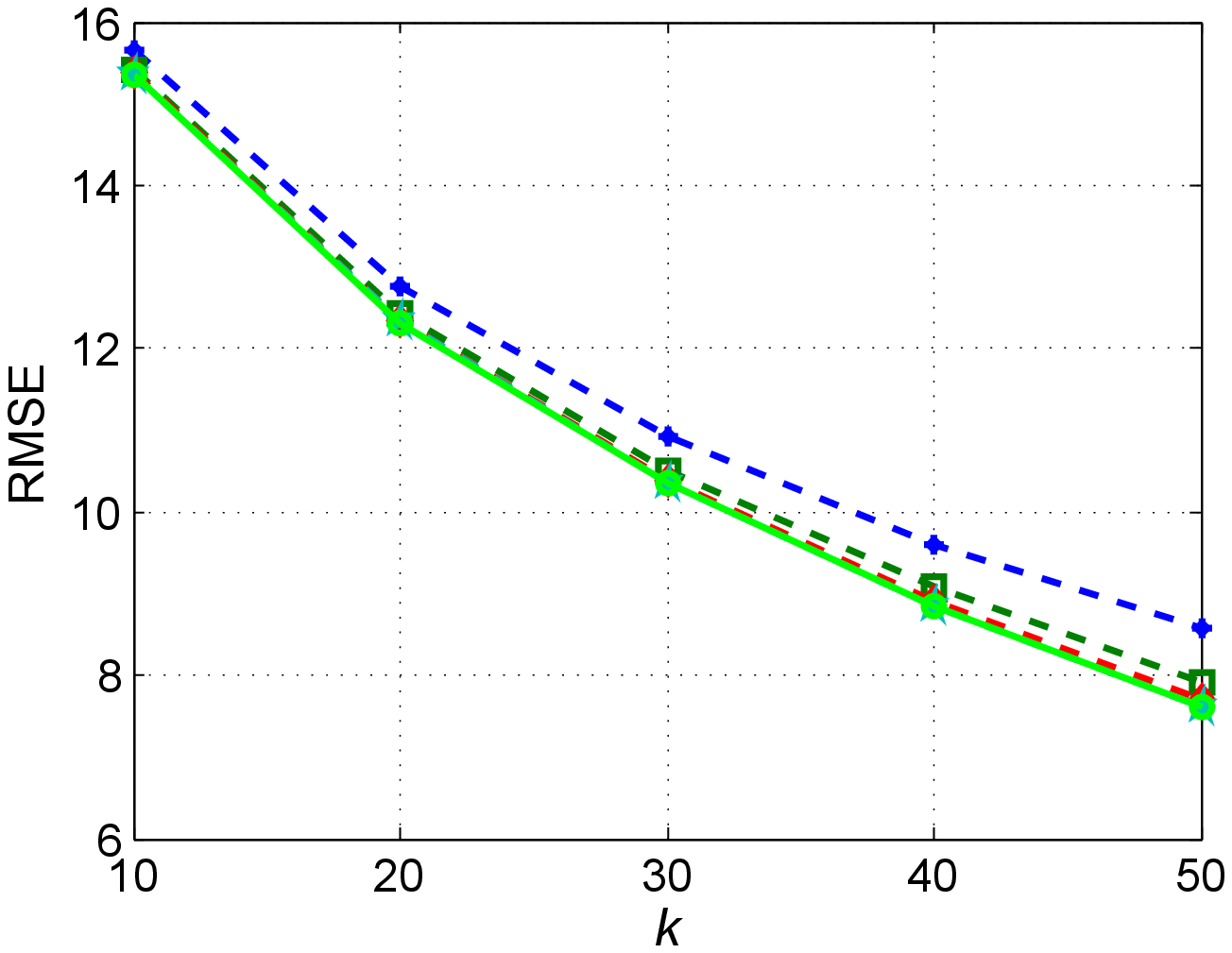}
\includegraphics[width=1.7in]{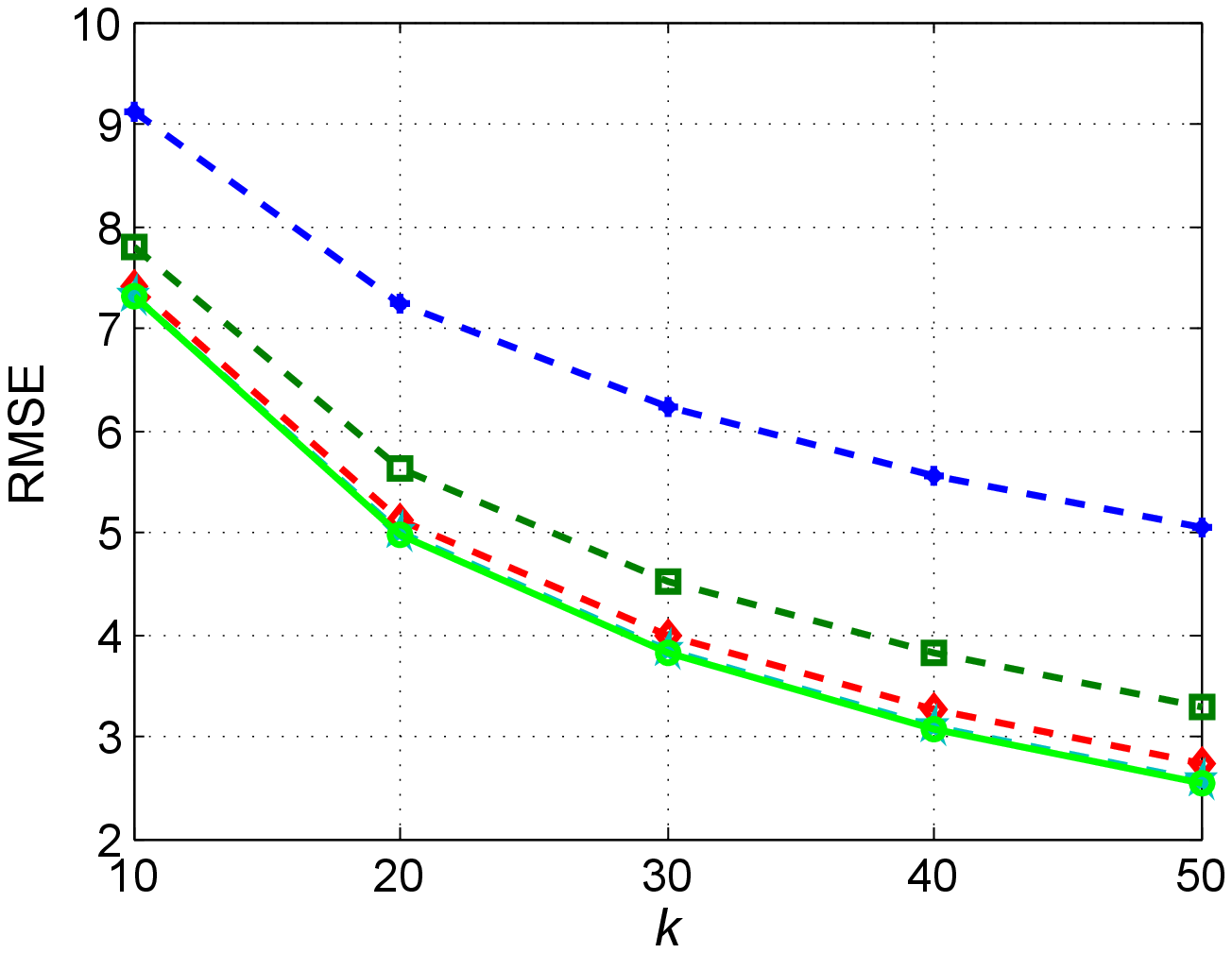}
\includegraphics[width=1.7in]{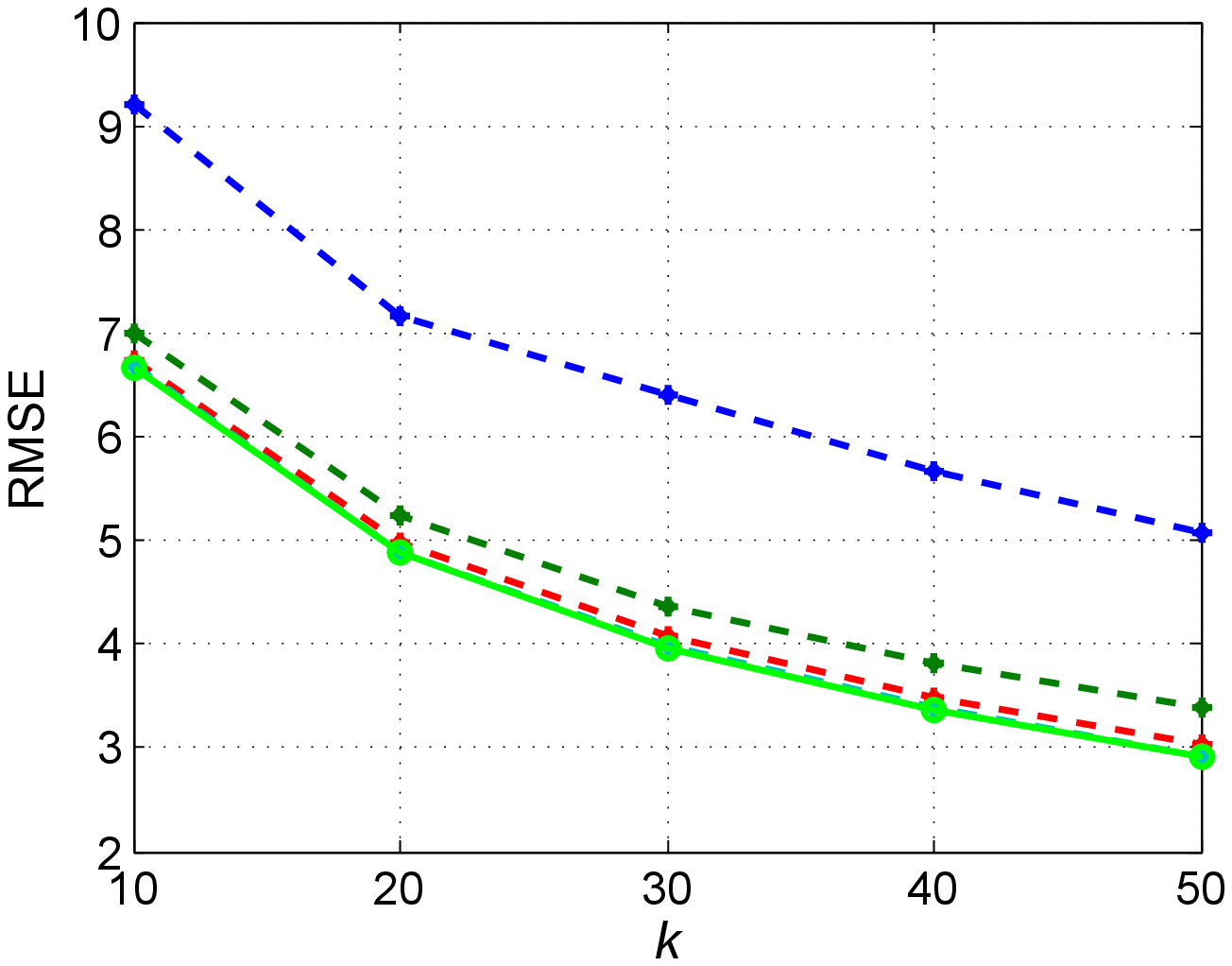}\\
\includegraphics[width=1.7in]{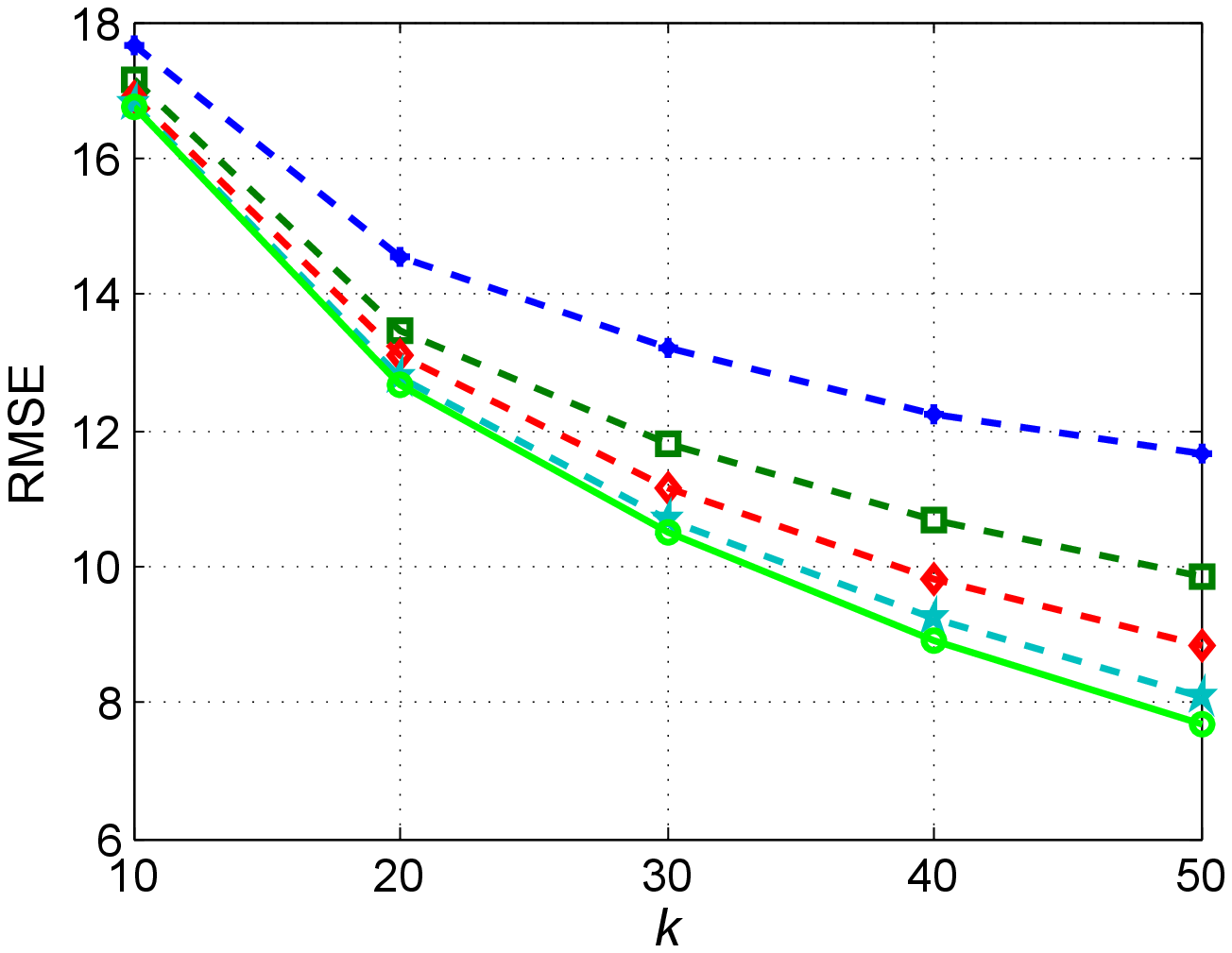}
\includegraphics[width=1.7in]{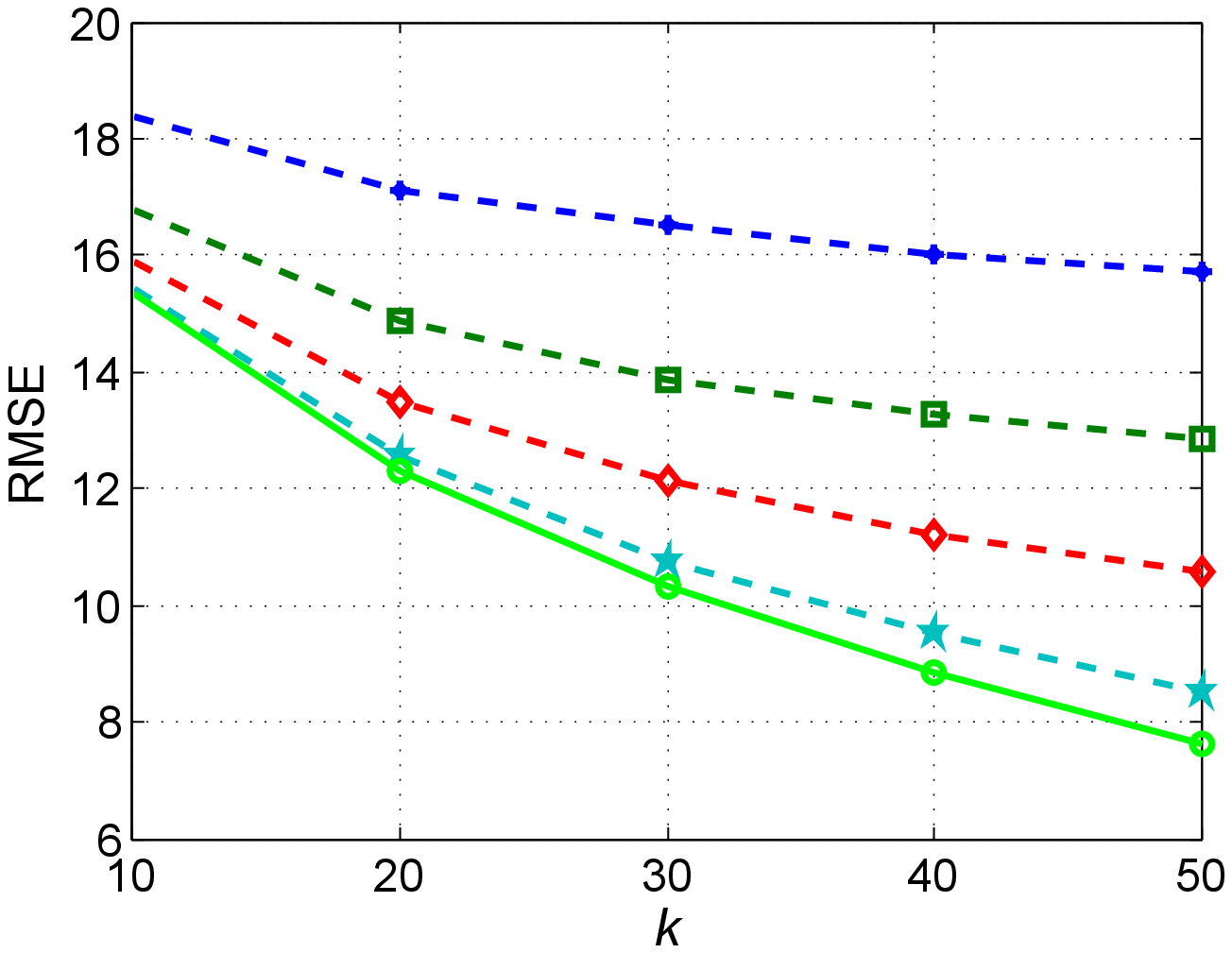}
\includegraphics[width=1.7in]{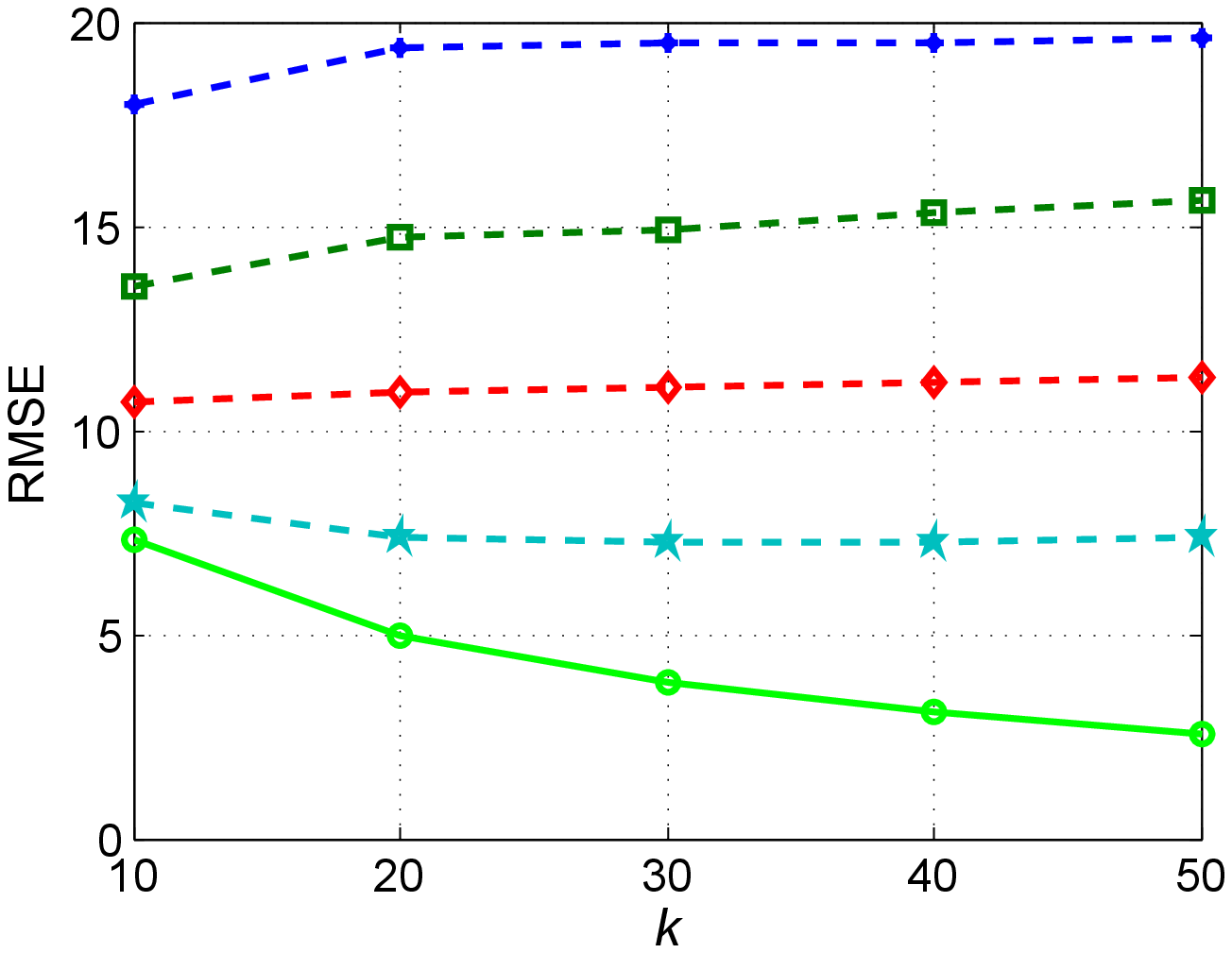}
\includegraphics[width=1.7in]{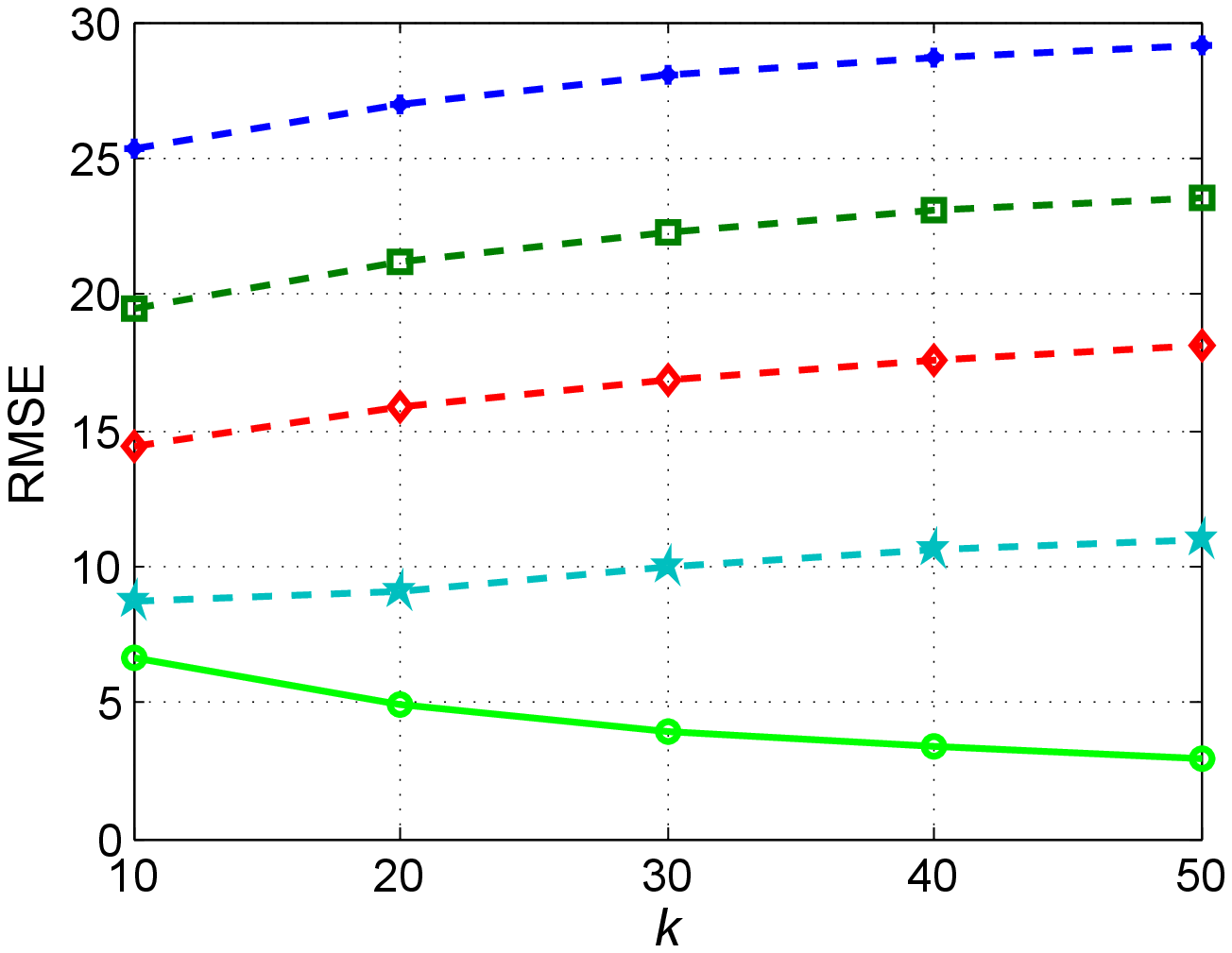}\\
\includegraphics[width=1.7in]{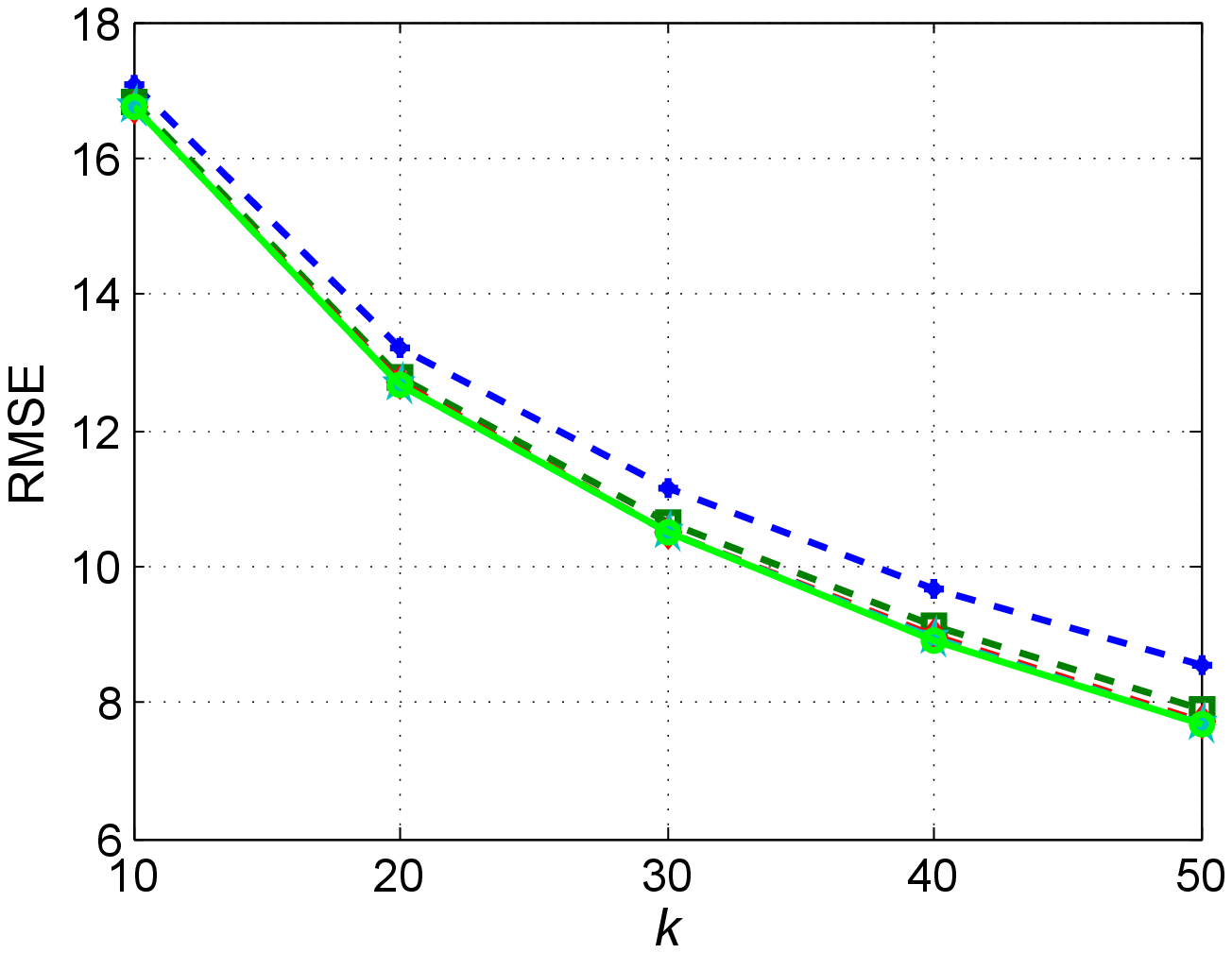}
\includegraphics[width=1.7in]{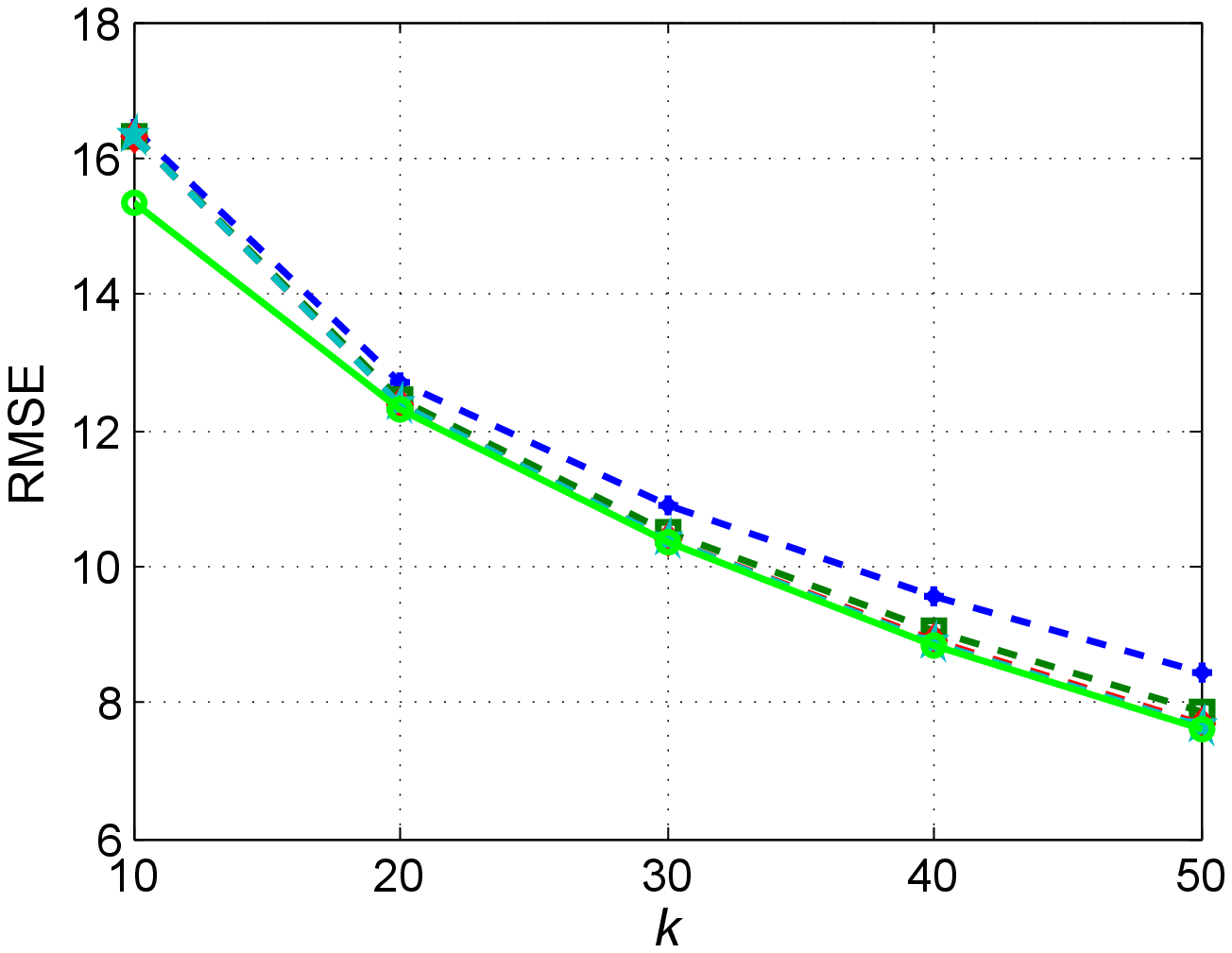}
\includegraphics[width=1.7in]{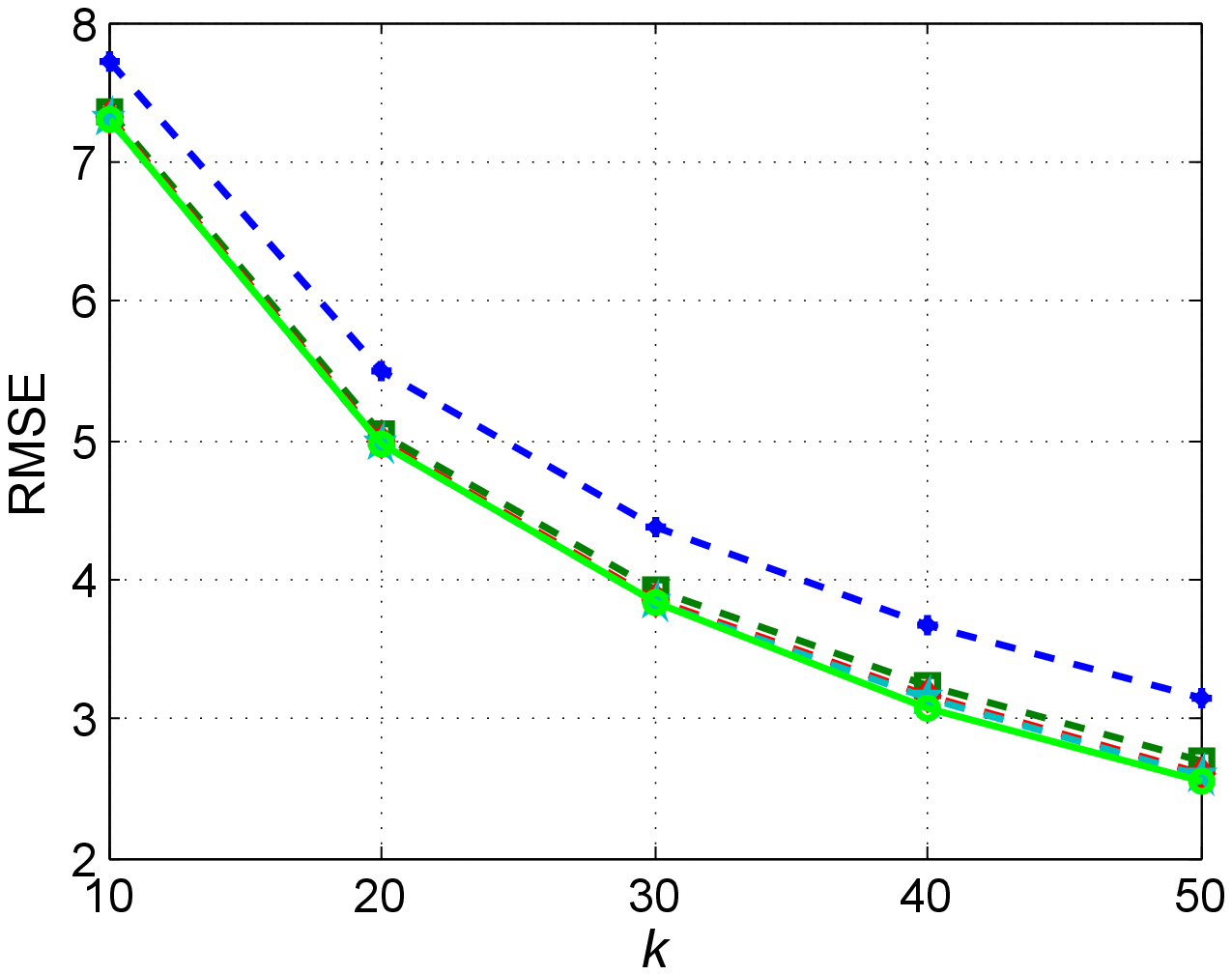}
\includegraphics[width=1.7in]{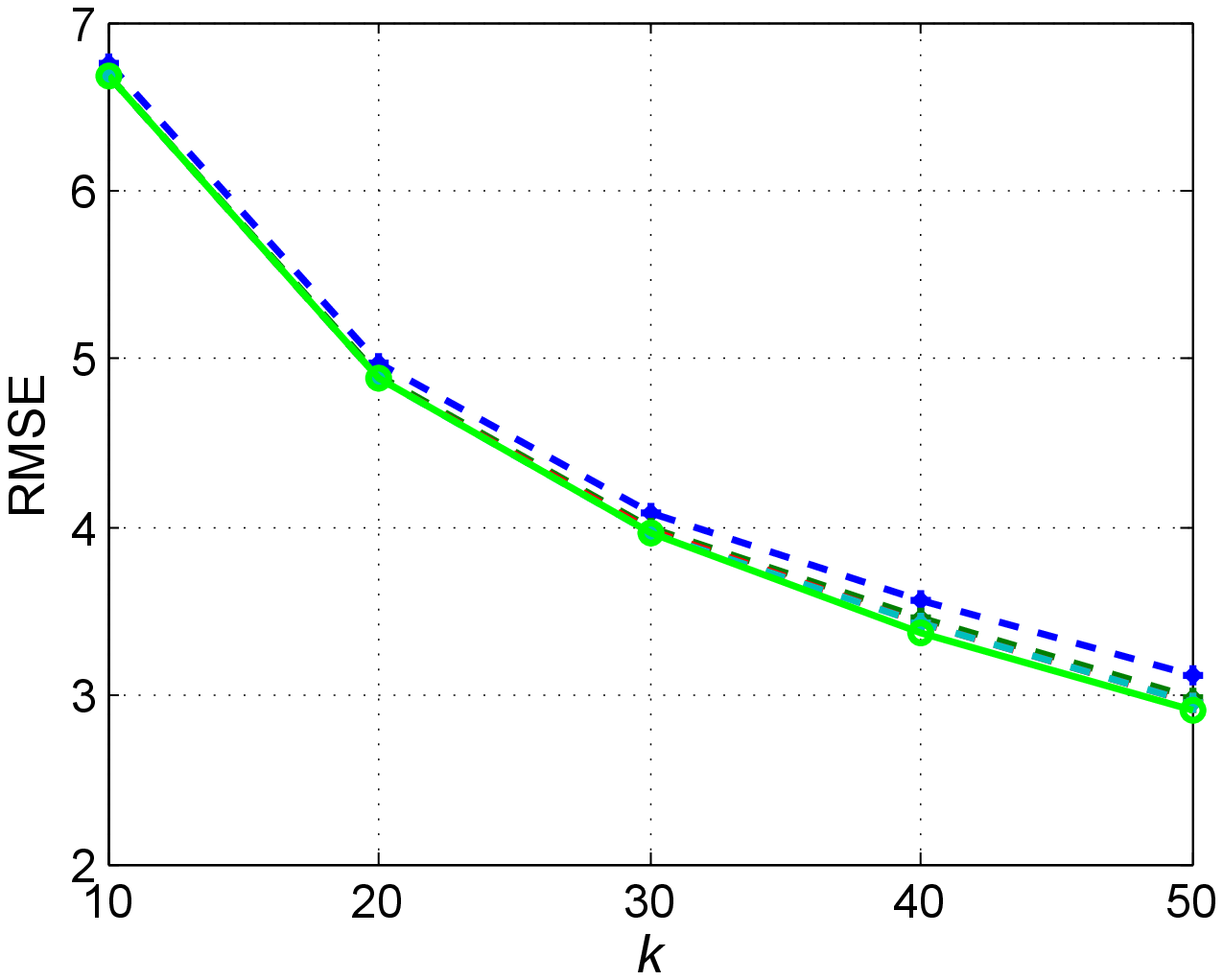}\\
\includegraphics[width=1.7in]{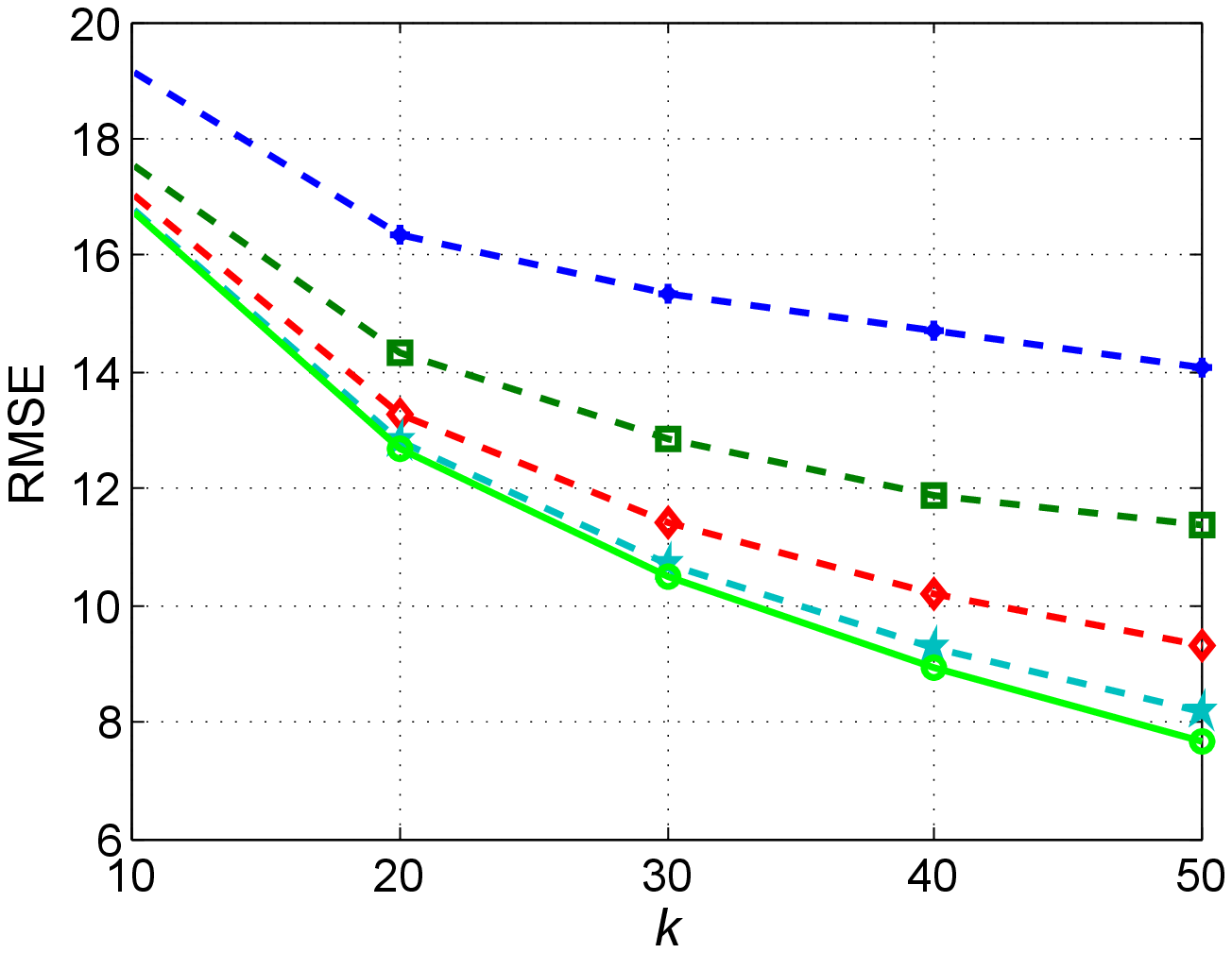}
\includegraphics[width=1.7in]{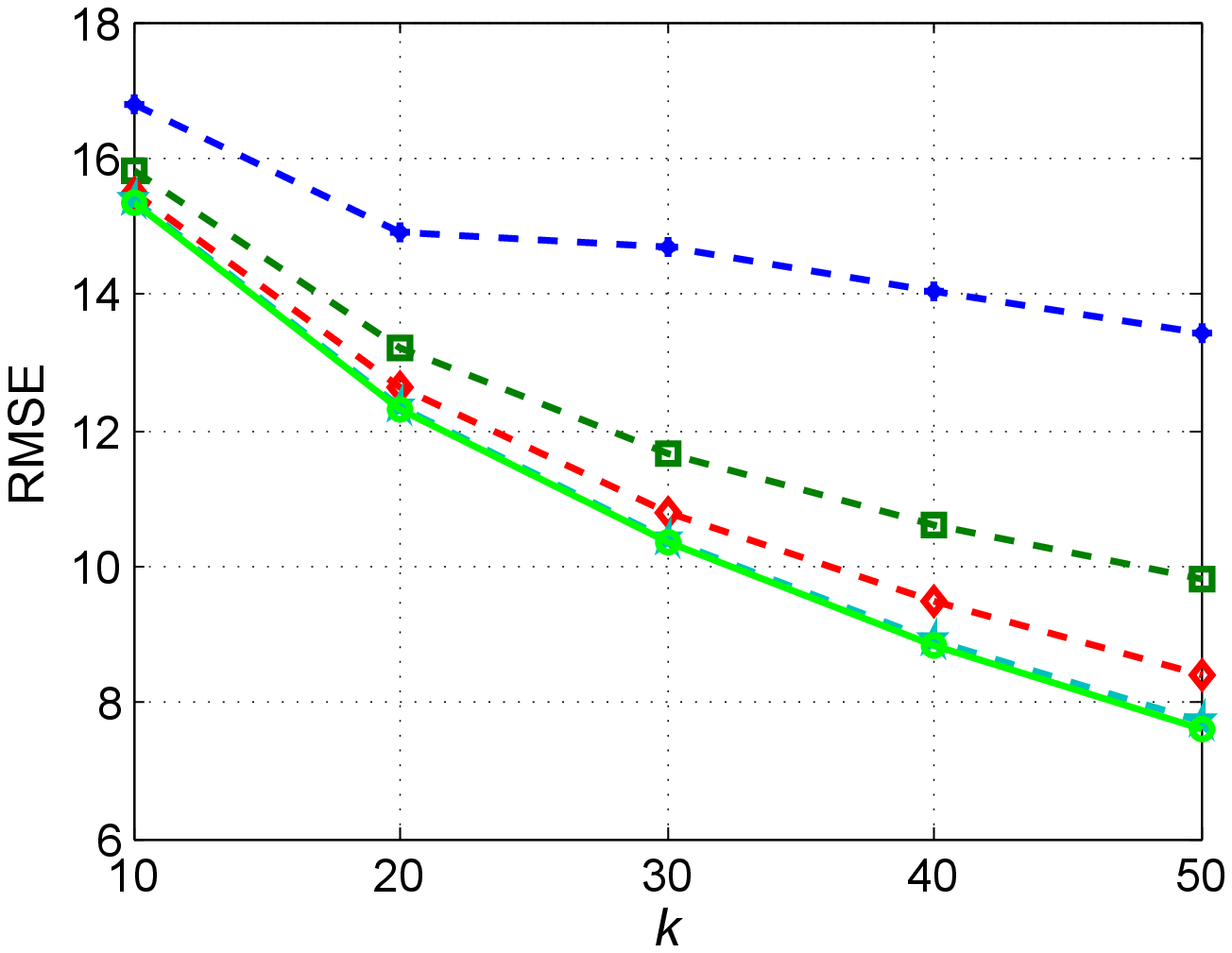}
\includegraphics[width=1.7in]{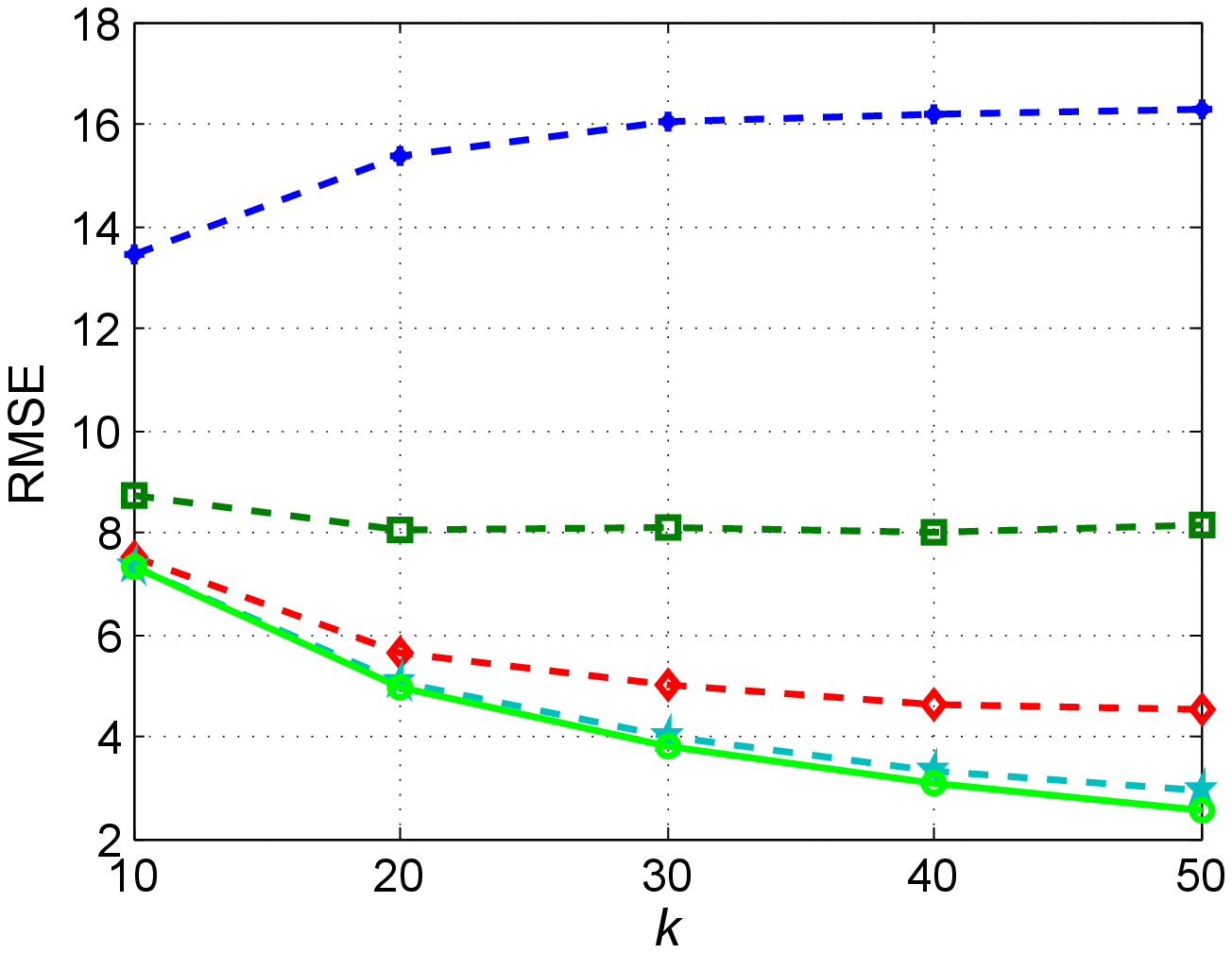}
\includegraphics[width=1.7in]{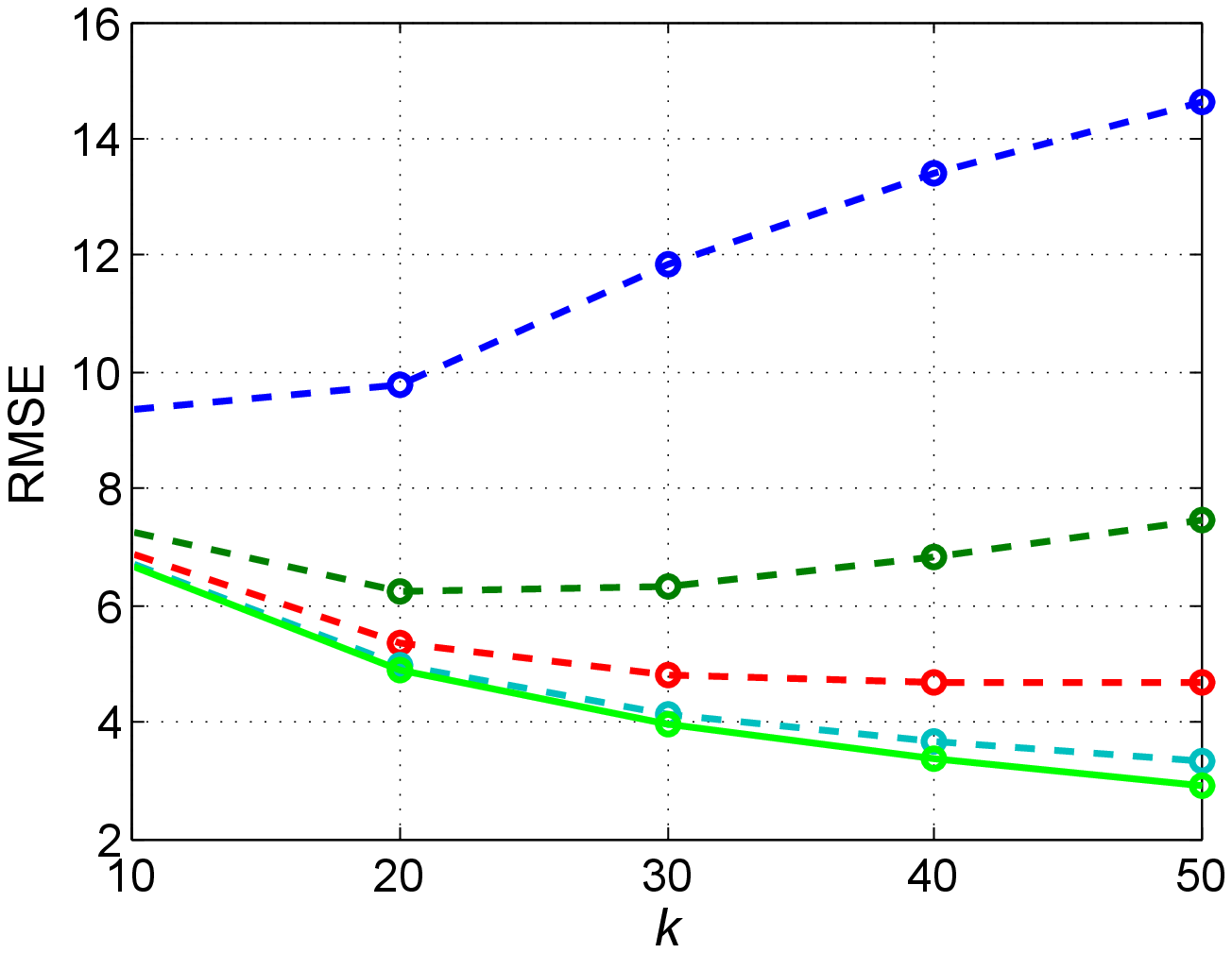}\\
\includegraphics[width=1.7in]{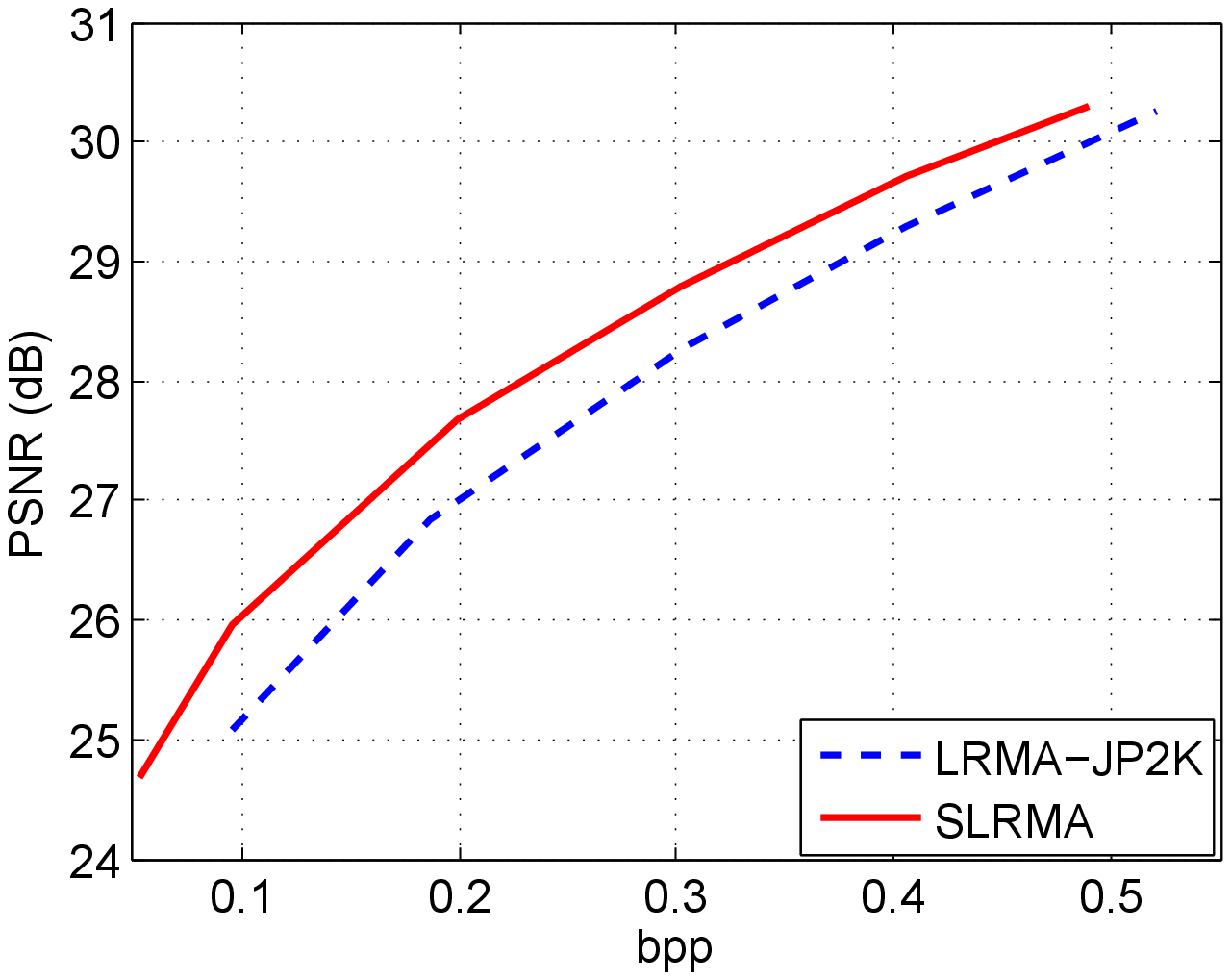}
\includegraphics[width=1.7in]{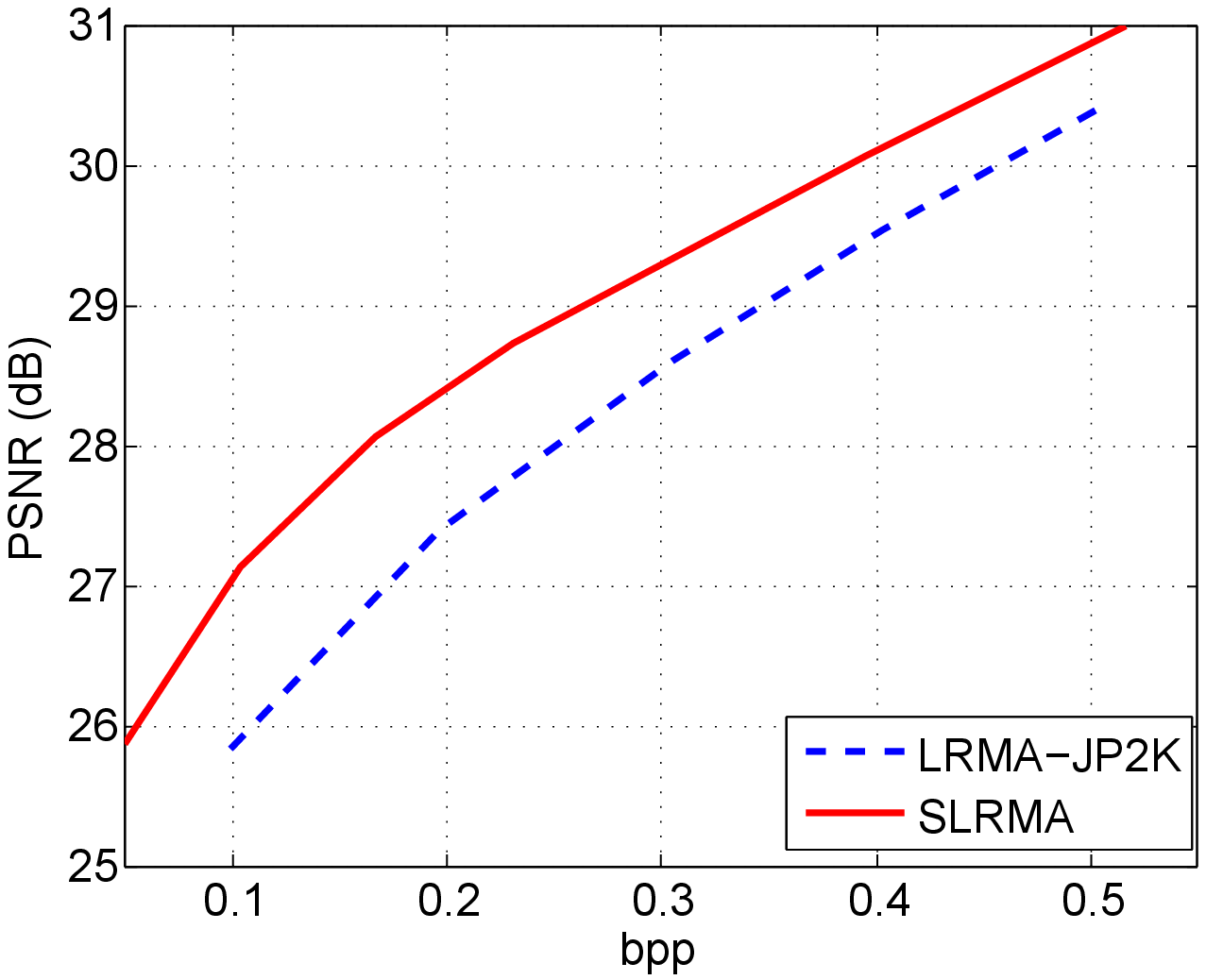}
\includegraphics[width=1.7in]{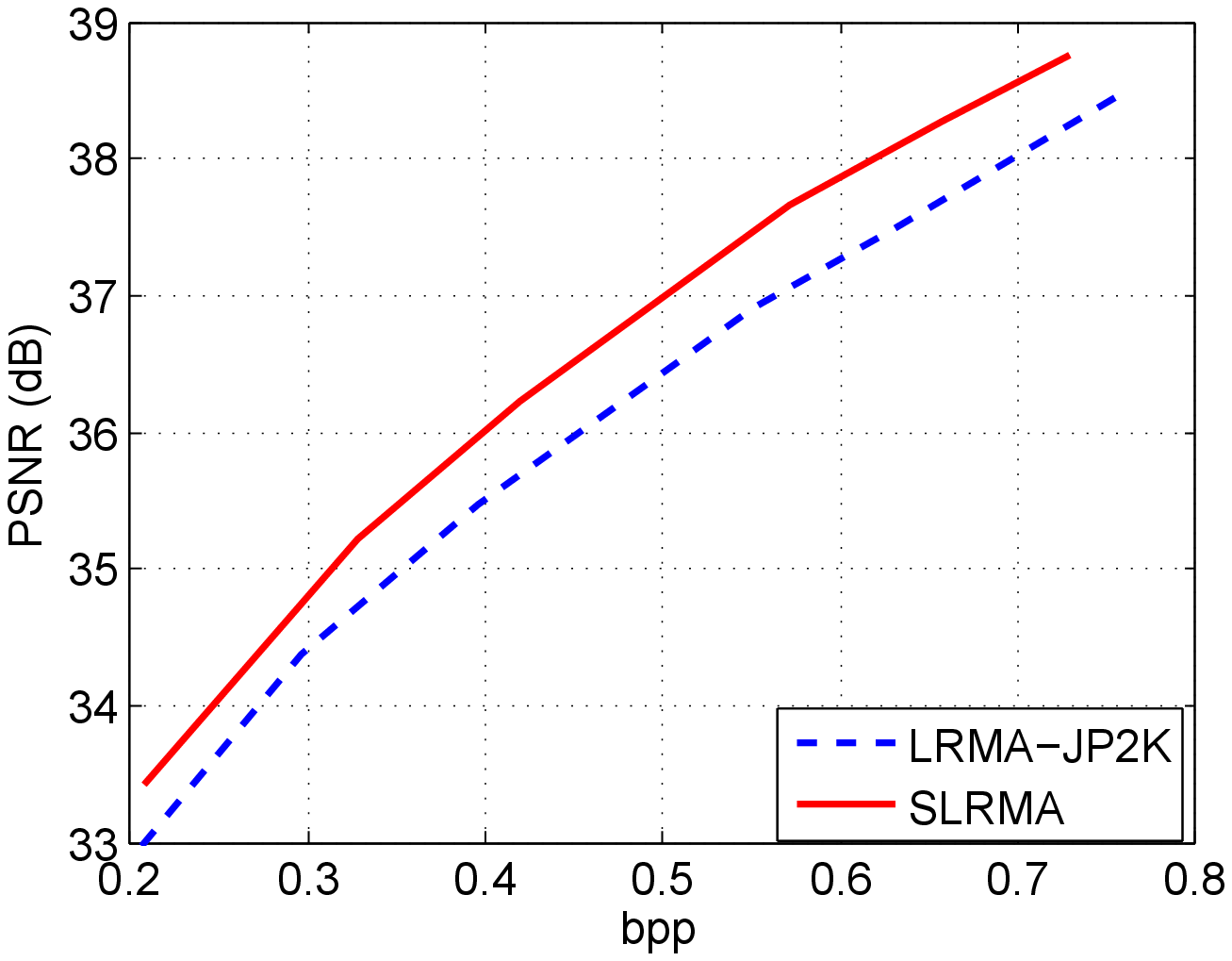}
\includegraphics[width=1.7in]{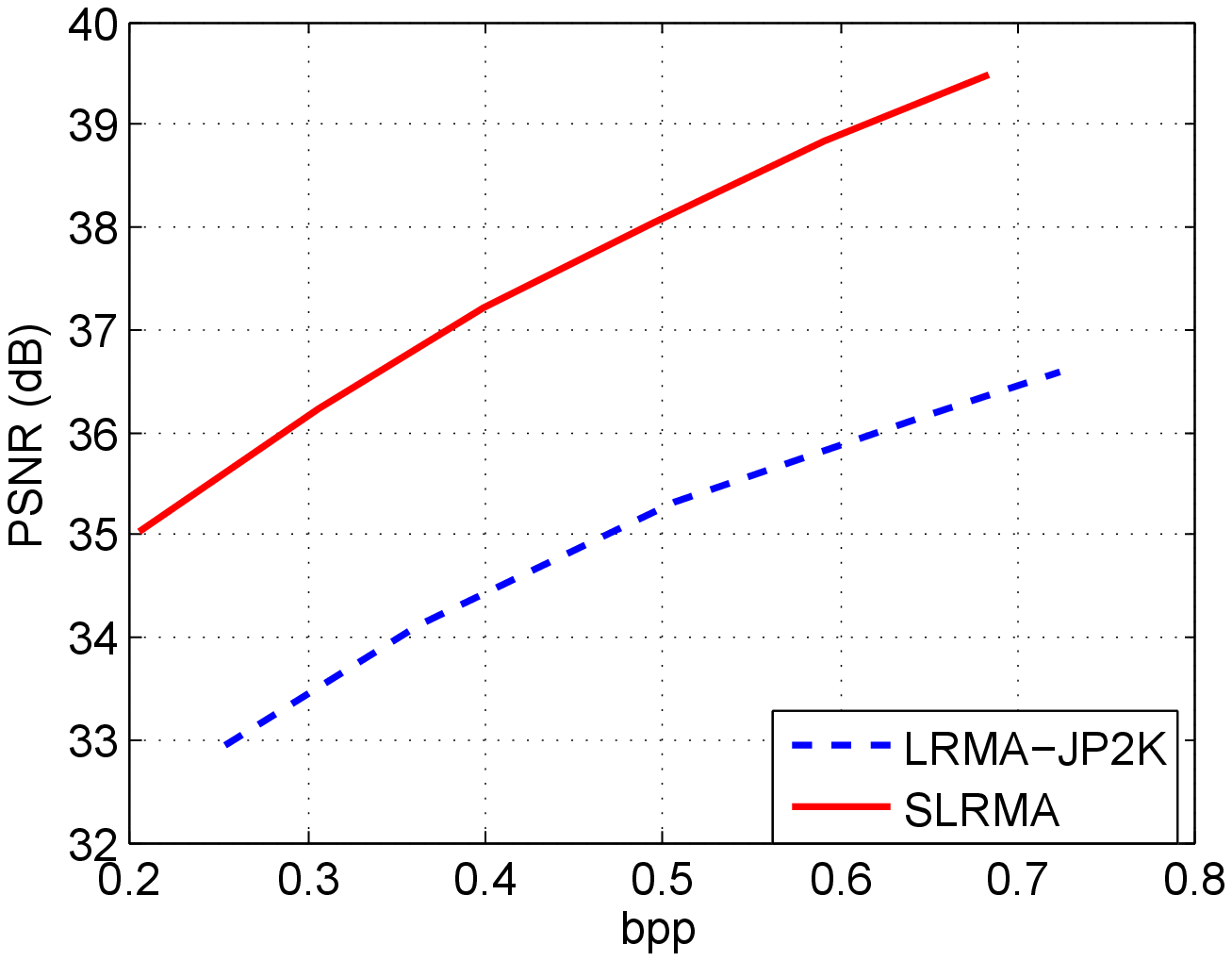}
\makebox[1.7in]{\footnotesize (a) Image set
\uppercase\expandafter{\romannumeral1}}\makebox[1.7in]{\footnotesize
(b) Image set \uppercase\expandafter{\romannumeral2}}
\makebox[1.7in]{\footnotesize (c) Image set
\uppercase\expandafter{\romannumeral3}}\makebox[1.7in]{\footnotesize
(d) Image set \uppercase\expandafter{\romannumeral4}}
\caption{Evaluation of SLRMA on image sets. $1^{st}$ row:
convergence verification of SLRMA ($p_B=0.6$ and $k=30$);
$2^{nd}$row: low-rank approximation performance of SLRMA (DCT) under
various $p_B$; $3^{rd}$ row: performance of the stepwise method
LRMA-DCT under various $p_B$; $4^{th}$ row: low-rank approximation
performance of SLRMA (DWT) under various $p_B$; $5^{th}$ row:
performance of the stepwise method LRMA-DWT under various $p_B$;
$6^{th}$ row: comparison of rate-distortion performance of
SLRMA-based and LRMA-JPEG2K, and $\bf\Phi$ is realized by DWT; Note
that rows 2-5 share the same legend.}\label{fig:result image}
\end{figure*}

Firstly, we evaluate the convergence and low-rank approximation
performance of SLRMA under various $p_B$, i.e., the relationship
among approximation error, $k$ and $p_B$. The approximation error is
measured by the Root Mean Square Error (RMSE) between the original
data $\mathbf{X}$ and approximate data $\widehat{\mathbf{X}}$, i.e.,
\begin{equation}
\centering {\rm
RMSE}=\sqrt{\frac{1}{mn}\left\|\mathbf{X}-\widehat{\mathbf{X}}\right\|_F^2}.
\label{equ:RMSE}
\end{equation}
The SLRAM is validated under two cases, which are denoted by:
(\lowercase\expandafter{\romannumeral1}) SLRMA (DCT) in which
$\bf{\Phi}$ is set as the 2D DCT matrix, i.e., the Kronecker product
of two 1D DCT matrices; (\lowercase\expandafter{\romannumeral2})
SLRMA (DWT) in which $\bf{\Phi}$ is realized by the 2D DWT matrix
obtained as the Kronecker product of two multi-level Haar matrices
(3-level for image sets \uppercase\expandafter{\romannumeral1} and
\uppercase\expandafter{\romannumeral2}, and 4-level for image sets
\uppercase\expandafter{\romannumeral3} and
\uppercase\expandafter{\romannumeral4}). As a comparison, we also
present the performance of stepwise methods, namely LRMA-DCT (resp.
LRMA-DWT), in which the same DCT (resp. DWT) matrix as SLRMA is
applied to columns of $\mathbf{B}$ obtained by LRMA, and then $p_B$
percentage of all transformed coefficients with smallest magnitude
are set to zero before carrying out the inverse transform.

As the first row of Figure \ref{fig:result image} shows, SLRMA
empirically converges well, i.e., the objective function is reduced
to a stable value after a few iterations. The second and forth rows
of Figure \ref{fig:result image} verify the excellent low-rank
approximation performance of SLRMA, that is, at the same $k$, the
approximation error by SLRMA is comparable to that by LRMA, which is
the lower bound. Moreover, SLRMA (DWT) is better than SLRMA (DCT) on
image sets \uppercase\expandafter{\romannumeral3} and
\uppercase\expandafter{\romannumeral4} since the multi-level DWT has
greater potential for sparsly representing images with complex
textures than DCT \cite{rao2000transform}. For image sets
\uppercase\expandafter{\romannumeral1} and
\uppercase\expandafter{\romannumeral2}, which contain relatively
smooth facial images, both DWT and DCT can decorrelate the data
well, so that the difference between SLRMA (DWT) and SLRMA (DCT) is
very slight. However, the stepwise methods LRMA-DCT and LRMA-DWT,
shown in the third and fifth rows, induce much larger approximation
error than SLRMA at the same $k$ and $p_B$. Finally, we show the
rate-distortion performance of the SLRMA-based compression scheme in
the sixth row of Figure \ref{fig:result image}, where we can see
that the SLRMA-based method consistently produces higher
peak-signal-noise-ratio (PSNR) at the same bit per pixel (bpp). The
improvement is up to 3 dB compared to the method in
\cite{du2007hyperspectral}, namely LRMA-JP2K, in which the DWT-based
JP2K image encoder is employed to encode the basis vectors after
LRMA is applied. Figure \ref{fig:visualhall} also compares visual
results by SLRMA and LRMA-JP2K, where we can observe that areas
around persons marked out by red ovals are obviously blurred for
LRMA-JP2K images, but the corresponding images by SLRMA are sharper
and visually closer to original images at the same bpp. Last but not
least, we believe that our SLRMA-based compression scheme can
achieve higher rate-distortion performance by adopting more advanced
entropy coding, e.g., context-adaptive binary arithmetic coding
(CABAC) \cite{marpe2003context} and  embedded block coding with
optimal truncation (EBCOT) \cite{skodras2001jpeg}.
\begin{figure*}
\centering
\includegraphics[width=7.0in]{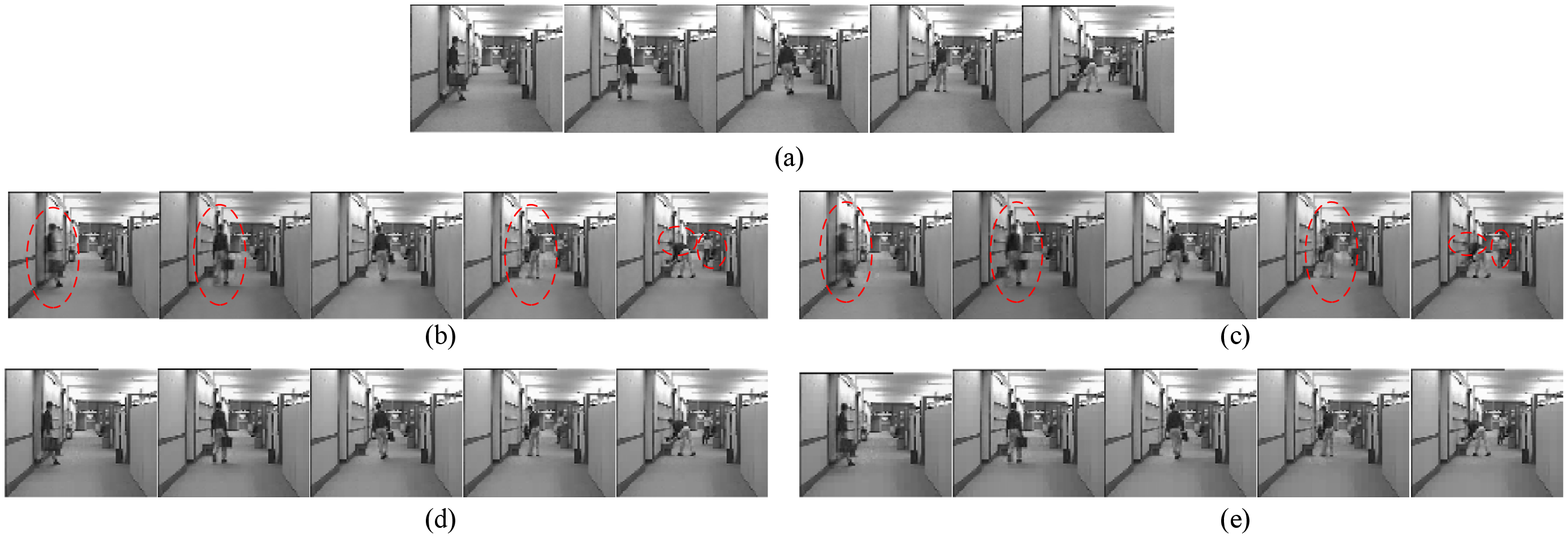}
\label{fig:visualhall} \caption{ Comparison of visual results on
image set. (a) original images from the Image set
\uppercase\expandafter{\romannumeral4}. (b) and (c) compressed ones
by the LRMA-JP2K method under 0.38 bpp and 0.21 bpp, respectively;
(d) and (e) compressed ones by our SLRMA-based method under 0.38 bpp
and 0.21 bpp, respectively.} \label{fig:visualhall}
\end{figure*}

\subsection{3D Dynamic Meshes}

\begin{figure}
\centering
\includegraphics[width=3.0in]{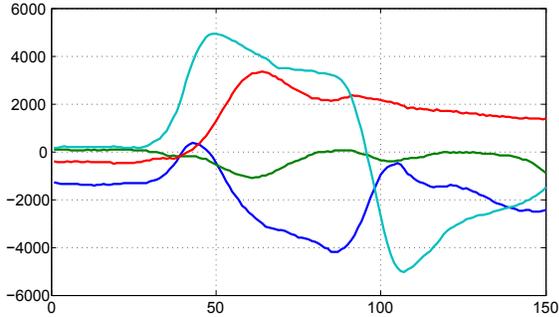}
\caption{Illustration of the coherence within rows of
$\mathbf{C}_d$.} \label{fig:row of C}
\end{figure}

Given a 3D dynamic mesh sequence with $m$ vertices and $n$ frames,
it can be represented as three matrices $\mathbf{X}_x\in
\mathbb{R}^{m\times n}$, $\mathbf{X}_y\in \mathbb{R}^{m\times n}$,
$\mathbf{X}_z\in \mathbb{R}^{m\times n}$, corresponding to the $x-$,
$y-$, and $z-$dimensions of vertices, respectively, and each column
of $\mathbf{X}_d~~(d:=\{x, y, z\})$ corresponds to the $d-$dimension
of vertices of one frame. The SLRMA-based compression scheme is
shown in Figure \ref{fig:flowchart}(b), where $\mathbf{X}_d$ is
firstly factored into $\mathbf{B}_d$ and $\mathbf{C}_d$ using SLRMA.
Different from image sets, 3D dynamic mesh sequences consist of
successive frames of objects in motion, so that each row of
$\mathbf{X}_d$, corresponding to the $d$-dimensional trajectory of
one vertex, is a relatively smooth curve, indicating strong
coherence. After the SLRMA, such coherence still exists in rows of
$\mathbf{C}_d$ \cite{karni2004compression}. For example, Figure
\ref{fig:row of C} plots several rows of $\mathbf{C}_d$, where their
relative smoothness is verified. This can be easily explained using
the analysis for intra-correlation of $\mathbf{B}$ of LRMA along the
row space of $\mathbf{X}$. To further reduce this temporal
coherence, 1D DCT is separately performed on rows of $\mathbf{C}_d$,
i.e.,
\begin{equation}
\centering
\widetilde{\mathbf{C}}_d=\mathbf{U}_{dct}^\textsf{T}\mathbf{C}_d^\textsf{T},
\end{equation}
where $\mathbf{U}_{dct}\in \mathbb{R}^{n\times n}$ stands for the 1D
DCT matrix, and each column corresponds to one frequency. Lastly,
the nonzero entries of uniformly quantized $\mathbf{B}_d$ and
$\widetilde{\mathbf{C}}_d$ as well as their locations are
entropy-coded using arithmetic coding.

We take four 3D dynamic mesh sequences from
\cite{dynamicmeshdataset}, including ``Wheel" (2501 vertices),
``Handstand" (2501 vertices), ``Skirt" (2502 vertices), and ``Dance"
(1502 vertices). Each sequence has 150 frames. Some samples are
shown in Figure \ref{fig:datasamples}(f). The values of $\rho$,
$\alpha$, and $\rho_{max}$ are set to $10^7$, 1.003, and $10^{12}$,
respectively. The orthogonal matrix $\bf{\Phi}$ is realized by the
GT, as explained in Section \ref{subsec:GT}, and can be computed
according to the topology of 3D meshes. Note that the three
dimensions are equally treated, i.e., the same $k$ and $p_B$ are
used.

\begin{figure*}
\centering
\includegraphics[width=1.7in]{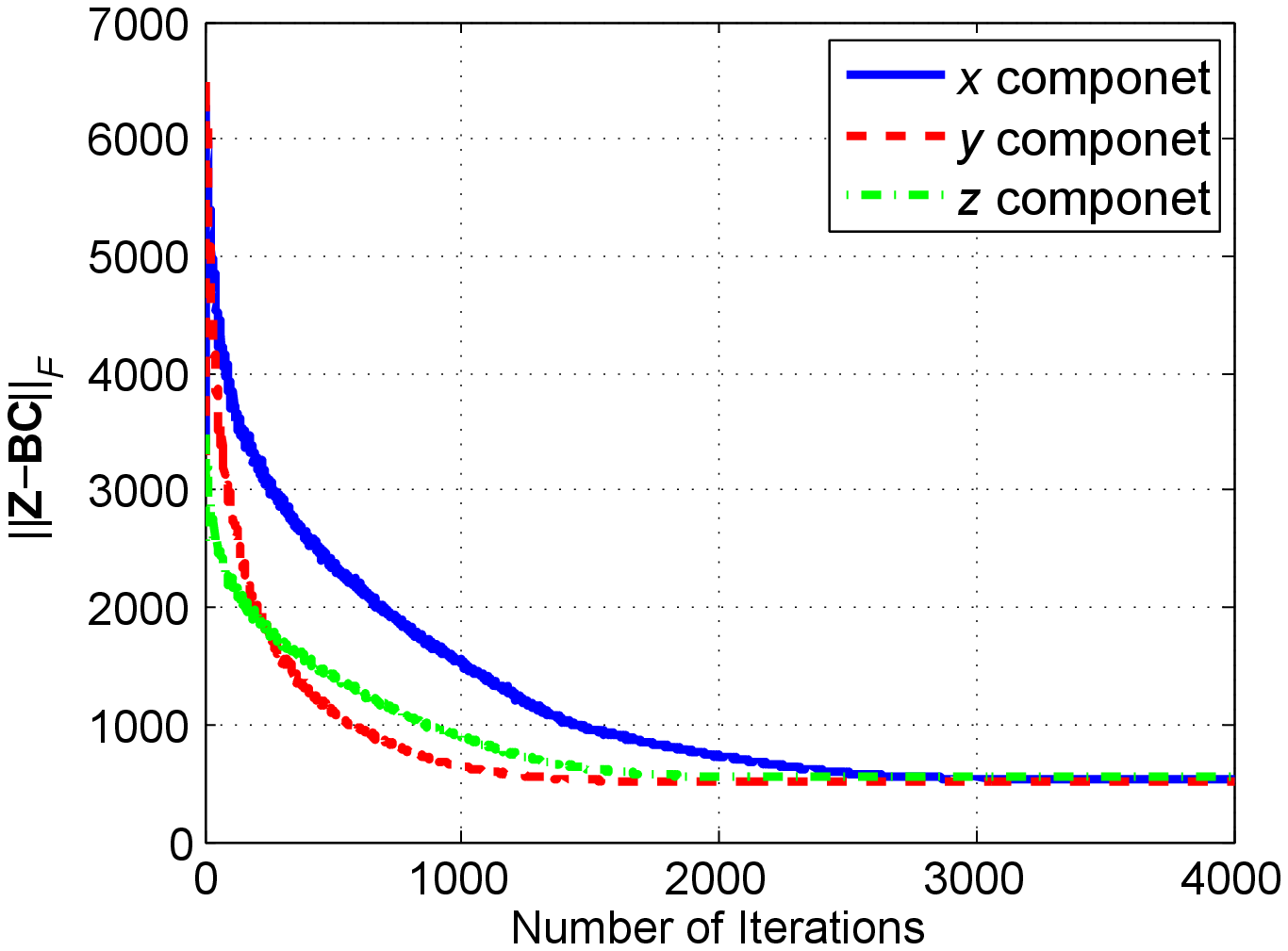}
\includegraphics[width=1.7in]{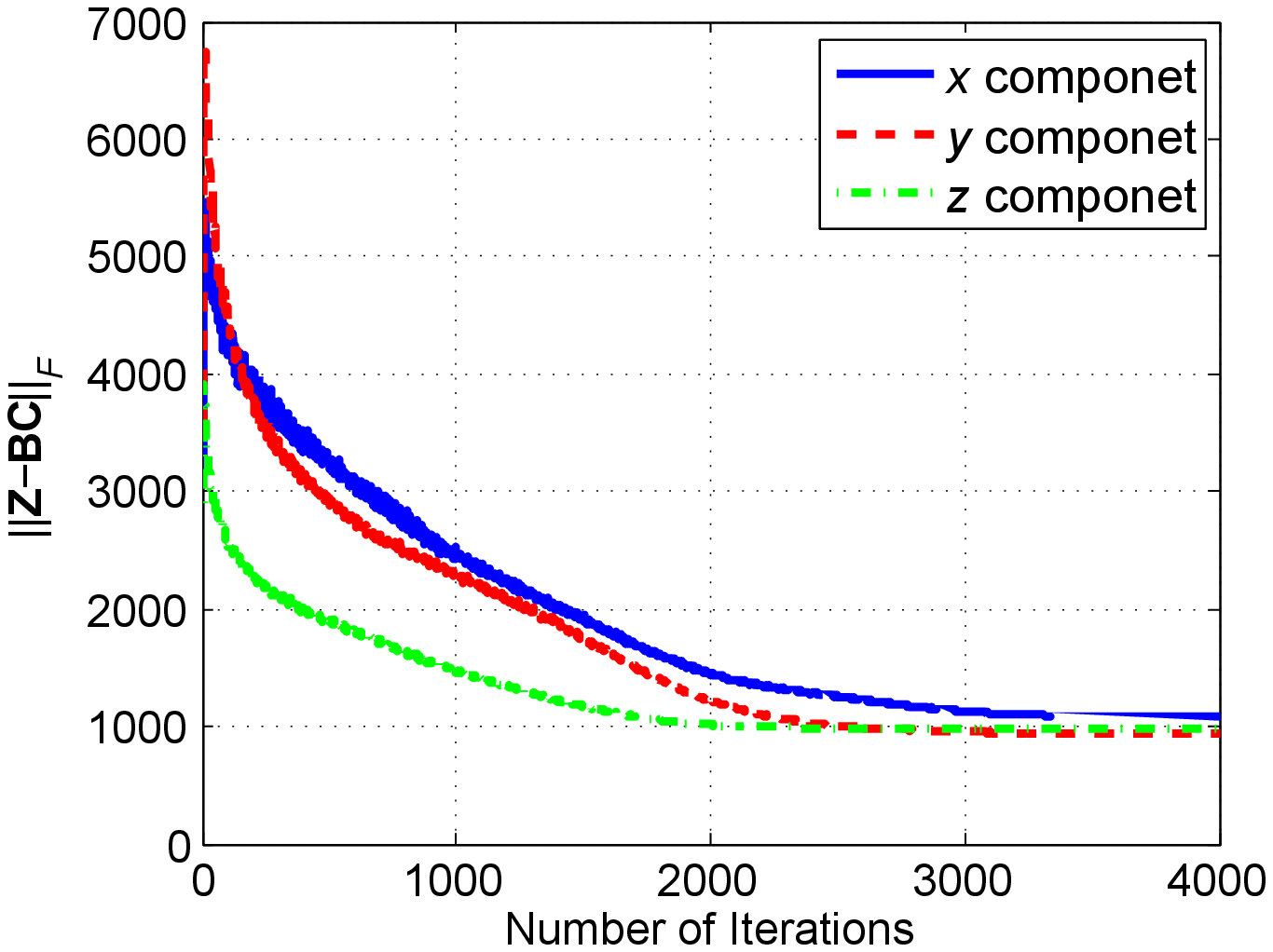}
\includegraphics[width=1.7in]{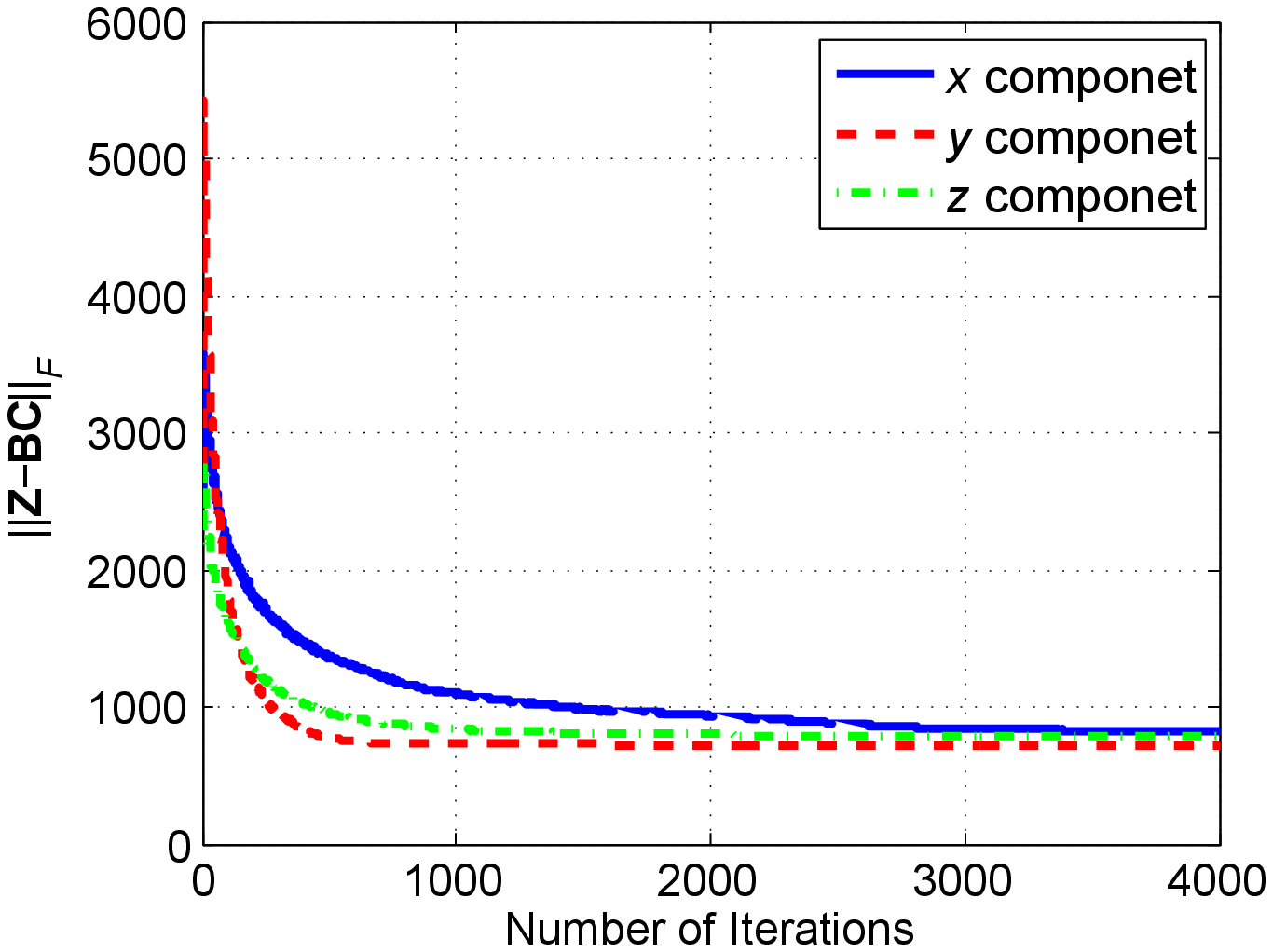}
\includegraphics[width=1.7in]{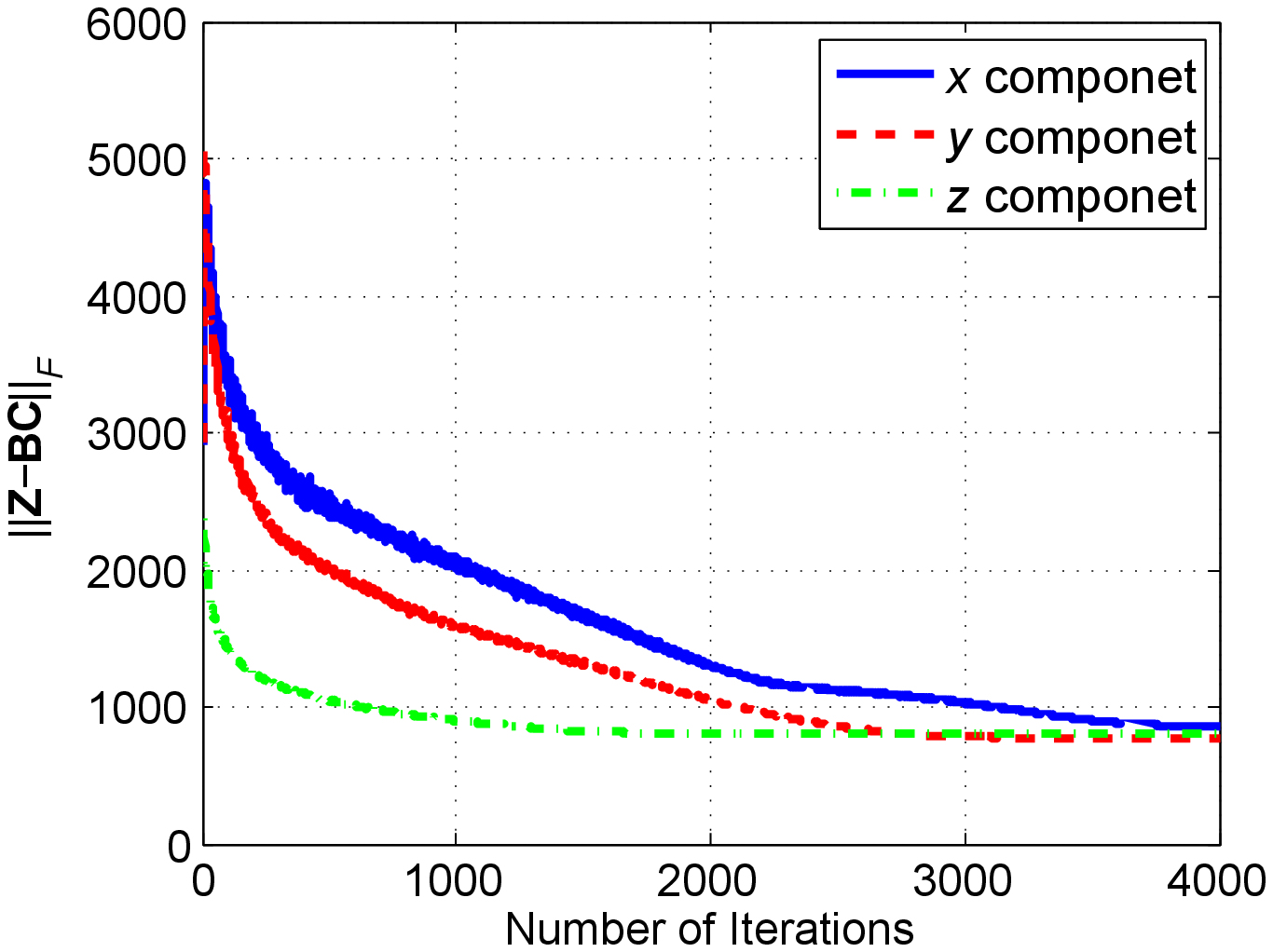}\\
\includegraphics[width=1.7in]{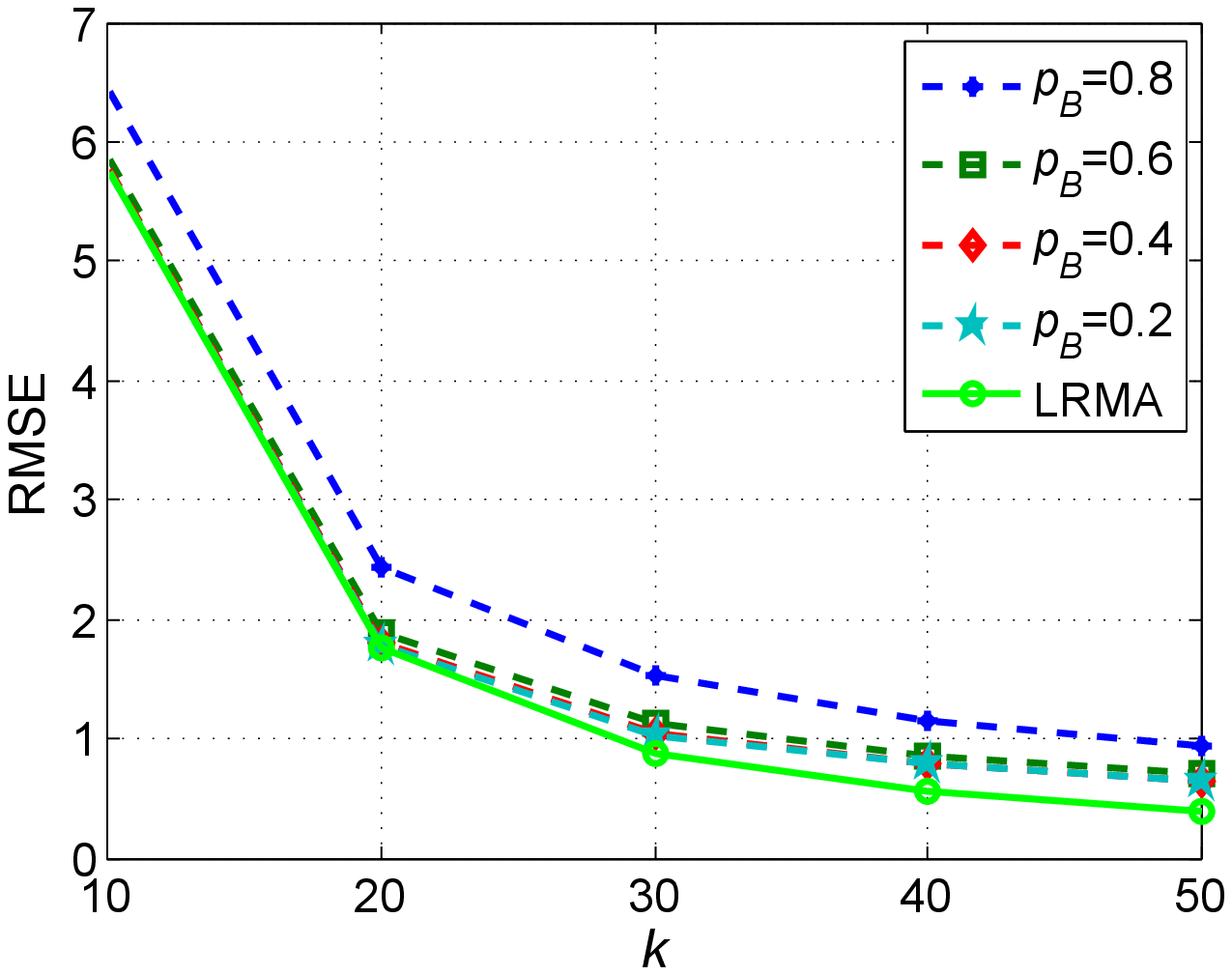}
\includegraphics[width=1.7in]{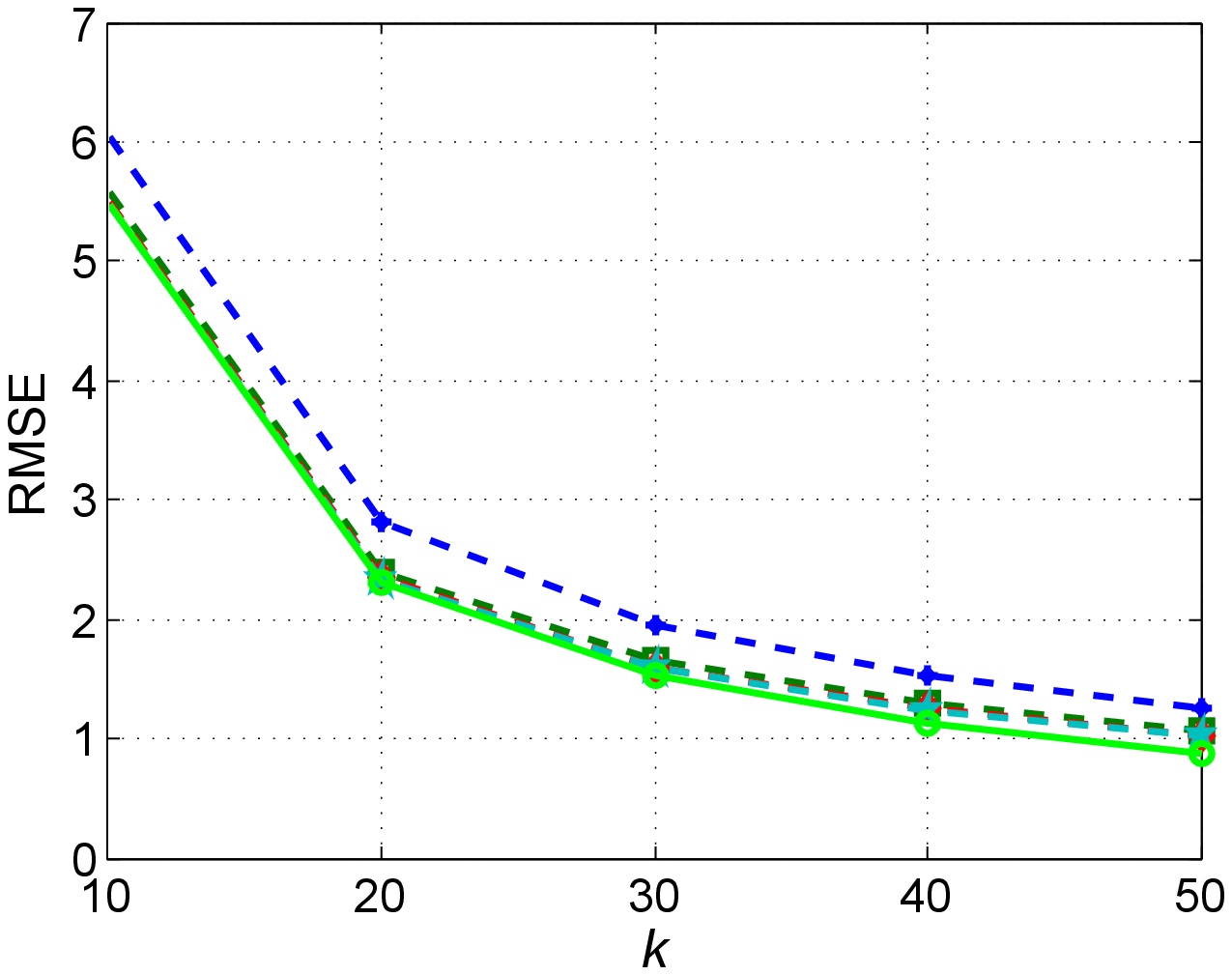}
\includegraphics[width=1.7in]{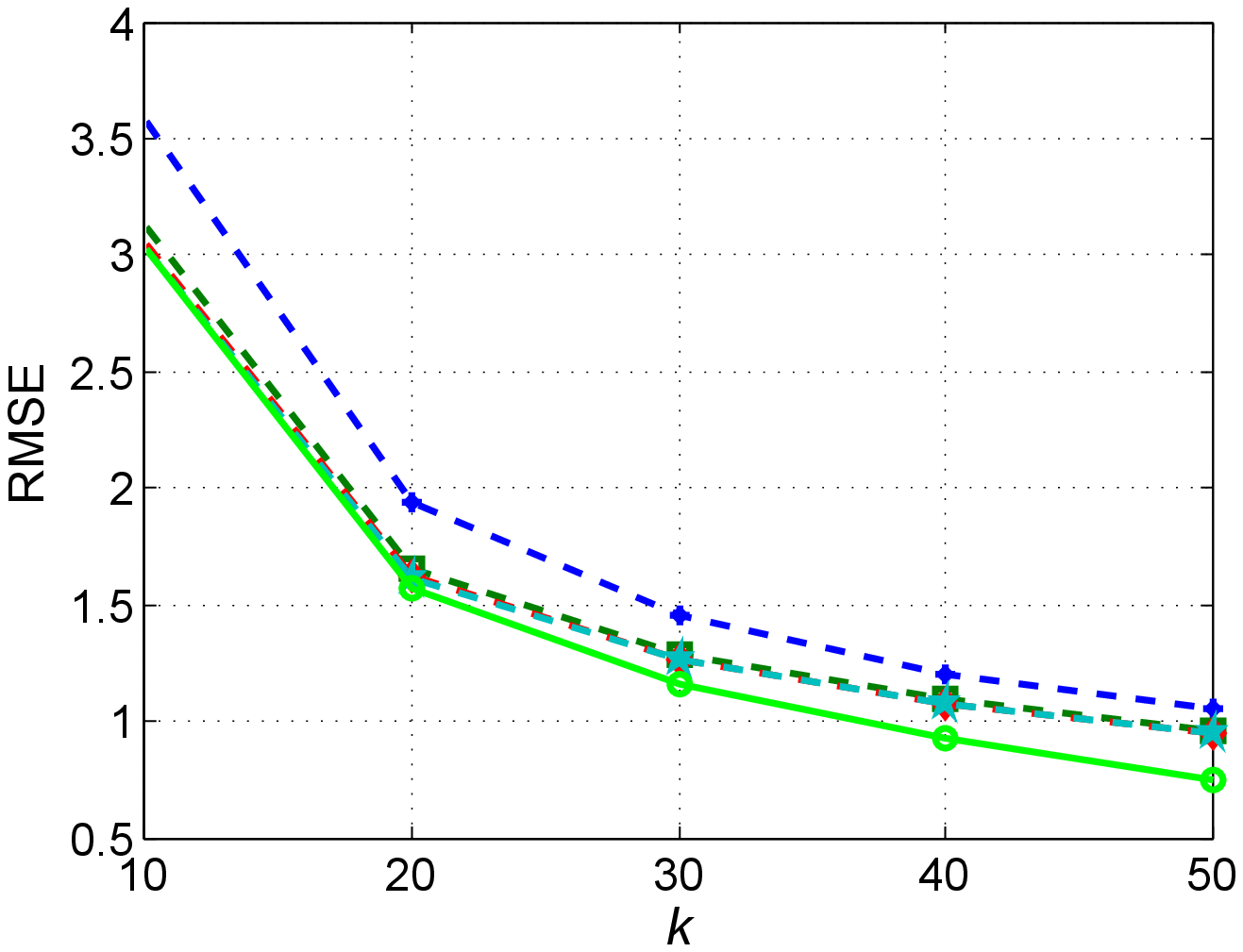}
\includegraphics[width=1.7in]{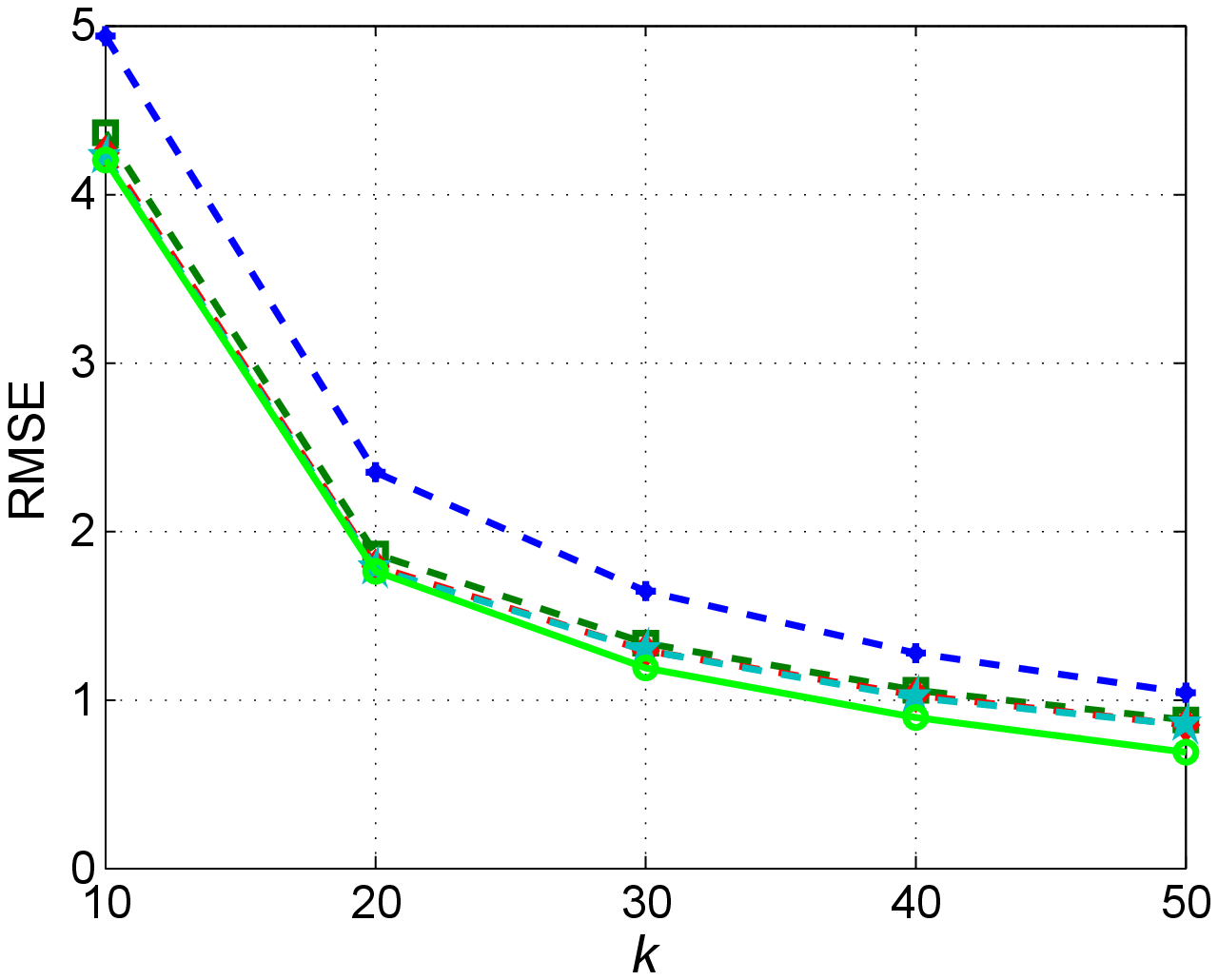}\\
\includegraphics[width=1.7in]{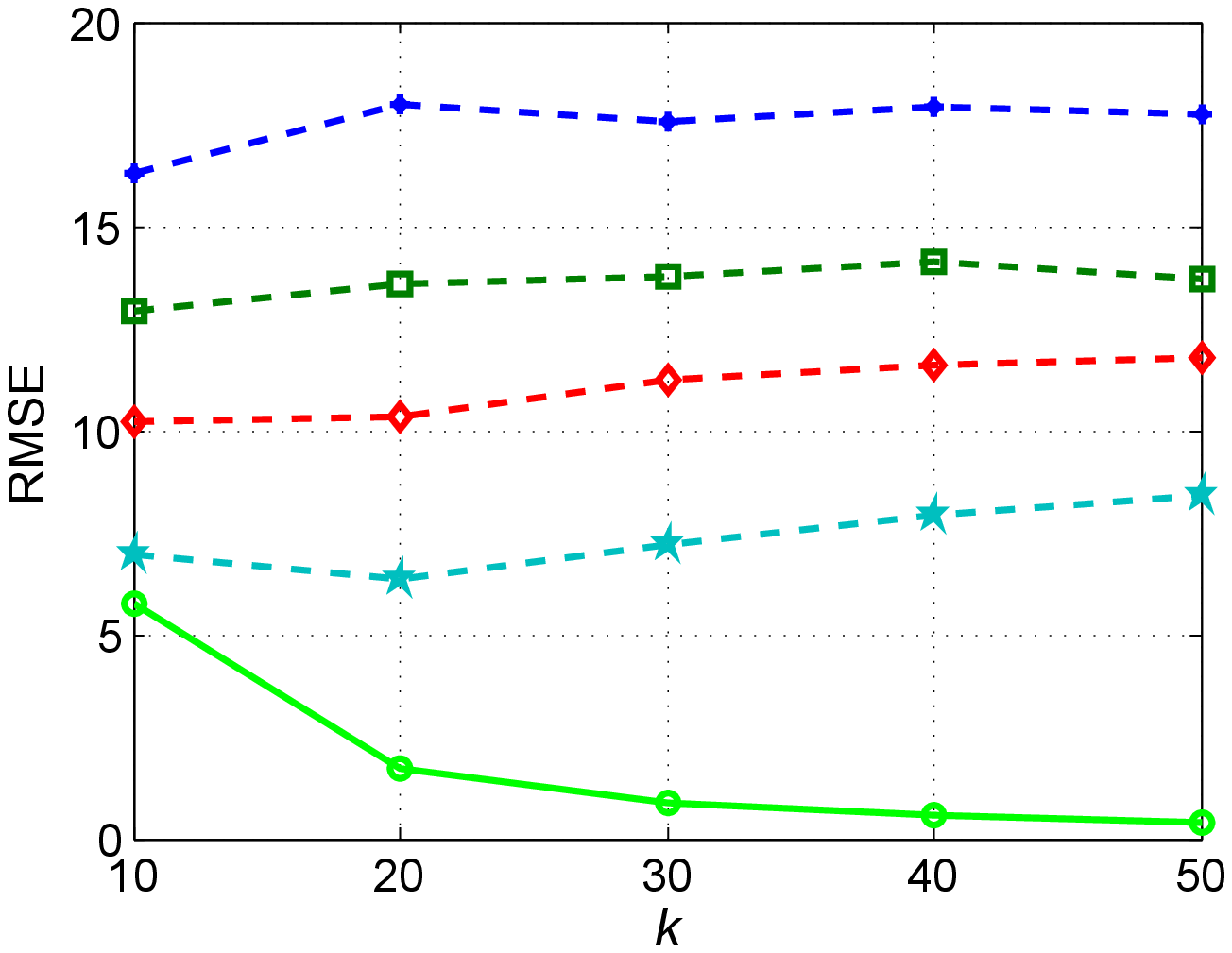}
\includegraphics[width=1.7in]{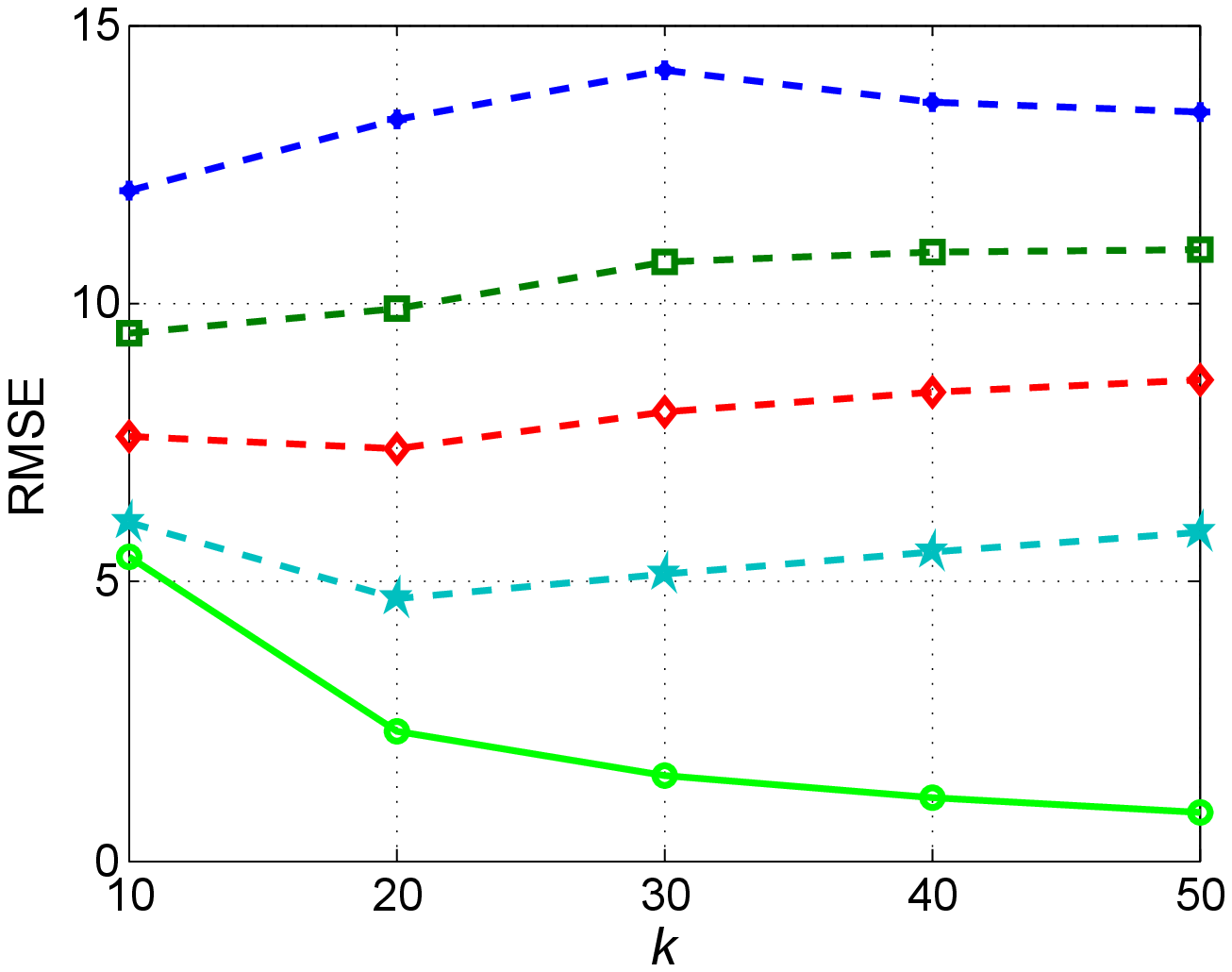}
\includegraphics[width=1.7in]{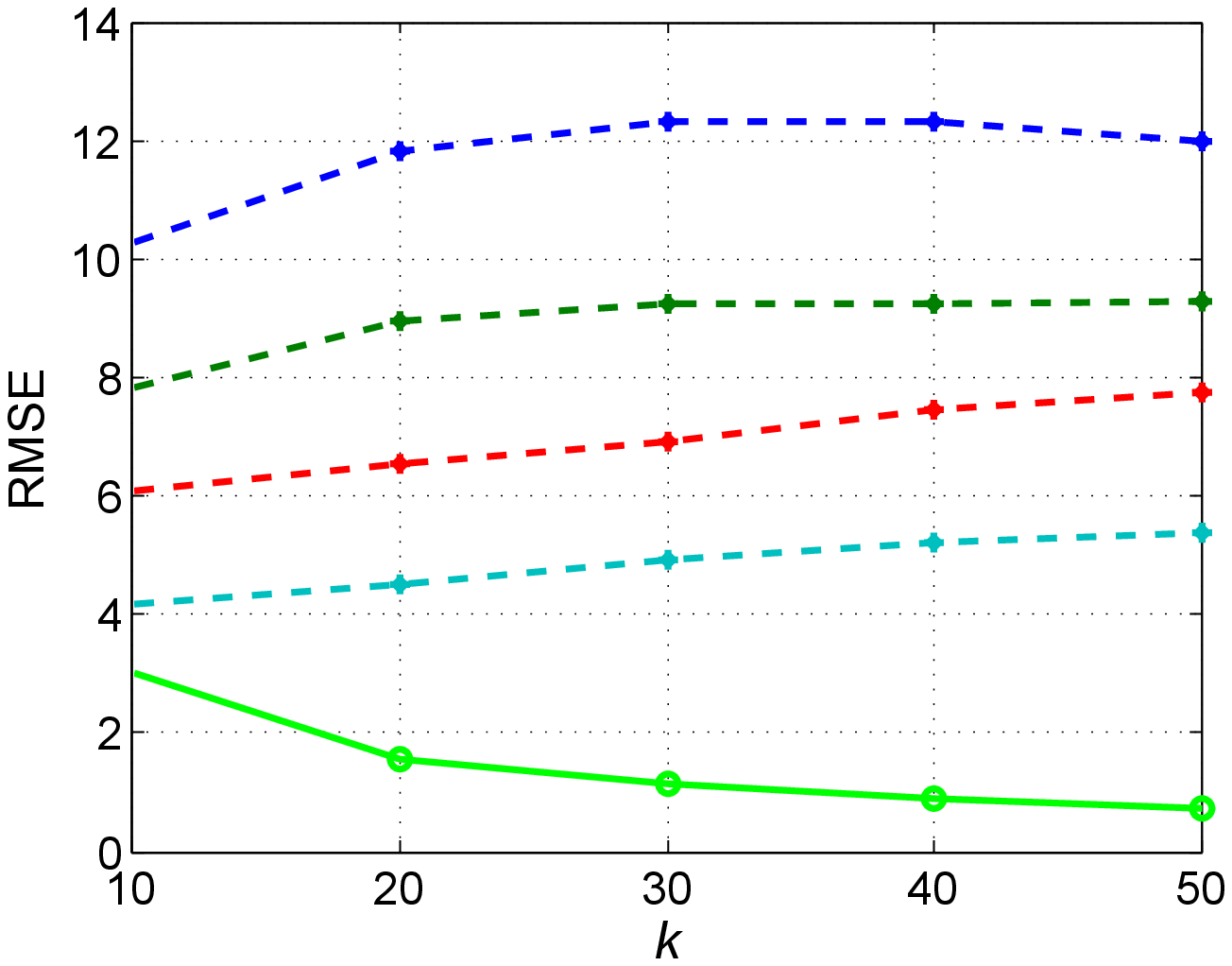}
\includegraphics[width=1.7in]{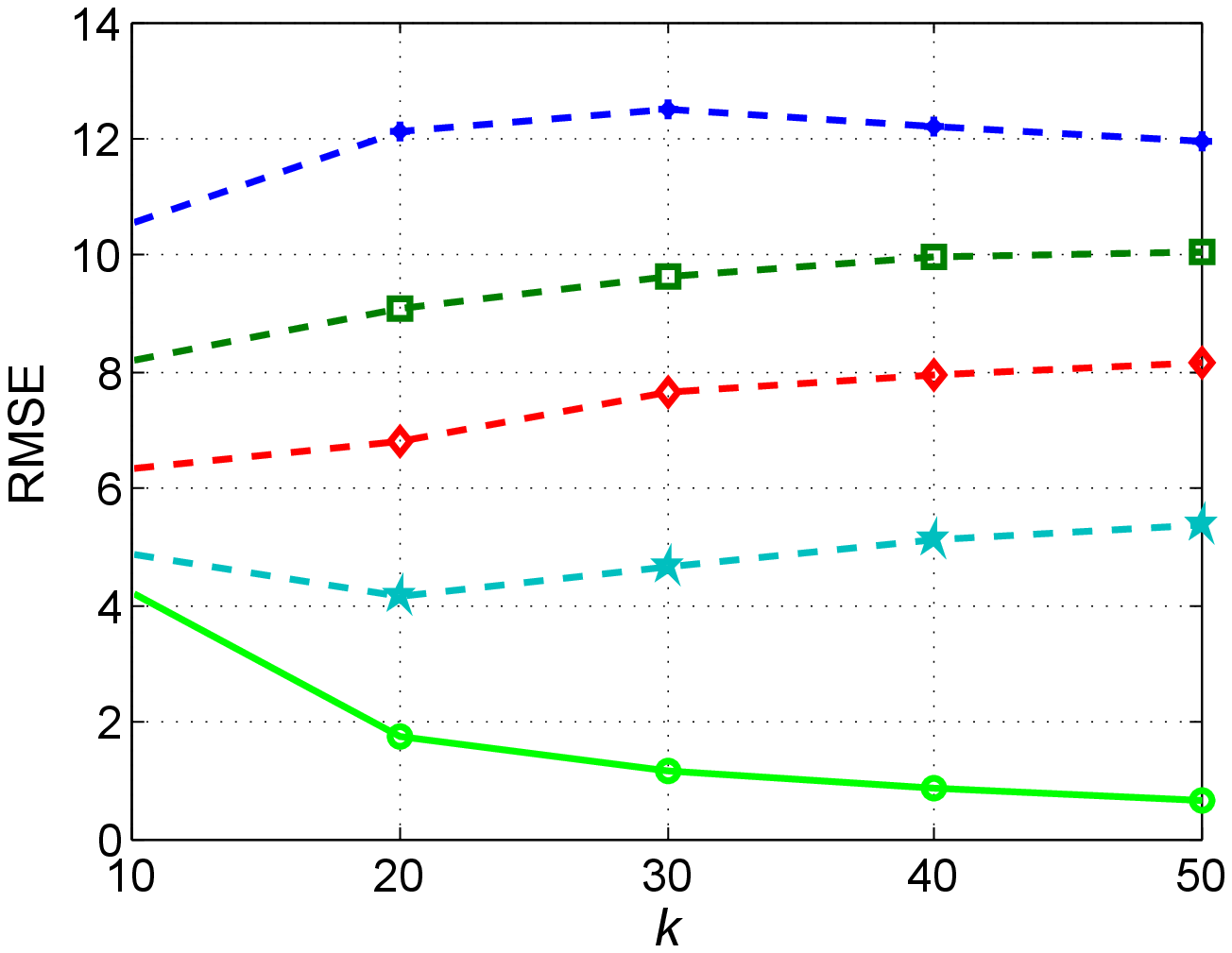}\\
\includegraphics[width=1.7in]{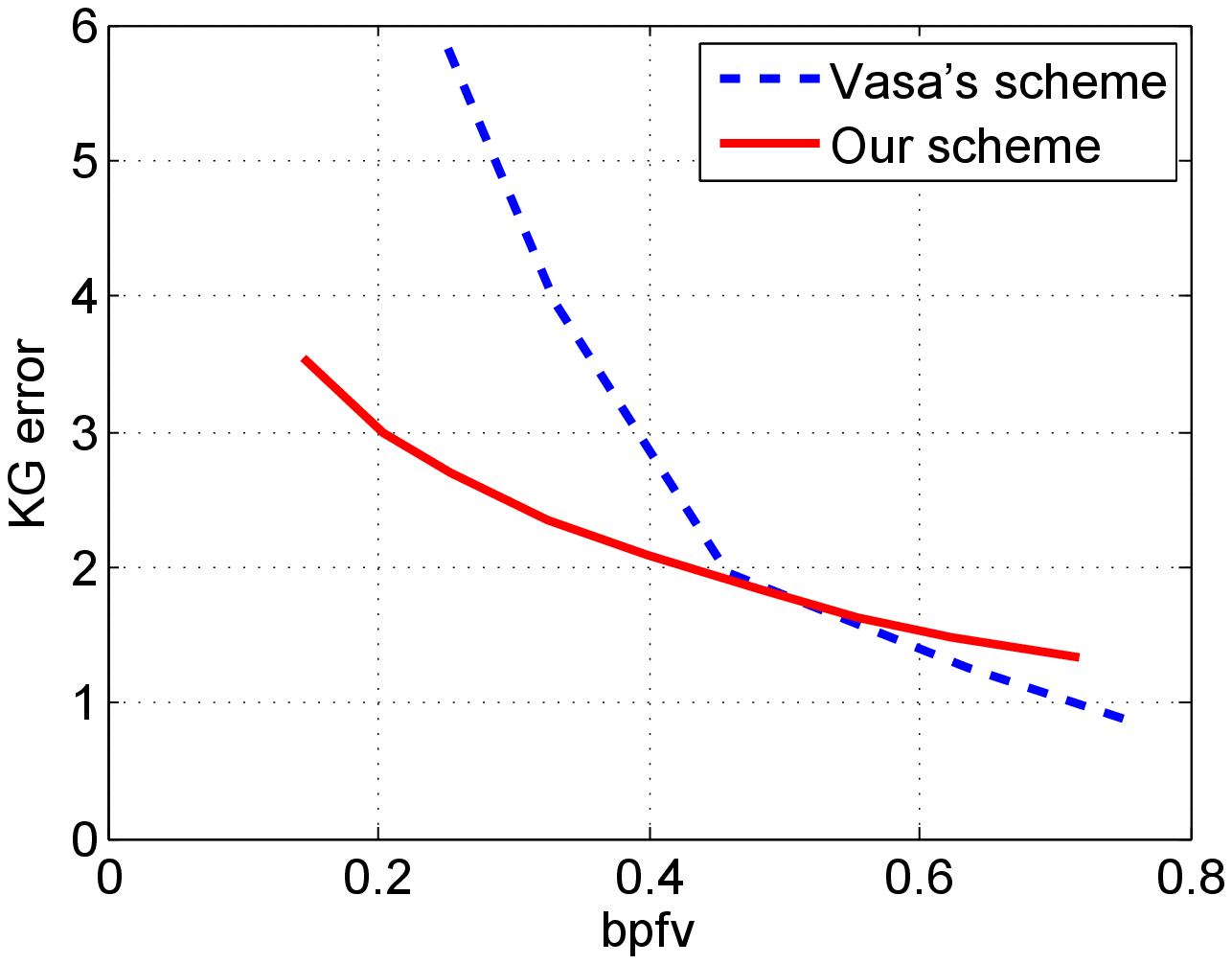}
\includegraphics[width=1.7in]{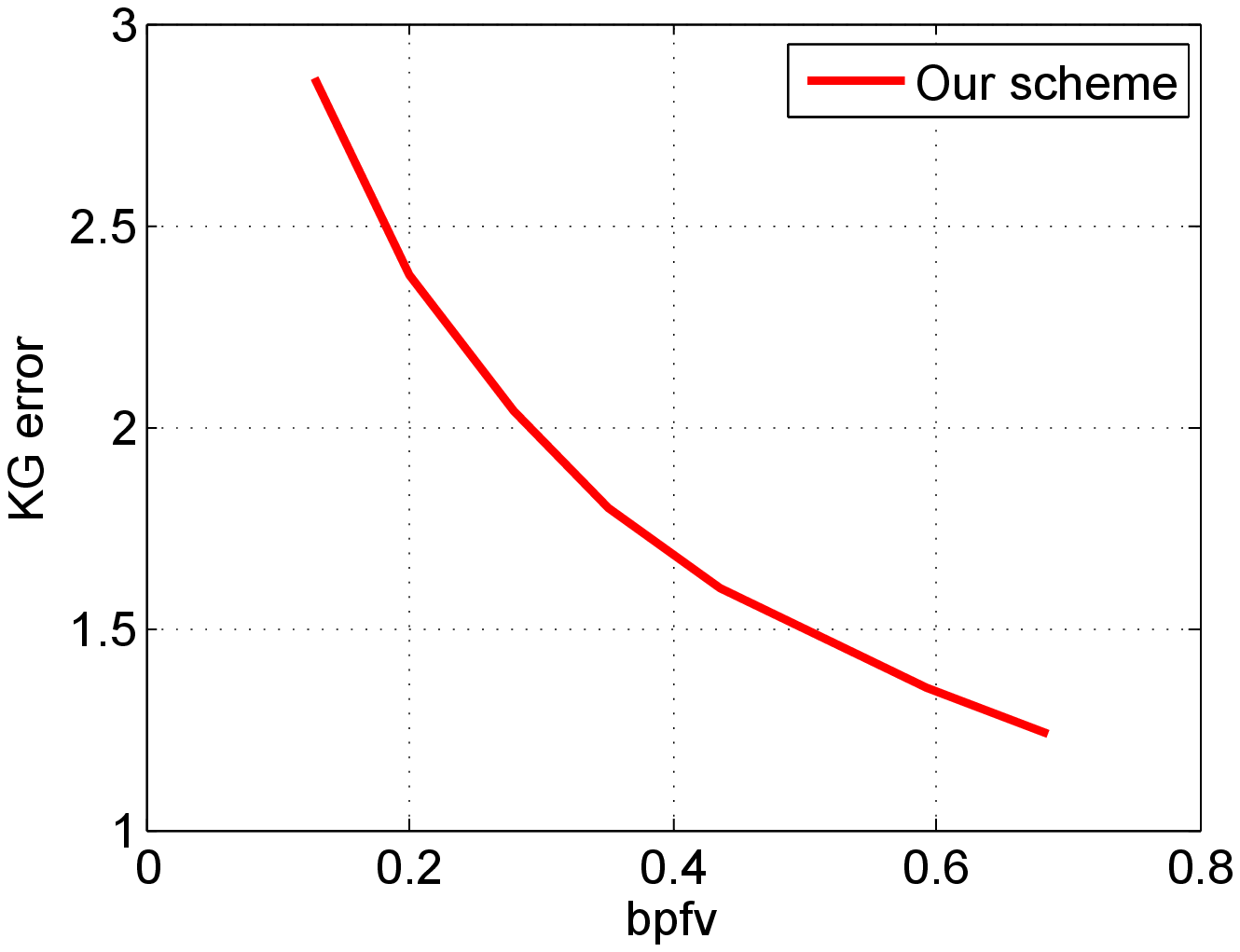}
\includegraphics[width=1.7in]{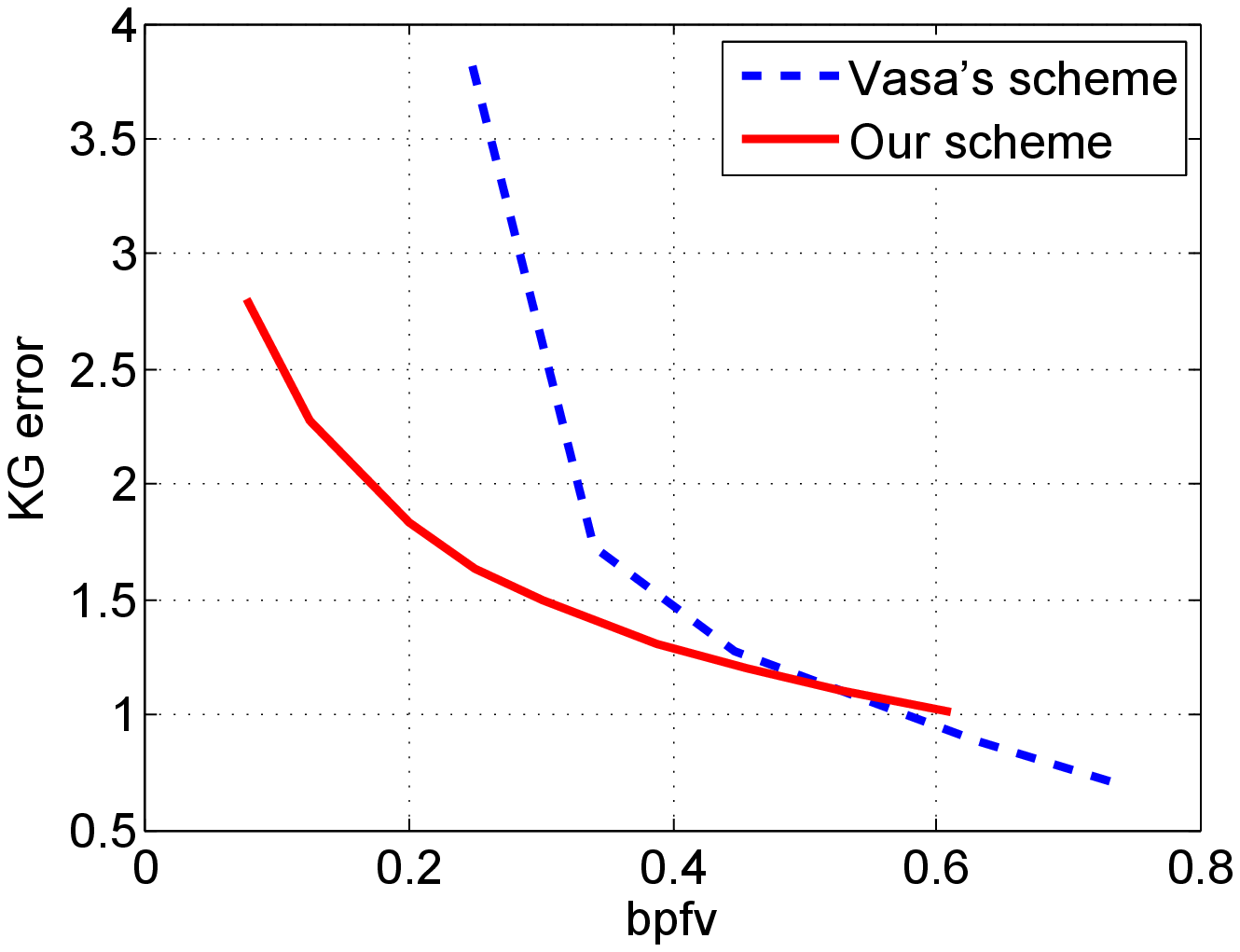}
\includegraphics[width=1.7in]{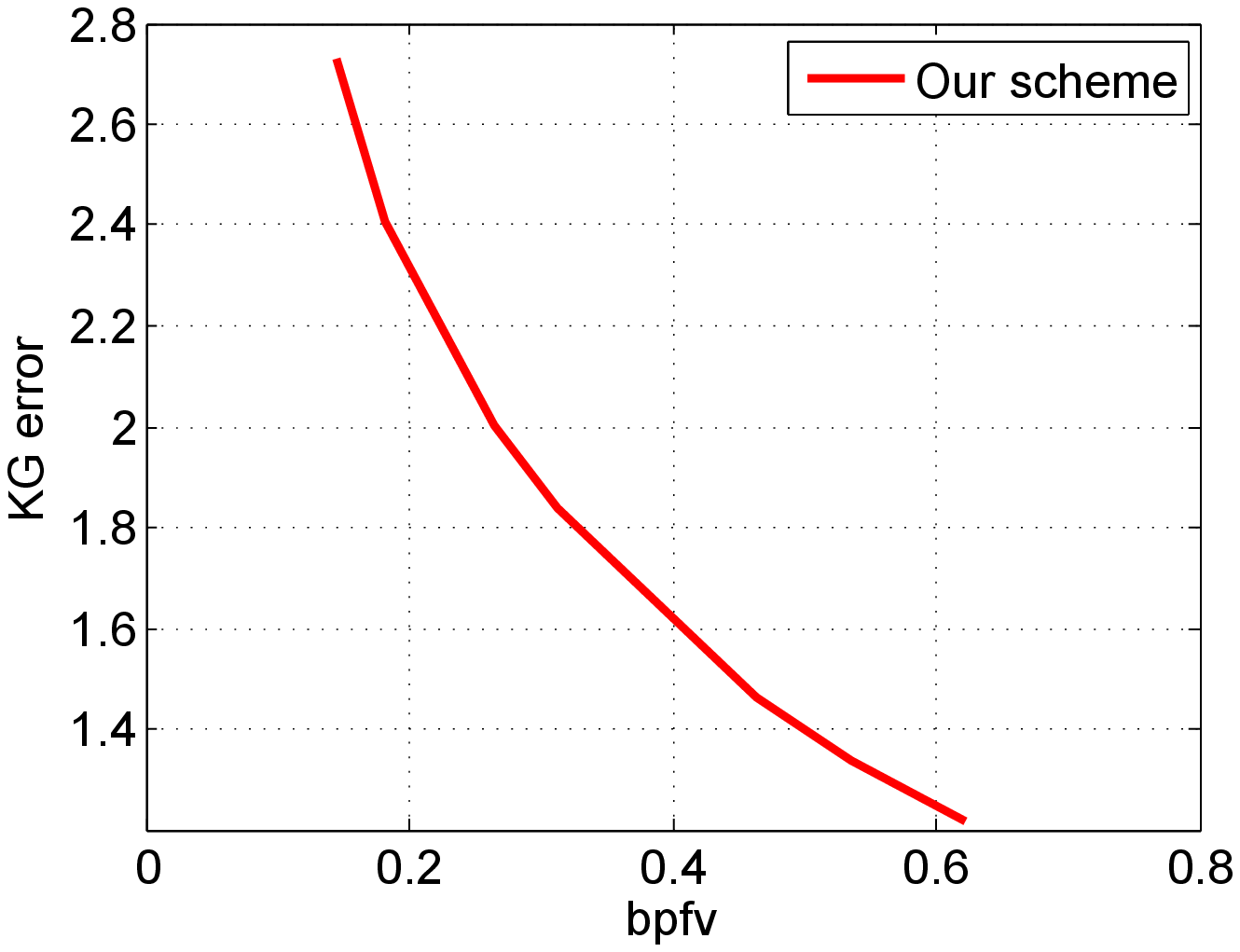}
\makebox[1.7in]{\footnotesize (a)
``Dance"}\makebox[1.7in]{\footnotesize (b) ``Handstand"}
\makebox[1.7in]{\footnotesize (c)
``Skirt"}\makebox[1.7in]{\footnotesize (d) ``Wheel"}
\caption{Evaluation of SLRMA on 3D dynamic meshes. $1^{st}$ row: the
convergence verification of SLRMA ($p_B=0.6$ and $k=30$); $2^{nd}$
row: low-rank approximation performance of SLRMA (GT) at various
$p_B$; $3^{rd}$ row: performance of the stepwise method LRMA-GT at
various $p_B$; $4^{th}$ row: rate-distortion performance of our
SLRMA-based method and the state-of-the-art method. Note that rows
2-3 share the same legend.}\label{fig:resultmeshandmocap}
\end{figure*}

The empirical convergence of SLRMA is verified again in the first
row of Figure \ref{fig:resultmeshandmocap}. The low-rank
approximation performance of SLRMA is shown in the second row of
Figure \ref{fig:resultmeshandmocap}, where we can see that at the
same $k$, SLRMA produces comparable RMSEs as LRMA even when the
value of $p_B$ increases up to 0.8. However, the stepwise method
denoted by LRMA-GT produces much larger RMSEs. See the third row of
Figure \ref{fig:resultmeshandmocap}. Note that
$\mathbf{X}=[\mathbf{X}_x; \mathbf{X}_y; \mathbf{X}_z]\in
\mathbb{R}^{3m\times n}$ and
$\widehat{\mathbf{X}}=[\widehat{\mathbf{X}}_x;
\widehat{\mathbf{X}}_y;
\widehat{\mathbf{X}}_z]\in\mathbb{R}^{3m\times n}$ when computing
the approximation error using (\ref{equ:RMSE}).

Finally, we evaluate the rate-distortion performance of the
SLRMA-based compression method. The compression distortion is
measured by the widely-adopted KG error \cite{karni2004compression},
defined as
\begin{equation}
\centering {\rm
KG~error}=100\times\frac{\left\|\mathbf{X}-\widehat{\mathbf{X}}\right\|_F}{\left\|\mathbf{X}-E(\mathbf{X})\right\|_F},
\end{equation}
where $E(\mathbf{X})$ is an average matrix of the same size as
$\mathbf{X}$, of which the $j$-th column is
$((\overline{\mathbf{x}}_x)_j[1~ \cdots~ 1]~
(\overline{\mathbf{x}}_y)_j[1~ \cdots~ 1]~
(\overline{\mathbf{x}}_z)_j[1~ \cdots~ 1])^\textsf{T}$ with
$(\overline{\mathbf{x}}_d)_j$ being the average of
$(\mathbf{x}_d)_j$. The bitrate is measured in bit per frame per
vertex (bpfv). The overall compression performance of our scheme is
affected by three parameters: $k$ (the number of basis vectors),
$p_B$ (the percentage of zero elements of $\mathbf{B}$), and the
quantization parameter. Currently, we use exhaustive search to
determine their optimal combination. One can also employ the method
in \cite{petvrik2010finding} or build rate and distortion models in
terms of the three parameters, like \cite{Hou2015compressing}, to
speed up this process in practice. We compare with V\'{a}\v{s}a's
scheme in \cite{vavsa2014dynamiccompressing}, which is LRMA-based
and represents the state-of-the-art performance.

\begin{figure*}
\centering
\includegraphics[width=7.0in]{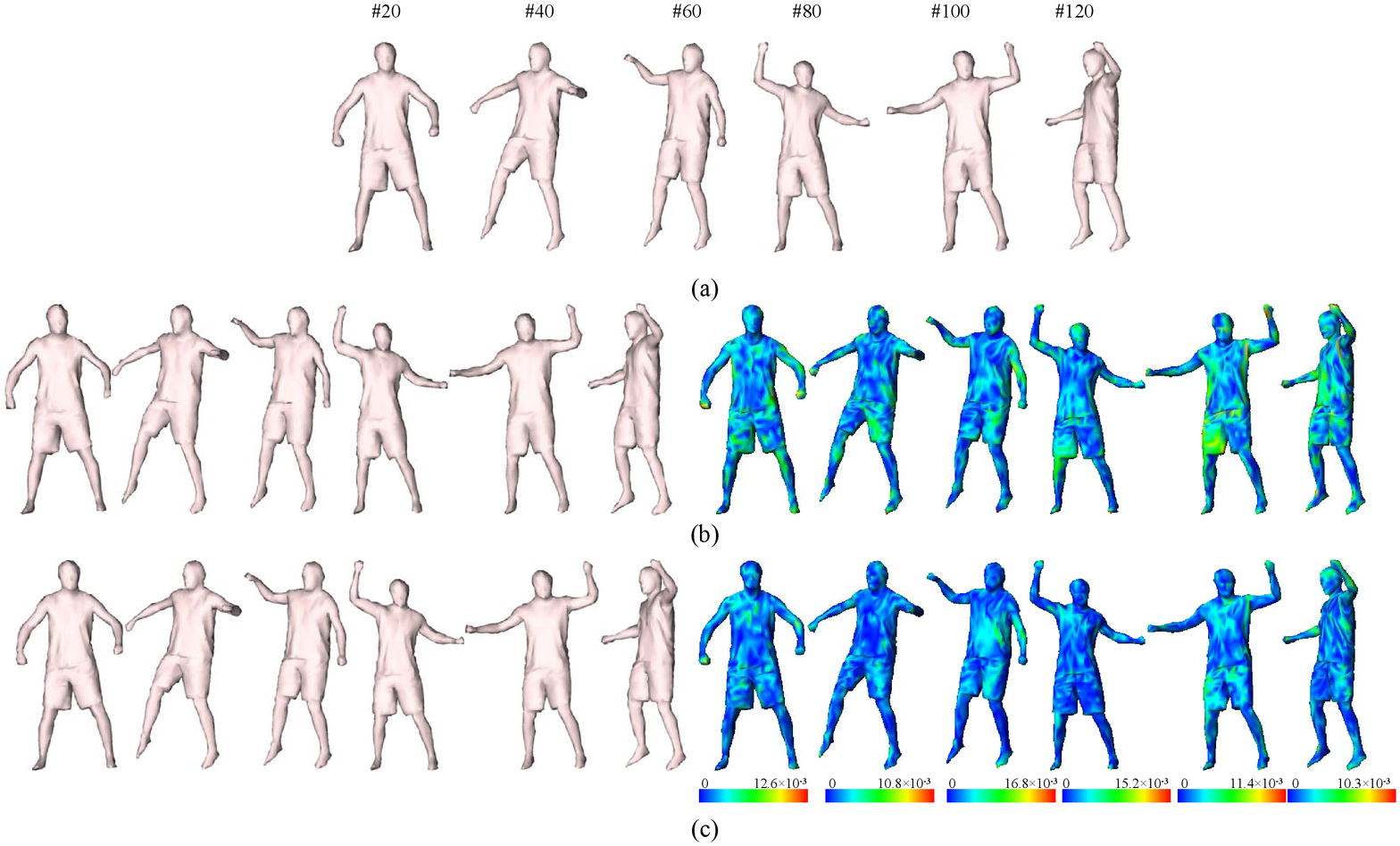}
\caption{Visual results on dynamic meshes. (a) shows six frames of the ``Dance'' dynamic mesh. (b) and (c) show decoded frames with bpfv 0.25 and 0.45, respectively.
The error is measured by RMS \cite{cignoni1998metro} with
respect to the bounding box diagonal and visualized using colormap.} \label{fig:visual result}
\end{figure*}

The forth row of Figure \ref{fig:resultmeshandmocap} shows typical
rate-distortion curves for our scheme as well as V\'{a}\v{s}a's,
where it can be seen that for ``Dance" and ``Skirt", our method
produces much smaller distortion than V\'{a}\v{s}a's at the same
bpfv, especially at relatively small bpfvs. With respect to
``Handstand" and ``Wheel", the rate-distortion curves of
V\'{a}\v{s}a's are not given since V\'{a}\v{s}a's method can only
work on manifold-meshes, but sequences ``Handstand" and ``Wheel"
consists of non-manifold-meshes. Our scheme is independent of
topology, which is an additional advantage. Finally, some visual
results are shown in Figures \ref{fig:visual result} and
\ref{fig:visual result2} to further demonstrate the performance of
our scheme, where we can observe that the decompressed frames are
still close to the original ones and their quality are reasonable
even when the bpfv is equal to 0.25.

\begin{figure*}
\centering
\includegraphics[width=7.0in]{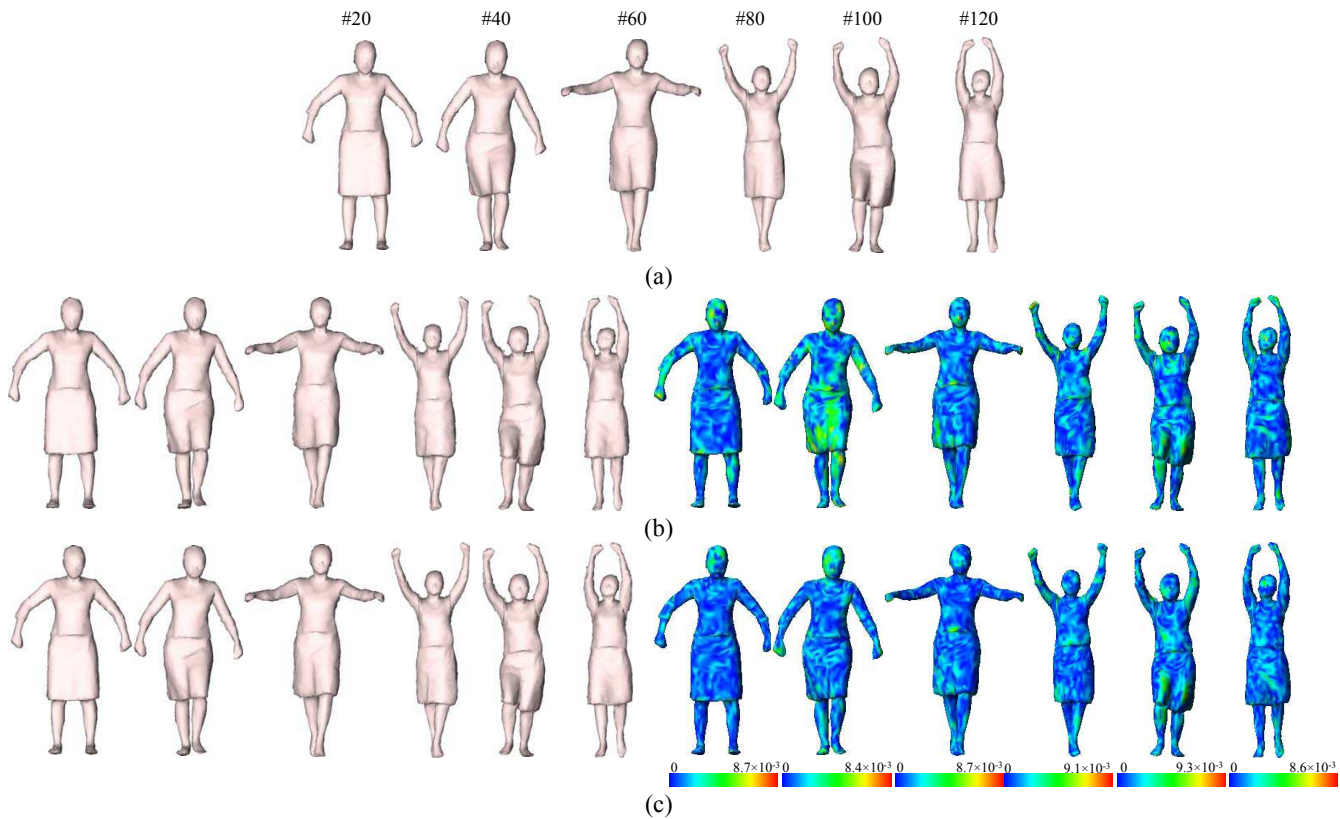}
\caption{Visual results on dynamic meshes. (a) shows six frames of the ``Skirt'' dynamic mesh. (b) and (c) show decoded frames with bpfv 0.25 and 0.45, respectively.
The error is measured by RMS \cite{cignoni1998metro} with
respect to the bounding box diagonal and visualized using colormap.} \label{fig:visual
result2}
\end{figure*}

\section{Conclusion and Future Work}
\label{sec:con}

  We presented sparse low-rank matrix approximation for effective data compression.
  In contrast to the conventional LRMA, SLRMA is able to explore both the intra- and inter-coherence among data \emph{simultaneously},
  producing extremely \emph{sparse} basis functions.
  We formulated the SLRMA problem as a constrained optimization problem and solved it using the inexact augmented Lagrangian multiplier method.
  Although the optimization problem is non-convex, we observed that our method empirically converges well on real-world data, such as image sets and 3D dynamic meshes.
  Also, SLRMA exhibits excellent low-rank approximation performance,
  i.e., at the same rank, comparable approximation error
  as LRMA is produced even when 80\% entries of basis vectors are
  zero.
  Moreover, at the same bitrate, the SLRMA-based compression schemes can reduce
  distortion by up to 53\% for 3D dynamic meshes and improve PSNR by up
  to 3 dB for image sets compared with existing methods.

  In the future, we would like to further investigate the potential of SLRMA to compress other types of data (e.g., EEG signals).
  We also believe that SLRMA can be applied to other applications, such as patch-based image and video denoising,
  where both the similarity among patches (i.e., low-rank characteristic) and spatial structure of patches
  (i.e., the sparseness of patches with respect to particular bases) can be jointly taken into account.

\ifCLASSOPTIONcompsoc
  \section*{Acknowledgments}
\else
  \section*{Acknowledgment}
\fi

The authors would like to thank Dr. Libor V\'{a}\v{s}a for providing
the rate-distortion data shown in Figure
\ref{fig:resultmeshandmocap} and the associate editor and the
anonymous reviewers for their valuable comments.

\ifCLASSOPTIONcaptionsoff
  \newpage
\fi

\bibliographystyle{IEEEtran}
\bibliography{refs}

\begin{thebibliography}{10}
\providecommand{\url}[1]{#1}
\csname url@rmstyle\endcsname
\providecommand{\newblock}{\relax}
\providecommand{\bibinfo}[2]{#2}
\providecommand\BIBentrySTDinterwordspacing{\spaceskip=0pt\relax}
\providecommand\BIBentryALTinterwordstretchfactor{4}
\providecommand\BIBentryALTinterwordspacing{\spaceskip=\fontdimen2\font plus
\BIBentryALTinterwordstretchfactor\fontdimen3\font minus
  \fontdimen4\font\relax}
\providecommand\BIBforeignlanguage[2]{{%
\expandafter\ifx\csname l@#1\endcsname\relax
\typeout{** WARNING: IEEEtran.bst: No hyphenation pattern has been}%
\typeout{** loaded for the language `#1'. Using the pattern for}%
\typeout{** the default language instead.}%
\else
\language=\csname l@#1\endcsname
\fi
#2}}

\bibitem{halko2011finding}
N.~Halko, P.-G. Martinsson, and J.~A. Tropp, ``Finding structure with
  randomness: Probabilistic algorithms for constructing approximate matrix
  decompositions,'' \emph{SIAM review}, vol.~53, no.~2, pp. 217--288, 2011.

\bibitem{gu20122dsvd}
Z.~Gu, W.~Lin, B.-S. Lee, and C.~Lau, ``Low-complexity video coding based on
  two-dimensional singular value decomposition,'' \emph{IEEE Trans. Image
  Processing}, vol.~21, no.~2, pp. 674--687, 2012.

\bibitem{Hou2015compressing}
J.~Hou, L.-P. Chau, N.~Magnenat-Thalmann, and Y.~He, ``Compressing 3-d human
  motions via keyframe-based geometry videos,'' \emph{IEEE Transactions on
  Circuits and Systems for Video Technology}, vol.~25, no.~1, pp. 51--62, Jan
  2015.

\bibitem{candes2011robust}
E.~J. Cand{\`e}s, X.~Li, Y.~Ma, and J.~Wright, ``Robust principal component
  analysis?'' \emph{Journal of the ACM}, vol.~58, no.~3, p.~11, 2011.

\bibitem{wen2014joint}
J.~Wen, Y.~Xu, J.~Tang, Y.~Zhan, Z.~Lai, and X.~Guo, ``Joint video frame set
  division and low-rank decomposition for background subtraction,'' \emph{IEEE
  Trans. Circuits and Systems for Video Technology}, vol.~24, no.~12, pp.
  2034--2048, 2014.

\bibitem{liu2013robust}
G.~Liu, Z.~Lin, S.~Yan, J.~Sun, Y.~Yu, and Y.~Ma, ``Robust recovery of subspace
  structures by low-rank representation,'' \emph{IEEE Trans. Pattern Analysis
  and Machine Intelligence}, vol.~35, no.~1, pp. 171--184, 2013.

\bibitem{zhang2013learning}
Y.~Zhang, Z.~Jiang, and L.~S. Davis, ``Learning structured low-rank
  representations for image classification,'' in \emph{Proc. IEEE Computer
  Vision and Pattern Recognition (CVPR)}.\hskip 1em plus 0.5em minus
  0.4em\relax IEEE, 2013, pp. 676--683.

\bibitem{ji2011robust}
H.~Ji, S.~Huang, Z.~Shen, and Y.~Xu, ``Robust video restoration by joint sparse
  and low rank matrix approximation,'' \emph{SIAM Journal on Imaging Sciences},
  vol.~4, no.~4, pp. 1122--1142, 2011.

\bibitem{wang2013robust}
Z.~Wang, H.~Li, Q.~Ling, and W.~Li, ``Robust temporal-spatial decomposition and
  its applications in video processing,'' \emph{IEEE Trans. Circuits and
  Systems for Video Technology}, vol.~23, no.~3, pp. 387--400, 2013.

\bibitem{peng2012rasl}
Y.~Peng, A.~Ganesh, J.~Wright, W.~Xu, and Y.~Ma, ``Rasl: Robust alignment by
  sparse and low-rank decomposition for linearly correlated images,''
  \emph{IEEE Trans. Pattern Analysis and Machine Intelligence}, vol.~34,
  no.~11, pp. 2233--2246, 2012.

\bibitem{cao2015image}
F.~Cao, M.~Cai, and Y.~Tan, ``Image interpolation via low-rank matrix
  completion and recovery,'' \emph{IEEE Trans. Circuits and Systems for Video
  Technology}, vol.~25, no.~8, pp. 1261--1270, 2015.

\bibitem{dai2013projective}
Y.~Dai, H.~Li, and M.~He, ``Projective multiview structure and motion from
  element-wise factorization,'' \emph{IEEE Trans. Pattern Analysis and Machine
  Intelligence}, vol.~35, no.~9, pp. 2238--2251, 2013.

\bibitem{dai2014simple}
------, ``A simple prior-free method for non-rigid structure-from-motion
  factorization,'' \emph{International Journal of Computer Vision}, vol. 107,
  no.~2, pp. 101--122, 2014.

\bibitem{meng2013cyclic}
D.~Meng, Z.~Xu, L.~Zhang, and J.~Zhao, ``A cyclic weighted median method for l1
  low-rank matrix factorization with missing entries.'' in \emph{Proc. AAAI},
  2013, pp. 1--7.

\bibitem{zhou2014low}
X.~Zhou, C.~Yang, H.~Zhao, and W.~Yu, ``Low-rank modeling and its applications
  in image analysis,'' \emph{ACM Computing Surveys (CSUR)}, vol.~47, no.~2,
  p.~36, 2014.

\bibitem{yang1995combined}
J.-F. Yang and C.-L. Lu, ``Combined techniques of singular value decomposition
  and vector quantization for image coding,'' \emph{IEEE Trans. Image
  Processing}, vol.~4, no.~8, pp. 1141--1146, 1995.

\bibitem{ochoa2003hybrid}
H.~Ochoa and K.~Rao, ``A hybrid dwt-svd image-coding system (hdwtsvd) for
  monochromatic images,'' in \emph{Proc. SPIE Image and Video Communications
  and Processing}, 2003, pp. 1056--1066.

\bibitem{du2007hyperspectral}
Q.~Du and J.~E. Fowler, ``Hyperspectral image compression using jpeg2000 and
  principal component analysis,'' \emph{IEEE Geoscience and Remote Sensing
  Letters}, vol.~4, no.~2, pp. 201--205, 2007.

\bibitem{chen2015incremental}
C.~Chen, J.~Cai, W.~Lin, and G.~Shi, ``Incremental low-rank and sparse
  decomposition for compressing videos captured by fixed cameras,''
  \emph{Journal of Visual Communication and Image Representation}, vol.~26, pp.
  338--348, 2015.

\bibitem{HoufacialGV}
J.~Hou, L.-P. Chau, M.~Zhang, N.~Magnenat-Thalmann, and Y.~He, ``A highly
  efficient compression framework for time-varying 3-d facial expressions,''
  \emph{IEEE Trans. Circuits and Systems for Video Technology}, vol.~24, no.~9,
  pp. 1541--1553, Sept 2014.

\bibitem{alexa2000representing}
M.~Alexa and W.~M{\"u}ller, ``Representing animations by principal
  components,'' \emph{Computer Graphics Forum}, vol.~19, no.~3, pp. 411--418,
  2000.

\bibitem{karni2004compression}
Z.~Karni and C.~Gotsman, ``Compression of soft-body animation sequences,''
  \emph{Computers \& Graphics}, vol.~28, no.~1, pp. 25--34, 2004.

\bibitem{vavsa2010geometry}
L.~V{\'a}{\v{s}}a and V.~Skala, ``Geometry-driven local neighbourhood based
  predictors for dynamic mesh compression,'' \emph{Computer Graphics Forum},
  vol.~29, no.~6, pp. 1921--1933, 2010.

\bibitem{vavsa2014dynamiccompressing}
L.~V{\'a}{\v{s}}a, S.~Marras, K.~Hormann, and G.~Brunnett, ``Compressing
  dynamic meshes with geometric laplacians,'' \emph{Computer Graphics Forum},
  vol.~33, no.~2, pp. 145--154, 2014.

\bibitem{Hou2014tvcg}
J.~Hou, L.~Chau, N.~Magnenat-Thalmann, and Y.~He, ``Human motion capture data
  tailored transform coding,'' \emph{IEEE Trans. Visualization and Computer
  Graphics}, vol.~21, no.~7, pp. 848--859, 2015.

\bibitem{li2007flow}
Q.~Li, H.~Jianming, and Z.~Yi, ``A flow volumes data compression approach for
  traffic network based on principal component analysis,'' in \emph{Proc. IEEE
  Intelligent Transportation Systems Conference (ITSC)}, 2007, pp. 125--130.

\bibitem{asif2013data}
M.~T. Asif, S.~Kannan, J.~Dauwels, and P.~Jaillet, ``Data compression
  techniques for urban traffic data,'' in \emph{Proc. IEEE Symposium on
  Computational Intelligence in Vehicles and Transportation Systems (CIVTS)},
  2013, pp. 44--49.

\bibitem{Asif2014near}
M.~Asif, K.~Srinivasan, N.~Mitrovic, J.~Dauwels, and P.~Jaillet,
  ``Near-lossless compression for large traffic networks,'' \emph{IEEE Trans.
  Intelligent Transportation Systems}, vol.~PP, no.~99, pp. 1--12, 2014.

\bibitem{skodras2001jpeg}
A.~Skodras, C.~Christopoulos, and T.~Ebrahimi, ``The jpeg 2000 still image
  compression standard,'' \emph{IEEE Signal Processing Magazine}, vol.~18,
  no.~5, pp. 36--58, 2001.

\bibitem{H264overview}
T.~Wiegand, G.~Sullivan, G.~Bjontegaard, and A.~Luthra, ``Overview of the
  h.264/avc video coding standard,'' \emph{IEEE Trans. Circuits and Systems for
  Video Technology}, vol.~13, no.~7, pp. 560--576, July 2003.

\bibitem{heu2009snr}
J.-H. Heu, C.-S. Kim, and S.-U. Lee, ``Snr and temporal scalable coding of 3-d
  mesh sequences using singular value decomposition,'' \emph{Journal of Visual
  Communication and Image Representation}, vol.~20, no.~7, pp. 439--449, 2009.

\bibitem{sattler2005simple}
M.~Sattler, R.~Sarlette, and R.~Klein, ``Simple and efficient compression of
  animation sequences,'' in \emph{Proc. ACM SIGGRAPH/Eurographics SCA}, 2005,
  pp. 209--217.

\bibitem{libor2007coddyac}
L.~V{\'a}{\v{s}}a and V.~Skala, ``Coddyac: Connectivity driven dynamic mesh
  compression,'' in \emph{Proc. IEEE 3DTV}, 2007, pp. 1--4.

\bibitem{vavsa2009cobra}
------, ``Cobra: Compression of the basis for pca represented animations,''
  \emph{Computer Graphics Forum}, vol.~28, no.~6, pp. 1529--1540, 2009.

\bibitem{kokiopoulou2011trace}
E.~Kokiopoulou, J.~Chen, and Y.~Saad, ``Trace optimization and eigenproblems in
  dimension reduction methods,'' \emph{Numerical Linear Algebra with
  Applications}, vol.~18, no.~3, pp. 565--602, 2011.

\bibitem{markovsky2012low}
I.~Markovsky and K.~Usevich, \emph{Low rank approximation}.\hskip 1em plus
  0.5em minus 0.4em\relax Springer, 2012.

\bibitem{zhang2010spectral}
H.~Zhang, O.~Van~Kaick, and R.~Dyer, ``Spectral mesh processing,''
  \emph{Computer Graphics Forum}, vol.~29, no.~6, pp. 1865--1894, 2010.

\bibitem{zhang2013analyzing}
C.~Zhang and D.~Flor{\^e}ncio, ``Analyzing the optimality of predictive
  transform coding using graph-based models,'' \emph{IEEE Signal Processing
  Letters}, vol.~20, no.~1, pp. 106--109, 2013.

\bibitem{wang2012introduction}
R.~Wang, \emph{Introduction to orthogonal transforms: with applications in data
  processing and analysis}.\hskip 1em plus 0.5em minus 0.4em\relax Cambridge
  University Press, 2012.

\bibitem{lin2010augmented}
Z.~Lin, M.~Chen, and Y.~Ma, ``The augmented lagrange multiplier method for
  exact recovery of corrupted low-rank matrices,'' \emph{arXiv preprint
  arXiv:1009.5055}, 2010.

\bibitem{lai2014splitting}
R.~Lai and S.~Osher, ``A splitting method for orthogonality constrained
  problems,'' \emph{Journal of Scientific Computing}, vol.~58, no.~2, pp.
  431--449, 2014.

\bibitem{martinez1998ar}
A.~M. Martinez, ``The ar face database,'' \emph{CVC Technical Report}, vol.~24,
  1998.

\bibitem{phillips2000feret}
P.~J. Phillips, H.~Moon, S.~A. Rizvi, and P.~J. Rauss, ``The feret evaluation
  methodology for face-recognition algorithms,'' \emph{IEEE Trans. Pattern
  Analysis and Machine Intelligence}, vol.~22, no.~10, pp. 1090--1104, 2000.

\bibitem{rao2000transform}
K.~R. Rao and P.~C. Yip, \emph{The transform and data compression
  handbook}.\hskip 1em plus 0.5em minus 0.4em\relax CRC press, 2000.

\bibitem{marpe2003context}
D.~Marpe, H.~Schwarz, and T.~Wiegand, ``Context-based adaptive binary
  arithmetic coding in the h. 264/avc video compression standard,'' \emph{IEEE
  Trans. Circuits and Systems for Video Technology}, vol.~13, no.~7, pp.
  620--636, 2003.

\bibitem{dynamicmeshdataset}
J.~Gall, C.~Stoll, E.~De~Aguiar, C.~Theobalt, B.~Rosenhahn, and H.-P. Seidel,
  ``Motion capture using joint skeleton tracking and surface estimation,'' in
  \emph{Proc. IEEE CVPR}, 2009, pp. 1746--1753.

\bibitem{petvrik2010finding}
O.~Pet{\v{r}}{\'\i}k and L.~V{\'a}{\v{s}}a, ``Finding optimal parameter
  configuration for a dynamic triangle mesh compressor,'' in \emph{Proc.
  Articulated Motion and Deformable Objects}.\hskip 1em plus 0.5em minus
  0.4em\relax Springer, 2010, pp. 31--42.

\bibitem{cignoni1998metro}
P.~Cignoni, C.~Rocchini, and R.~Scopigno, ``Metro: measuring error on
  simplified surfaces,'' \emph{Computer Graphics Forum}, vol.~17, no.~2, pp.
  167--174, 1998.

\end{thebibliography}

\end{document}